\documentclass[preprint,3p,times]{elsarticle}

\usepackage{amsmath,amssymb,amsthm,amsfonts}
\usepackage{mathabx}
\usepackage{mathrsfs}
\usepackage{bm}
\usepackage{graphicx}
\usepackage{float}
\usepackage{subfig}
\usepackage{color,xcolor}
\usepackage{tabularx}
\usepackage{microtype}
\biboptions{numbers,sort&compress}
\usepackage{epsfig,psfrag}
\usepackage{tikz}
\usepackage{todonotes}
\usepackage{pgfplots}
\pgfplotsset{compat=1.18}
\usepackage{pgfplotstable}
\makeatletter
\def\ds@nanUC{NaN}
\def\ds@nanlc{nan}
\newcommand{\relcell}[1]{%
	\edef\ds@cell{#1}%
	\ifx\ds@cell\ds@nanUC $\sim$\else
	\ifx\ds@cell\ds@nanlc $\sim$\else
		\pgfmathprintnumber[fixed,fixed zerofill,precision=2]{#1}%
	\fi\fi}
\makeatother
\definecolor{matlabblue}{RGB}{0,114,189}
\pgfplotsset{colormap={shine}{%
	rgb(0cm)=(0.20,0.20,0.21)
	rgb(3cm)=(0.39,0.39,0.41)
	rgb(6cm)=(0.60,0.60,0.62)
	rgb(8cm)=(0.78,0.78,0.80)
	rgb(9.3cm)=(0.90,0.90,0.91)
	rgb(10cm)=(0.99,0.99,1.00)}}
\usepackage{todonotes}
\usepackage{comment}
\usepackage{graphicx,psfrag}
\usepackage{hyperref}
\usepackage[nameinlink,capitalise]{cleveref}
\usepackage{xspace}
\usepackage{enumitem}
\usepackage{booktabs}

\usetikzlibrary{shapes}
\usetikzlibrary{arrows.meta}

\setcitestyle{authoryear,round,semicolon}

\usepackage[commandnameprefix=always,commentmarkup=footnote]{changes}
\makeatletter
\@namedef{Changes@AuthorColor}{blue}
\colorlet{Changes@Color}{blue}
\makeatother

\makeatletter
\hypersetup{colorlinks=true}
\AtBeginDocument{\@ifpackageloaded{hyperref}
	{\def\@linkcolor{magenta}
		\def\@anchorcolor{black}
		\def\@citecolor{teal}
		\def\@filecolor{cyan}
		\def\@urlcolor{magenta}
		\def\@menucolor{red}
		\def\@pagecolor{cyan}
		\begingroup
		\@makeother\`%
		\@makeother\=%
		\edef\x{%
			\edef\noexpand\x{%
				\endgroup
				\noexpand\toks@{%
					\catcode 96=\noexpand\the\catcode`\noexpand\`\relax
					\catcode 61=\noexpand\the\catcode`\noexpand\=\relax
				}%
			}%
			\noexpand\x
		}%
		\x
		\@makeother\`
		\@makeother\=
	}{}}
\makeatother

\makeatletter \oddsidemargin 0.0cm \evensidemargin 0.0cm
\newcommand{\be}{\begin{equation}}
\newcommand{\en}{\end{equation}}

\def\bm#1{\mbox{\boldmath{$#1$}}}

\numberwithin{equation}{section}
\theoremstyle{plain}
\newtheorem{theorem}{Theorem}[section]
\newtheorem{theorem*}{Theorem}

\theoremstyle{definition}

\newtheorem{remark}[theorem]{Remark}

\newcommand{\rd}{\mathrm{d}}
\newcommand{\ri}{\mathrm{i}}

\DeclareMathOperator{\tr}{tr}
\DeclareMathOperator{\Div}{Div}
\DeclareMathOperator{\ddiv}{div}

\DeclareMathOperator{\cof}{cof}

\usetikzlibrary{external}
\newif\iffigconvert
\ifdefined\REBUILDFIGS
	\ifdefined\pdfshellescape
		\ifnum\pdfshellescape=1 \figconverttrue \fi
	\fi
\fi
\iffigconvert
	\tikzset{external/system call={lualatex \tikzexternalcheckshellescape
		-halt-on-error -interaction=batchmode -jobname "\image" "\texsource"}}
\else
\fi

\newlength{\panelhalf}
\newcommand{\rowlab}[1]{\raisebox{\panelhalf}[0pt][0pt]{\footnotesize$\xi = #1$}}

\newsavebox{\scaleaxisbox}

\newsavebox{\keyshapebox}
\newsavebox{\keycurvebox}

\newcommand{\bifref}{\cref{fig:bif_page}\xspace}

\newif\ifrevmarks

\newenvironment{fixblock}{\ifrevmarks\color{blue}\fi}{}

\begin{document}
\newcommand{\beq}{\begin{equation}}
\newcommand{\beql}[1]{\begin{equation}\label{#1}}
\newcommand{\eeq}{\end{equation}}
\newcommand{\bea}{\begin{eqnarray}}
\newcommand{\beal}[1]{\begin{eqnarray}\label{#1}}
\newcommand{\eea}{\end{eqnarray}}
\newcommand{\bean}{\begin{eqnarray*}}
\newcommand{\eean}{\end{eqnarray*}}

\newcommand{\del}{\delta}

\newcommand{\mbf}[1]{\mathbf{#1}}
\newcommand{\sbf}[1]{\boldsymbol{#1}}

\newcommand{\intL}{\int_{\cal L}}

\newcommand{\beps}{\sbf{\varepsilon}}
\newcommand{\bsig}{\sbf{\sigma}}
\newcommand{\blam}{\sbf{\lambda}}
\newcommand{\bmu}{\sbf{\mu}}
\newcommand{\ba}{\mbf{a}}
\newcommand{\bi}{\mbf{i}}
\newcommand{\bB}{\mbf{B}}
\newcommand{\bS}{\mbf{S}}
\newcommand{\bV}{\mbf{V}}
\newcommand{\bA}{\mbf{A}}
\newcommand{\bE}{\mbf{E}}
\newcommand{\bU}{\mbf{U}}
\newcommand{\bR}{\mbf{R}}
\newcommand{\bX}{\mbf{X}}
\newcommand{\bK}{\mbf{K}}
\newcommand{\bd}{\mbf{d}}
\newcommand{\bs}{\mbf{s}}
\newcommand{\bff}{\mbf{f}}
\newcommand{\bF}{\mbf{F}}
\newcommand{\bg}{\mbf{g}}
\newcommand{\bG}{\mbf{G}}
\newcommand{\bL}{\mbf{L}}
\newcommand{\bM}{\mbf{M}}
\newcommand{\bH}{\mbf{H}}
\newcommand{\bkron}{\mbf{1}}
\newcommand{\sigm}{\sigma_{\rm m}}
\newcommand{\epsm}{\varepsilon_{\rm m}}
\newcommand{\bI}{\mbf{I}}
\newcommand{\bIv}{\mbf{I}_{\rm K}}
\newcommand{\bId}{\mbf{I}_{\rm D}}
\newcommand{\bIs}{\mbf{I}_{\rm S}}
\newcommand{\bCe}{\mbf{C}_{\rm e}}
\newcommand{\bDe}{\mbf{D}_{\rm e}}
\newcommand{\bC}{\mbf{C}}
\newcommand{\bD}{\mbf{D}}
\newcommand{\bu}{\mbf{u}}
\newcommand{\bv}{\mbf{v}}
\newcommand{\bw}{\mbf{w}}
\newcommand{\bz}{\mbf{z}}
\newcommand{\bT}{\mbf{T}}
\newcommand{\bt}{\mbf{t}}
\newcommand{\bn}{\mbf{n}}
\newcommand{\bN}{\mbf{N}}
\newcommand{\bx}{\mbf{x}}
\newcommand{\bq}{\mbf{q}}
\newcommand{\bh}{\mbf{h}}

\newcommand{\cE}{{\cal E}}
\newcommand{\cH}{{\cal H}}
\newcommand{\cA}{{\cal A}}
\newcommand{\cD}{{\cal D}}

\newcommand{\G}{{\rm I\!G}}

\newcommand{\bnab}{\sbf{\nabla}}
\newcommand{\bnabs}{\sbf{\nabla}_{\rm s}}

\newcommand{\half}{\mbox{$\frac 1 2$}}
\newcommand{\dV}{\,\mbox{d}V}
\newcommand{\dx}{\,\mbox{d}x}
\newcommand{\dxi}{\,\mbox{d}\xi}
\newcommand{\deta}{\,{\rm d}\eta}
\newcommand{\dS}{\,\mbox{d}S}
\newcommand{\ds}{\,\mbox{d}s}
\newcommand{\dA}{\,\mbox{d}A}
\newcommand{\dt}{\,\mbox{d}t}
\newcommand{\pard}[2]{\frac{\partial #1}{\partial #2}}
\newcommand{\parder}[2]{\frac{\partial #1}{\partial #2}}

\begin{frontmatter}

\title{Finite-strain second-gradient elasticity with an energetic surface:
existence of minimizers, incremental theory and elastocapillary beading}

\author[mysecondaryaddress]{Michal \v{S}mejkal\corref{cor1}}
\ead{michal.smejkal@fsv.cvut.cz}
\cortext[cor1]{Corresponding author}

\author[mysecondaryaddress,myteritaryaddress]{Martin Hor\'{a}k}
\ead{martin.horak@cvut.cz}

\author[myteritaryaddress,mysecondaryaddress]{Martin Kru\v{z}\'{i}k}
\ead{kruzik@utia.cas.cz, kruzik@fsv.cvut.cz}

\address[mysecondaryaddress]{Czech Technical University in Prague, Faculty of Civil Engineering, Th\'{a}kurova 2077/7, 166 29 Prague 6, Czechia}

\address[myteritaryaddress]{Czech Academy of Sciences, Institute of Information Theory and Automation, Pod vod\'{a}renskou v\v{e}\v{z}\'{i} 4, 182 00 Prague 8, Czechia}

\begin{abstract}
We develop a finite-strain theory of second-gradient elasticity for solids with a material surface endowed with its own constitutive energy. The bulk second gradient plays two complementary roles. First, under suitable coercivity assumptions and convexity with respect to the second gradient, we prove the existence of minimizers for nonnegative, lower semicontinuous surface energy densities, without requiring convexity, polyconvexity, positive one-homogeneity, or an upper growth bound on the surface energy. Second, it introduces an intrinsic bulk length scale that affects the onset of instability and wavelength selection. To investigate these effects, we derive an incremental small-on-large formulation in the finitely deformed configuration and apply it to the elastocapillary beading of an axially stretched soft cylinder. For an infinite cylinder and the parameters considered, increasing the second-gradient modulus delays bifurcation and shifts the critical mode towards longer wavelengths. Above a finite threshold of this modulus, the critical stretch saturates in the long-wave limit.

The model is implemented in the open-source finite element code OOFEM using $C^1$-continuous isogeometric analysis. Finite-cylinder simulations yield onset stretches in close agreement with the analytical predictions. Selected post-bifurcation branches traced by arc-length continuation exhibit beaded states and supercritical bifurcation at small second-gradient moduli, whereas larger moduli are associated with subcritical bifurcation and the formation of distinct thick and thin segments.

\end{abstract}


\end{frontmatter}

\section{Introduction}

The mechanical response of soft materials, including biological tissues, hydrogels, and synthetic polymers, is governed not only by bulk elasticity but also, often strongly, by surface stresses \citep{style2017elastocapillarity}. Surface effects become particularly important when a characteristic dimension of the body is comparable to the elastocapillary length, defined as the ratio of a characteristic surface stress to a bulk elastic modulus. Depending on the convention, this modulus is Young's modulus \citep{style2017elastocapillarity} or the shear modulus \citep{mora2010capillarity}. The simplest surface-tension model assumes a constant, isotropic surface stress, as at a liquid interface. More generally, the continuum theory of \citet{gurtin1975continuum,gurtin1978surface} treats the surface as a zero-thickness elastic membrane with its own constitutive response, perfectly bonded to the underlying bulk. Surface stress may then depend on surface strain through the Shuttleworth effect \citep{shuttleworth1950surface}, demonstrated experimentally in soft gels \citep{xu2017direct}. The interplay between bulk and surface mechanics can give rise to instabilities. A canonical example is elastocapillary beading of a soft cylinder, a solid analogue of the classical Plateau--Rayleigh instability in fluids. \citet{mora2010capillarity} demonstrated experimentally that soft gel cylinders undergo beading when capillary forces overcome the elastic resistance of the bulk. Subsequent work extended the analysis to prestressed cylindrical structures, showing how axial stretch shifts the critical bifurcation threshold \citep{ciarletta2012peristaltic,taffetani2015beading,fu2021necking}. Weakly nonlinear and energetic analyses show that the beading transition can be subcritical, with homogeneous and bifurcated states coexisting over a finite range of loading, and can be interpreted as phase separation between thin and thick states \citep{taffetani2015beading,taffetani2015elastocapillary,xuan2017plateau,fu2021necking}. Furthermore, the constitutive assumptions of the surface model---such as area incompressibility, curvature dependence, or negative surface tension---can also qualitatively change the instability \citep{bakiler2023surface,taffetani2024curvature,ru2023circumferential}. Incremental equations for curvature-dependent surface elasticity have also been derived recently \citep{YU2026104577}.

The present work combines finite-strain second-gradient bulk elasticity with an independently prescribed surface energy to address two questions: how the intrinsic bulk length scale affects elastocapillary instability and how the additional bulk regularity changes the conditions for the existence of minimizers. Classical first-gradient elasticity contains no intrinsic bulk material length scale and cannot, by itself, capture bulk size effects associated with strongly nonuniform deformation. Higher-order theories originating from the works of \citet{toupin1962elastic} and \citet{mindlin1964microstructure} address this limitation by allowing the bulk energy to depend on higher spatial derivatives of the deformation \citep{mindlin1968first,askes2011gradient}. In particular, second-gradient hyperelasticity introduces such a length scale by penalizing spatial variations of the deformation gradient. This mechanism is especially relevant to instability problems, where modes of different wavelengths may compete and the intrinsic length scale can therefore influence mode selection.

Recent finite-strain computations by \citet{wang2025size} illustrate this role of strain-gradient elasticity in highly deformable materials. Their analyses of buckling, wrinkling, and period-doubling instabilities showed that size effects become significant when structural dimensions are comparable to the intrinsic material length scale. Increasing the strain-gradient contribution can delay the onset of instability, alter mode transitions and post-critical behavior, and even suppress particular instability modes. These findings demonstrate the physical role of higher-gradient terms in instability and pattern selection, beyond their use for numerical regularization. Here, we examine the role of bulk strain gradients in an elastocapillary instability driven by surface stresses. Specifically, we investigate how the second-gradient modulus affects the bifurcation spectrum and the critical wavelength.

Combining second-gradient elasticity with a surface energy requires distinguishing independent surface constitutive contributions from boundary terms generated by the bulk energy. Both occur in the present formulation: the bulk energy depends on the deformation gradient and its gradient, while the independently prescribed surface energy depends on the surface deformation gradient. Variation of the bulk energy also produces higher-order boundary terms through integration by parts; these are determined by the bulk constitutive law. Connections between higher-gradient bulk elasticity and surface effects were explored in \citet{casal1961capillarite} and \citet{mindlin1965second}, and subsequently by Vardoulakis, Exadaktylos, and Aifantis \citep{vardoulakis1996gradient,exadaktylos1998surface}. In particular, \citet{exadaktylos1998surface} extended Biot's stability analysis of a prestressed elastic half-space \citep{biot1963surface} to gradient elasticity. Although the resulting wrinkling-type instability differs from the cylindrical beading mode studied here, it provides a precedent for the influence of higher-gradient bulk elasticity on surface instability.

By contrast, the present formulation follows the theory of material surfaces of \citet{gurtin1975continuum,gurtin1978surface}, in which the surface is endowed with its own constitutive response and material parameters. Geometrically nonlinear formulations of such energetic surfaces are well established \citep{javili2013thermomechanics,javili2018variational,javili2018aspects}. Moreover, \citet{javili2013higher} developed a higher-gradient framework with independent bulk, surface, and edge energies, and related the admissible boundary kinematics to the gradient order in the bulk. When the bulk energy depends on \(\bm H=\nabla\bm F\), the surface energy may depend on the \emph{full} deformation gradient at the surface and not only on its tangential projection, i.e., including its action in the normal direction. Here, we retain the tangential dependence, that is, the surface energy keeps the Gurtin--Murdoch form \(\varPhi(\bm F_s)\) with \(\bm F_s=\bm F\bm I_s\), which a first-gradient bulk already admits. The second gradient supplies an intrinsic bulk length scale and the regularity needed for the existence result in \cref{sec:existence}. Since the bulk and surface energies are prescribed independently, their material parameters can be varied separately to distinguish their effects on instability.

On the variational side, a classical existence framework for finite-strain second-gradient elasticity combines suitable coercivity assumptions with convexity in the highest-order deformation gradient \citep[see, e.g.,][]{kruzik2019mathematical}. This requirement can be relaxed within more general notions of higher-order polyconvexity, as introduced by \citet{ballcurrieolver1981} and subsequently employed in nonlinear higher-gradient elasticity \citep{healey2009injective}. In the present setting, we adopt the simpler assumption of convexity with respect to the second deformation gradient $\bm H$, while imposing neither convexity nor polyconvexity with respect to the first deformation gradient $\bm F$. A conceptually different regularization is provided by gradient polyconvexity, in which the stored energy depends on gradients of selected minors of $\bm F$, rather than on the full second deformation gradient $\nabla\bm F$ \citep{benesova2018note,horak2020gradient}.

The additional regularity provided by the second-gradient bulk is particularly useful when an independent surface energy is present. In a first-gradient theory, bulk energy alone generally does not provide a boundary trace of \(\bm F\), so the surface term requires suitable surface regularity or additional constitutive structure. One approach is surface polyconvexity, originating in the interface polyconvexity of \citet{silhavy2010phase,silhavy2011equilibrium} and used for energetic surfaces in our previous work \citep{HorakSurface25}. In that measure-theoretic formulation, convexity and positive one-homogeneity underpin the lower semicontinuity of the surface contribution. In the present second-gradient setting, existence can be established without these structural restrictions on the surface energy. Under the assumptions of our existence theorem, bounded-energy sequences have subsequences whose boundary traces of \(\bm F\) converge strongly in \(L^1\), and hence almost everywhere along a further subsequence. For nonnegative, lower semicontinuous surface densities, this gives the lower semicontinuity needed for the direct method. No surface convexity, polyconvexity, positive one-homogeneity, or upper growth bound is required, provided the admissible class contains a deformation of finite total energy. The second-gradient bulk thus serves two complementary purposes: it introduces a physical length scale that affects instability and wavelength selection, and it provides the boundary compactness needed to accommodate a broader class of independently prescribed surface energies.

This framework is developed at the levels of variational formulation, existence, incremental theory, and nonlinear computation, and the paper is organized as follows. \Cref{sec:2} introduces the bulk and surface kinematics, while \cref{sec:VarFormulation} derives the equilibrium equations together with the associated higher-order bulk, surface, and edge conditions. \Cref{sec:existence} establishes existence of minimizers and discusses what its assumptions require of the constitutive laws. The incremental small-on-large formulation is developed in \cref{sec:incremental} and applied to an axially stretched cylinder in \cref{sec:app}. \Cref{sec:results} examines the effect of the second-gradient modulus on the bifurcation threshold and critical wavelength, and \cref{sec:num} compares finite-cylinder computations with the infinite-cylinder predictions and follows selected post-bifurcation branches by arc-length continuation. Finally, \cref{sec:con} summarizes the conclusions. In addition, the push-forward derivation and intermediate cylinder calculations are given for convenience in \ref{app:pushforward} and \ref{app:cylinder}, respectively.

\section{Kinematics}\label{sec:2}
The geometric and kinematic notation employed in what follows is collected in this section; fuller expositions are available in \cite{HorakSurface25}, \cite{YU2026104577}, and \cite{vsilhavy2013direct}.

Let the reference configuration occupy a bounded domain $\Omega\subset\mathbb{R}^3$ with piecewise-smooth boundary $\Gamma=\partial\Omega$. Surface energy is carried by a portion $S\subseteq\Gamma$. We partition $\Gamma$ into finitely many smooth faces $\Gamma_\ell$, with the index $\ell$ labeling the individual faces. The partition is chosen so that $S$ is a union of faces, up to their boundary curves. This may require subdividing an otherwise smooth part of $\Gamma$ along the rim of $S$.

Let $\mathcal{C}$ denote the union of the piecewise-smooth curves separating adjoining faces. It includes the geometric edges of the body, the rim of $S$, and any additional subdivision curves. Away from vertices, exactly two faces meet at each point of $\mathcal{C}$. The vertices have zero arc-length measure and therefore do not contribute to line integrals.

Boundary fields, such as the outward unit normal $\bm{N}$ and the tangential projector $\bm{I}_s$, are defined on each face separately and may jump across $\mathcal{C}$. On $\mathcal{C}$, $[\,\cdot\,]_\ell$ denotes the one-sided limit from $\Gamma_\ell$, whereas bulk fields remain single-valued. At a given point of $\mathcal{C}$ away from vertices, let $\mathcal{F}$ denote the set of indices of the two adjoining faces, and let $\mathcal{F}_S\subseteq\mathcal{F}$ identify those carrying surface energy.

Material points of the reference configuration are labeled by their position $\bm{X}$. Under a deformation  $\bm{\phi}:{\Omega} \to  \mathbb{R}^3$, the material point $\bm{X}$ is taken to a point $\bm{y}$ of the deformed configuration  $\bm{\phi}(\Omega)$,
\begin{align}
\bm{y}=\bm{\phi}(\bm{X}),\quad \bm{X}\in\Omega.
\end{align}
For the classical kinematic, equilibrium and incremental formulas, $\bm{\phi}$ is assumed sufficiently smooth, injective, and orientation-preserving. It is assumed further that $\bm{\phi}$, together with its derivatives, extends continuously to the closure $\overline{\Omega}$, with $\det\nabla\bm{\phi}>0$ there; no notational distinction is drawn between $\bm{\phi}$ and that extension. Restriction of $\bm{\phi}$ to the boundary furnishes the surface deformation  ${\bm{\phi}}_s:\Gamma \to \bm{\phi}(\Gamma)$,
\begin{align}
{\bm{\phi}}_s(\bm{X}) = {\bm{\phi}}(\bm{X}), \quad \bm{X} \in \Gamma.
\end{align}

\subsection{Bulk deformation}  
In the bulk, the gradient of the deformation is
\begin{align}
\bm{F} = \nabla {\bm{\phi}}=\frac{\partial\bm{y}}{\partial\bm{X}}, 
\end{align}
the operator $\nabla$ being the gradient relative to the reference coordinates. This tensor transports infinitesimal vectors of the reference configuration into the deformed one, while the change of volume is controlled by the Jacobian determinant $J = \det(\bm{F})$; the reference element $\mathrm{d}V$ and the current element $\mathrm{d}v$ are tied by $\mathrm{d}v = J \,\mathrm{d}V$. Additionally, preservation of orientation imposes $J > 0$.

With line and volume elements accounted for, there remains the transport of area elements. This is the content of Nanson's formula, which ties the oriented reference element $\bm{N}\,\mathrm{d}A$ to its deformed counterpart $\bm{n}\,\mathrm{d}a$ through
\begin{align}\label{eq:cof}
\bm{n} \,\rd a = \cof(\bm{F})\bm{N}\,\rd A, 
\end{align}
where the unit outward normals of the reference and the current configuration are written $\bm{N}$ and $\bm{n}$, respectively. As for the cofactor tensor of $\bm{F}$, it is characterized by the universal property
\begin{align}\label{eq:cofF}
\bm{F}\bm{u}\times\bm{F}\bm{v}=\cof(\bm{F})(\bm{u}\times\bm{v}),\qquad \forall\ \bm{u},\bm{v}\in\mathbb{R}^3,
\end{align}
with $\times $ denoting the cross product; whenever $\bm{F}$ is invertible it possesses the explicit representation $\cof(\bm{F})= J \bm{F}^{-T}$.

A higher-order measure of the bulk deformation is provided by the second gradient of the deformation map,
\begin{equation}
\bm{H} = \nabla^2 \bm{\phi} = \nabla ( \nabla \bm{\phi}).
\end{equation}

\subsection{Surface deformation}\label{surfDef}
Deformation measures attached to the boundary presuppose a notion of differentiation for fields that live on the boundary surface alone. We therefore begin from the differential of a mapping on a manifold.

Consider a tensor field $\bm{D}_s: \Gamma \to V$ on the boundary surface $\Gamma$ with values in a finite-dimensional vector space $V$. The cases required below are $V = \mathbb{R}$, $\mathbb{R}^3$ and $\mathbb{R}^3 \otimes \mathbb{R}^3$, for which $\bm{D}_s$ is a scalar, a vector, or a second-order tensor field, respectively. At a point $\bm{X}$ the \emph{surface gradient} of $\bm{D}_s$ is that linear transformation $\nabla_s\bm{D}_s(\bm{X}): \mathbb{R}^3 \to V$ for which
\begin{equation}
\lim_{\bm{Y}\to \bm{X},{\bm{Y}\in \Gamma} }
\frac{| \bm{D}_s(\bm{Y}) - \bm{D}_s(\bm{X}) - \nabla_s \bm{D}_s(\bm{X})(\bm{Y}-\bm{X}) |}{|\bm{Y}-\bm{X}|} = 0,
\end{equation}
as well as 
\begin{equation}
\nabla_s \bm{D}_s(\bm{X}) \bm{I}_s = \nabla_s \bm{D}_s(\bm{X}),
\label{eq26}
\end{equation}
with $\bm{I}_s = \bm{I} - \bm{N}\otimes \bm{N}$ the orthogonal projector onto $T_{\boldsymbol{X}}\Gamma$, the tangent space at $\bm{X}$ of the surface $\Gamma$, while $\bm{I}$ stands for the identity tensor in $\mathbb{R}^3$ and $|\cdot|$ for the Euclidean norm. Of the two requirements, the first carries over the Fr\'echet derivative to the manifold setting; the second widens the domain of the operator from $T_{\boldsymbol{X}}\Gamma$ to all of $\mathbb{R}^3$, the widening being fixed by annihilation of the normal direction, $\nabla_s \bm{D}_s(\bm{X})\bm{N}=0$. The convention adopted is that of \cite{vsilhavy2013direct}, and it departs slightly from the classical treatment of \cite{gurtin1975continuum}, in which the surface gradient is declared on the tangent plane alone. 
The advantage is that $\nabla_s$ acts as a linear map on one and the same space $\mathbb{R}^3$ at every point of $\Gamma$, so that expressions may be combined without regard for the local tangent plane. Should $\bm{D}_s$ arise as the trace of a bulk field $\bm{D}:\Omega \to V$, its surface gradient is obtained from the bulk gradient by tangential projection,
\begin{equation}
\nabla_s \bm{D}_s =(\nabla \bm{D})\bm{I}_s.
\end{equation}

A companion operator is the \emph{surface divergence}. For a tensor field $\bm{R}_s:\Gamma \to V \otimes \mathbb{R}^3$ on the boundary, with $V=\mathbb{R}$ or $\mathbb{R}^3$, it is the unique field $\Div_s(\bm{R}_s):\Gamma\to V$ for which
\begin{equation}
\Div_s (\bm R_s)\cdot\bm{v}=\tr\big(\nabla_s(\bm R_s^T \bm v)\big)
\qquad \forall\,\bm v\in V, \label{eq:divs_def_Silhavy}
\end{equation}
with $\bm{R}_s^T:\Gamma \to \mathbb{R}^3 \otimes V$ the transpose of $\bm{R}_s$. The scalar case $V=\mathbb{R}$, in which $\bm{R}_s:\Gamma\to \mathbb{R}\otimes\mathbb{R}^3\cong\mathbb{R}^3$ is nothing but a vector field over $\Gamma$, recovers the familiar $\Div_s (\bm R_s)=\tr(\nabla_s\bm{R}_s)$. A product (Leibniz) rule accompanies the definition; for arbitrary \(\bm a:\Gamma\to V\) it reads
\begin{equation}
\Div_s(\bm R_s^T\bm a)=\Div_s (\bm R_s)\cdot \bm{a}+ \bm R_s \bm\cdot \nabla_s \bm a,
\label{eq:divs_product_Silhavy}
\end{equation}
the pairing being understood as $\bm R_s \cdot \nabla_s \bm a :=\tr (\bm R_s^T\nabla_s \bm a)$.

On each face $\Gamma_\ell$, a superficial field, $\bm R_s=\bm R_s\bm I_s$, satisfies the surface divergence theorem \citep{steinmann2008boundary}
\begin{equation}\label{eq:surf_div_thm}
\int_{\Gamma_\ell}\Div_s(\bm R_s)\,\rd A=\int_{\partial\Gamma_\ell}\bm R_s\bm V\,\rd s,
\end{equation}
where $\bm V$ is the in-plane unit normal of the boundary curve $\partial\Gamma_\ell$, lying in the face and pointing out of it. For a field that is not superficial, an additional curvature term would appear on the right-hand side.

\subsection{Surface deformation gradient}
Specialization of the foregoing to the surface deformation map $\bm{\phi}_s$, that is, to the choice $\bm{D}_s = \bm{\phi}_s$ and $V = \mathbb{R}^3$, produces the surface deformation gradient $\bm{F}_s:\mathbb{R}^3 \to \mathbb{R}^3$,
\begin{equation*}
\bm{F}_s = \nabla_s \bm{\phi}_s.
\end{equation*}
Tangent vectors of the reference surface are thereby carried to tangent vectors of the deformed surface. Since $\bm{F}_s$ is of rank two, no inverse exists in the classical sense; that role is taken over by the \emph{pseudo-inverse} $\bm{F}_s^{-1}:\mathbb{R}^3 \to \mathbb{R}^3$, characterized by
\begin{equation}
\bm{F}_s^{-1}\bm{F}_s = \bm{I}_s, 
\qquad
\bm{F}_s \bm{F}_s^{-1} = \bm{i}_s,\qquad \bm{F}_s^{-1}\bm{i}_s=\bm{F}_s^{-1},
\label{eq39}
\end{equation}
in which $\bm{i}_s = \bm{I} - \bm{n}\otimes \bm{n}$ projects onto the tangent space of $\bm{\phi}(\Gamma)$, the deformed boundary surface, and $\bm{n}$ is the current normal vector. The third of these relations plays for $\bm{F}_s^{-1}$ the part that \eqref{eq26} plays for surface gradients: it confines the action of the pseudo-inverse to tangent vectors of the deformed surface. A further set of identities follows at once,
\begin{equation}\label{eq:FIs}
\bm{F}_s \bm{I}_s = \bm{F}_s, \qquad
\bm{i}_s \bm{F}_s = \bm{F}_s, \qquad
\bm{I}_s \bm{F}_s^{-1} = \bm{F}_s^{-1}.
\end{equation}

Since $\bm{\phi}_s$ is the restriction of $\bm{\phi}$ and $\bm{F}$ is invertible up to the boundary, the surface deformation gradient, together with its pseudo-inverse, follows from the bulk deformation gradient,
\begin{equation}
\bm{F}_s = \bm{F} \bm{I}_s, 
\qquad 
\bm{F}_s^{-1} = \bm{F}^{-1} \bm{i}_s.
\label{eq41}
\end{equation}
The identities \eqref{eq:FIs} yield, in particular,
\begin{equation}\label{eq:FsTn}
\bm{F}_s \bm{N} = \bm{0}, \qquad \bm{F}^T_s \bm{n} = \bm{0},
\end{equation}
so that the referential normal $\bm{N}$ is annihilated by the surface deformation gradient, and the current normal $\bm{n}$ by its transpose; both statements express the tangential character of the map.

Changes of infinitesimal area may be described purely in surface terms. To that end one introduces the \emph{surface Jacobian determinant}, namely the current area of an infinitesimal surface element divided by its area in the reference configuration; in view of \eqref{eq:cof}, and since $\cof(\bm{F}_s)\bm{N}=\cof(\bm{F})\cof(\bm{I}_s)\bm{N}=\cof(\bm{F})\bm{N}$, this quantity is expressible through $\bm{F}_s$ as
\begin{equation}
{J}_s:= |\cof(\bm{F}_s) \bm{N}|.
\end{equation}
For vectors lying in the tangent space, $J_s$ is readily seen to obey a relation patterned after \eqref{eq:cofF},
\begin{align}
|\bm{F}_s\bm{u}\times \bm{F}_s\bm{v}|=J_s|\bm{u}\times\bm{v}|,\qquad \forall\ \bm{u},\bm{v}\in T_{\boldsymbol{X}}\Gamma,
\end{align}
and this relation may itself be taken as a definition of the surface Jacobian determinant, in the spirit of exterior algebra \citep{winitzki2009linear}. Nanson's formula \eqref{eq:cof} yields directly a further expression for $J_s$, this time in terms of the bulk deformation gradient,
\bea \label{eq:Js_bulk}
    J_s = |\cof(\bm{F}) \bm{N}| = J |\bm{F}^{-T} \bm{N}|.
\eea

The Piola identities transfer divergences to the current configuration:
\bea \label{eq:piola_t}
\Div(\bm R)=J\ddiv(J^{-1}\bm R\bm F^T), \qquad
\Div_s (\bm R_s) =J_s \ddiv_s( J_s^{-1} \bm R_s \bm{F}_s^T).
\eea
Here $\Div$ is the standard bulk divergence, acting on the last index, and $\ddiv$ and $\ddiv_s$ are its bulk and surface counterparts on the deformed body $\bm{\phi}(\Omega)$ and deformed surface $\bm{\phi}(\Gamma)$. The fields $\bm R$ and $\bm R_s$ take values in $V\otimes\mathbb{R}^3$, with $V=\mathbb R$, $\mathbb R^3$, or $\mathbb R^3\otimes\mathbb R^3$. 
\section{Variational formulation}\label{sec:VarFormulation}

\subsection{\texorpdfstring{Total energy and equilibrium equations}{Total energy and equilibrium equations}}

Three contributions enter the total energy of the system: the energy stored in the bulk, the energy stored on the surface, and the work expended by the prescribed external bulk force ${\bm{f}}$ and surface traction ${\bm{t}}$. Accordingly,
\begin{equation}
E[\bm{\phi}]= \int_\Omega
   W(\bm{F},\bm{H}) \,\rd V
          + \int_S \varPhi(\bm{F}_s) \,\rd A
          - \int_\Omega \bm{\phi}\cdot {\bm{f}} \,\rd  V
          - \int_{\Gamma_N} \bm{\phi}\cdot {\bm{t}} \,\rd A,
\end{equation}
the arguments of the two energy densities being
\begin{equation}\label{eq:FHs}
\bm{F} = \nabla \bm{\phi}, 
\qquad 
\bm{H} = \nabla \bm{F},
\qquad 
\bm{F}_s = \nabla_s \bm{\phi}.
\end{equation}
The last of these involves the surface gradient $\nabla_s(\cdot) = \nabla(\cdot)\,\bm{I}_s$, understood as a linear map on $\mathbb{R}^3$ and built from $\bm{I}_s = \bm{I} - \bm{N}\otimes\bm{N}$, the projector onto the tangent space of $S$.

The placement is prescribed, $\bm\phi=\bm\phi_D$, on a closed portion $\Gamma_D\subseteq\Gamma$, and the traction $\bm t$ acts on $\Gamma_N=\Gamma\setminus\Gamma_D$; it is extended by zero to $\Gamma_D$ wherever it appears under an integral over $\partial\Omega$. Admissible variations vanish on $\Gamma_D$, while their normal derivatives $(\nabla\delta\bm\phi)\bm N$ remain free.

Both densities are referential: $W$ measures the bulk energy stored per unit volume of the reference configuration, and $\varPhi$ the surface energy stored per unit area of the reference surface. Because $\bm{H}$ appears among its arguments, $W$ is taken to respond to the first and to the second gradient of the deformation alike, so that the bulk material is of second-gradient type.

Frame indifference requires
\begin{equation*}
W(\bm R\bm F,\bm R\bm H)=W(\bm F,\bm H),
\qquad
\varPhi(\bm R\bm F_s)=\varPhi(\bm F_s)
\end{equation*}
for every proper orthogonal tensor $\bm R$. For the compatible deformation measures used here, these conditions are equivalent to the representations
\begin{align}
W=\widetilde W(\bm C,\nabla\bm C),
\qquad
\varPhi=\widetilde\varPhi(\bm C_s),
\end{align}
where $\bm C=\bm F^T\bm F$ and $\bm C_s=\bm F_s^T\bm F_s$. For notational convenience, we retain $(\bm F,\bm H)$ and $\bm F_s$ as the arguments of the bulk and surface energies, respectively. The constitutive models adopted in \cref{sec:app} satisfy these invariance requirements.

Differentiation of the two densities supplies the work-conjugate stress measures.  The first Piola--Kirchhoff stress $\bm{P}$, the referential higher-order stress $\bm{A}$, and the referential surface first Piola--Kirchhoff  stress $\bm{P}_s$ are
\begin{equation}
\label{eq:constitutiveBulk}
\bm{P}   = \frac{\partial W}{\partial\bm{F}}, 
\qquad
\bm{A} = \frac{\partial W }{\partial \bm{H}},
\qquad
\bm{P}_s = \frac{\partial \varPhi }{\partial \bm{F}_s}.
\end{equation}
Since $\varPhi$ is defined on superficial arguments alone,
$\bm{F}_s=\bm{F}_s\bm{I}_s$, the last of these is fixed only up to its normal
column; $\bm{P}_s$ is understood throughout as the superficial representative
$\bm{P}_s=(\partial\varPhi/\partial\bm{F}_s)\bm{I}_s$, which is what makes the
surface divergence theorem applicable.
Using definitions \eqref{eq:FHs} for $\bm{H}$ and $\bm{F}_s$, the first variation of the energy functional becomes
\begin{equation}\label{eq:deltaE-compact}
\delta E 
= \int_\Omega (\bm{P} : \nabla\delta \bm{\phi}+ \bm{A} \;\vdots\; \nabla^2\delta\bm{\phi}) \,\mathrm{d}V 
 + \int_S \bm{P}_s : \nabla_s \delta\bm{\phi} \, \mathrm{d}A
 - \int_\Omega \delta\bm{\phi}\cdot \bm{f} \,\mathrm{d}V
 - \int_{\Gamma_N} \delta\bm{\phi}\cdot \bm{t} \,\mathrm{d}A,
\end{equation}
in which $\delta$ denotes a variation, while $:$ and $\vdots$ stand for the double and the triple contraction of tensors, respectively.

Green's identity transfers the derivative in the first bulk term onto the stress,
\begin{equation}
\int_\Omega\bm{P} : \nabla\delta \bm{\phi}\,\mathrm{d}V
= - \int_\Omega \delta\bm{\phi}\cdot\Div (\bm{P}) \,\mathrm{d}V
  + \int_{\partial\Omega} \delta\bm{\phi}\cdot \bm{P}\bm{N} \,\mathrm{d}A.
\end{equation}
Two successive applications of the same identity recast the second bulk term as
\bea
\int_\Omega \bm{A} \;\vdots\; \nabla^2\delta\bm{\phi} \,\mathrm{d}V
&=&  - \int_\Omega \nabla \delta \bm{\phi} : \Div (\bm{A}) \,\mathrm{d}V
  + \int_{\partial\Omega} \nabla \delta \bm{\phi} : \bm{A}\bm{N} \,\mathrm{d}A \nonumber \\
  &=&
 \int_\Omega  \delta \bm{\phi} \cdot \Div\left(\Div(\bm{A})\right)\,\mathrm{d}V
  - \int_{\partial \Omega} \delta \bm{\phi} \cdot \Div (\bm{A})\bm{N} \,\mathrm{d}A
  + \int_{\partial \Omega} \nabla \delta\bm{\phi} : \bm{A} \bm{N} \,\mathrm{d}A. \label{eq:term_A}
\eea
Resolving the remaining boundary integral into contributions normal and tangential to $\partial \Omega$ and invoking the surface divergence theorem \eqref{eq:surf_div_thm}, which the superficiality of $\left(\bm{A}\bm{N}\right)\bm{I}_s$ permits, on each smooth face of the boundary separately, one arrives at
\bea
\int_{\partial \Omega} \nabla \delta\bm{\phi} : \bm{A}\bm{N} \,\mathrm{d}A
&=& \int_{\partial \Omega} \nabla \delta\bm{\phi} : \left[\left(\bm{A}\bm{N} \right)(\bm{I}_s + \bm{N}\otimes \bm{N}) \right]  \,\mathrm{d}A = 
\int_{\partial \Omega} \nabla \delta\bm{\phi} : \left[\left(\bm{A}\bm{N} \right)\bm{I}_s  \right]   + (\nabla \delta\bm{\phi} \cdot \bm{N}) \cdot \left[\left(\bm{A}\bm{N} \right) \cdot\bm{N} \right]\,\mathrm{d}A 
\nonumber \\
&=&
\int_{\partial \Omega} - \delta\bm{\phi} \cdot \Div_s\left[\left(\bm{A}\bm{N} \right)\bm{I}_s  \right]   + (\nabla \delta\bm{\phi} \cdot \bm{N}) \cdot \left[\left(\bm{A}\bm{N} \right) \cdot\bm{N} \right]\,\mathrm{d}A 
+ \int_{\mathcal{C}} \delta\bm{\phi} \cdot \sum_{\ell\in\mathcal{F}} \left[\left(\bm{A}\bm{N}\right)\bm{V}\right]_{\ell}\,\mathrm{d}s.
\eea
Here $[\,\cdot\,]_\ell$ indicates that the enclosed quantities are evaluated on the face $\Gamma_\ell$, and $\bm{V}$ is the in-plane unit normal of its bounding curve: it lies in that face, is orthogonal to the curve and points out of the face. Each face therefore enters with its own $\bm{N}$ and $\bm{V}$, while $\bm{A}$ is common to them; the projector is absent because $\bm{I}_s\bm{V}=\bm{V}$, and $\delta\bm{\phi}$, single-valued on $\mathcal{C}$, has been taken outside the sum.
The two results combine to give the integral in \eqref{eq:term_A} in the form
\bea
\int_\Omega \bm{A} \;\vdots\; \nabla^2\delta\bm{\phi} \,\mathrm{d}V
&=&
 \int_\Omega  \delta \bm{\phi} \cdot \Div\left(\Div(\bm{A})\right)\mathrm{d}V
  - \int_{\partial \Omega} \delta \bm{\phi} \cdot \left(\Div (\bm{A})\bm{N} +\Div_s\left[\left(\bm{A}\bm{N} \right)\bm{I}_s  \right] \right)\mathrm{d}A \nonumber \\
&&  + \int_{\partial \Omega}  (\nabla \delta\bm{\phi} \cdot \bm{N}) \cdot \left[\left(\bm{A}\bm{N} \right) \cdot\bm{N} \right]\mathrm{d}A 
  + \int_{\mathcal{C}} \delta\bm{\phi} \cdot \sum_{\ell\in\mathcal{F}} \left[\left(\bm{A}\bm{N}\right)\bm{V}\right]_{\ell}\,\mathrm{d}s.
\eea

Superficiality of the surface stress $\bm{P}_s $ likewise permits the surface divergence theorem \eqref{eq:surf_div_thm} to act on the surface integral,
\begin{equation}
\int_S \bm{P}_s  : \nabla_s\delta \bm{\phi} \,\mathrm{d}A
= - \int_S \delta \bm{\phi} \cdot \Div_s(\bm{P}_s ) \,\mathrm{d}A
  + \int_{\mathcal{C}} \delta \bm{\phi} \cdot \sum_{\ell\in\mathcal{F}_S} \left[\bm{P}_s\bm{V}\right]_{\ell} \,\mathrm{d}s,
\end{equation}
the theorem being applied to each face of $S$ in turn, so that the line contributions again collect on $\mathcal{C}$; the faces outside $S$ carry no $\varPhi$ and hence no $\bm{P}_s$, and are absent from $\mathcal{F}_S$.

Let the effective stress be denoted by $\bm{Q} = \bm{P}-\Div\bm{A}$.  Substituting the identities established above into \eqref{eq:deltaE-compact} then gives
\begin{align}
\delta E
&= - \int_\Omega \delta \bm{\phi}\cdot (\Div\bm{Q}+{\bm{f}})\,\mathrm{d}V
   + \int_{\partial\Omega} \delta \bm{\phi}\cdot(\bm{Q}\bm{N} - \Div_s\left[\left(\bm{A}\bm{N} \right)\bm{I}_s  \right] - {\bm{t}}) +(\nabla \delta\bm{\phi} \cdot \bm{N}) \cdot \left[\left(\bm{A}\bm{N} \right) \cdot\bm{N} \right] \,\mathrm{d}A \nonumber\\
&\quad - \int_S \delta \bm{\phi} \cdot \Div_s\left(\bm{P}_s \right)\mathrm{d}A
   + \int_{\mathcal{C}} \delta \bm{\phi}\cdot \left(\sum_{\ell\in\mathcal{F}} \left[\left(\bm{A}\bm{N}\right)\bm{V}\right]_{\ell} + \sum_{\ell\in\mathcal{F}_S} \left[\bm{P}_s\bm{V}\right]_{\ell}\right) \,\mathrm{d}s .
\end{align}
Since $\delta E$ must vanish for all admissible variations, the fundamental lemma of the calculus of variations yields the equilibrium equations and natural boundary conditions,
\begin{alignat}{2}
& \Div\bm{Q} + \bm{f} = \bm{0}
&& \qquad \text{in } \Omega,
\label{eq:bulk-eq}
\\
& \bm{A}:(\bm{N}\otimes\bm{N}) = \bm{0}
&& \qquad \text{on } \partial\Omega,
\label{eq:edge-bc-add}
\\
& \bm{Q}\bm{N}
  - \Div_s\left[\left(\bm{A}\bm{N}\right)\bm{I}_s\right]
  = \bm{t}
&& \qquad \text{on } \Gamma_N\setminus S,
\label{eq:bulk-bc}
\\
& \bm{Q}\bm{N}
  - \Div_s\left[\left(\bm{A}\bm{N}\right)\bm{I}_s+\bm{P}_s\right]
  = \bm{t}
&& \qquad \text{on } S\cap\Gamma_N,
\label{eq:surf-eq}
\\
& \sum_{\ell\in\mathcal{F}} \left[\left(\bm{A}\bm{N}\right)\bm{V}\right]_{\ell} + \sum_{\ell\in\mathcal{F}_S} \left[\bm{P}_s\bm{V}\right]_{\ell} = \bm{0}
&& \qquad \text{on } \mathcal C\cap\Gamma_N .
\label{eq:edge-eq}
\end{alignat}
The last of these is the edge condition. Its first sum, the line force carried by a second-gradient bulk, survives only at geometric edges: where two adjoining faces meet smoothly, with a common $\bm{N}$ along the curve, their in-plane normals are opposite and their terms cancel.
Its second sum, the rim traction of the energetic surface, enters only where a face of $S$ adjoins the curve. Without geometric edges, the edge condition thus reduces to the familiar $\bm{P}_s\bm{V}=\bm{0}$ on the rim of $S$. Note that condition \eqref{eq:edge-bc-add} holds on the whole boundary, including $\Gamma_D$.

\section{Existence of minimizers}\label{sec:existence}

We now show that the total energy of \cref{sec:VarFormulation} attains
its minimum among orientation-preserving Sobolev deformations with the
prescribed placement on \(\Gamma_D\), provided that the surface energy
density is nonnegative and that at least one such deformation has finite
energy. The key is the second-gradient bulk energy, which controls not only
the deformation gradient but also its material gradient. This additional
regularity provides a compact boundary trace of the gradient, so that the
surface deformation gradient converges almost everywhere along a
subsequence of every minimizing sequence. Fatou's lemma then yields lower
semicontinuity of the surface energy, without convexity of the surface
density and without any upper bound on its growth. In the bulk, as is
standard for second-gradient (nonsimple) materials
\citep{kruzik2019mathematical}, convexity is needed only with respect to
the second gradient.

\subsection{Functional setting and assumptions}

Let \(\Omega\subset\mathbb R^3\) be a bounded connected Lipschitz domain,
with boundary \(\Gamma=\partial\Omega\). Let \(S\subseteq\Gamma\) be the
relatively open portion carrying surface energy, and let the part
\(\Gamma_D\subseteq\Gamma\) with prescribed placement be
\(\mathcal H^2\)-measurable with positive area,
\(\mathcal H^2(\Gamma_D)>0\), where \(\mathcal H^2\) denotes the
two-dimensional Hausdorff (area) measure. As in
\cref{sec:VarFormulation}, \(\Gamma_N=\Gamma\setminus\Gamma_D\).
For \(p,q>1\), the energy space is
\begin{equation}\label{eq:energy_space}
 \mathcal X
 :=W^{2,q}(\Omega;\mathbb R^3)\cap W^{1,p}(\Omega;\mathbb R^3),
 \qquad
 \|\bm\phi\|_{\mathcal X}
 :=\|\bm\phi\|_{W^{2,q}}+\|\bm\phi\|_{W^{1,p}}.
\end{equation}
We assume that the placement \(\bm\phi_D\) prescribed on \(\Gamma_D\) is
the trace of a map in \(\mathcal X\), denoted again by \(\bm\phi_D\), and
define the admissible class
\begin{equation}\label{eq:admissible}
 \mathcal A
 :=\left\{\bm\phi\in\mathcal X:
 \bm\phi=\bm\phi_D\ \text{on }\Gamma_D,
 \quad \det\nabla\bm\phi>0\ \text{a.e. in }\Omega\right\}.
\end{equation}
The boundary condition is understood in the trace sense. Thus, for the
existence argument, the smooth kinematics of the preceding sections are
replaced by Sobolev regularity and almost-everywhere preservation of
orientation; global injectivity is not imposed.

As before, \(\bm F=\nabla\bm\phi\),
\(\bm H=\nabla^2\bm\phi\), and
\(\bm I_s=\bm I-\bm N\otimes\bm N\). Since
\(\bm F\in W^{1,q}(\Omega;\mathbb R^{3\times3})\), its boundary trace is
well defined. Writing the trace operator as \(\operatorname{Tr}_{\Gamma}\),
we define the surface deformation gradient by
\begin{equation}\label{eq:Fs_trace_definition}
 \bm F_s:=\bigl(\operatorname{Tr}_{\Gamma}\bm F\bigr)\bm I_s
 \qquad\text{a.e. on }\Gamma.
\end{equation}
We write \(\varPhi(\bm X,\bm F_s)\) rather than \(\varPhi(\bm F_s)\) as in
\cref{sec:VarFormulation}: the explicit dependence on \(\bm X\in S\)
allows the surface normal \(\bm N(\bm X)\), the projector
\(\bm I_s(\bm X)\), and material inhomogeneity to enter the surface
constitutive law. The total potential energy is
\begin{equation}\label{eq:energy_existence}
 \begin{aligned}
 E[\bm\phi]
 &=\int_\Omega W(\bm F,\bm H)\,\rd V
   +\int_S\varPhi(\bm X,\bm F_s)\,\rd A-\mathcal W_{\mathrm{ext}}(\bm\phi),\\
 \mathcal W_{\mathrm{ext}}(\bm\phi)
 &:=\int_\Omega\bm\phi\cdot\bm f\,\rd V
   +\int_{\Gamma_N}\bm\phi\cdot\bm t\,\rd A.
 \end{aligned}
\end{equation}
Here \(\bm f\) and \(\bm t\) are the body force and the traction of
\cref{sec:VarFormulation}. We impose the following assumptions.

\begin{itemize}
\item[\textnormal{(H1)}] \emph{Bulk energy:}
The density
\(W:\mathbb R^{3\times3}\times\mathbb R^{3\times3\times3}
\to[0,+\infty]\)
is continuous for \(\det\bm F>0\). It is extended by \(+\infty\) when
\(\det\bm F\le0\), and this extension is jointly lower semicontinuous.
For every fixed \(\bm F\), the map \(\bm H\mapsto W(\bm F,\bm H)\)
is convex. Moreover, there exist \(c>0\) and \(C\ge0\) such that
\begin{equation}\label{eq:coercivity}
 W(\bm F,\bm H)\ge c\bigl(|\bm F|^p+|\bm H|^q\bigr)-C.
\end{equation}

\item[\textnormal{(H2)}] \emph{Nonnegative surface energy:}
The density
\(\varPhi:S\times\mathbb R^{3\times3}\to[0,+\infty]\)
is measurable with respect to the product of the
\(\mathcal H^2\)-measurable subsets of \(S\) and the Borel subsets of
\(\mathbb R^{3\times3}\), and \(\varPhi(\bm X,\cdot)\) is lower
semicontinuous for almost every \(\bm X\in S\).

\item[\textnormal{(H3)}] \emph{Loads:}
The body force and traction satisfy
\begin{equation}\label{eq:load_assumptions}
 \bm f\in L^{p'}(\Omega;\mathbb R^3),
 \qquad
 \bm t\in L^{p'}(\Gamma_N;\mathbb R^3),
 \qquad p'=\frac{p}{p-1}.
\end{equation}
Consequently, \(\mathcal W_{\mathrm{ext}}\) is a bounded linear functional on
\(W^{1,p}(\Omega;\mathbb R^3)\), and hence on \(\mathcal X\).
\end{itemize}

No convexity, polyconvexity, or homogeneity of \(\varPhi\) is required.
Under these assumptions, the surface integral is well defined in
\([0,+\infty]\); allowing the value \(+\infty\) accommodates densities
with unrestricted growth and surface constraints.

\subsection{Existence theorem}

\begin{theorem}[Existence of minimizers]\label{thm:existence_nonneg}
Let \emph{(H1)}, \emph{(H2)}, and \emph{(H3)} hold with
\(p,q>1\), and suppose that there exists
\(\overline{\bm\phi}\in\mathcal A\) with
\(E[\overline{\bm\phi}]<+\infty\).
Then \(E\) attains its minimum on \(\mathcal A\).
Every minimizing sequence has a subsequence converging weakly in
\(\mathcal X\) to a minimizer.
\end{theorem}

\begin{proof}
The proof follows the direct method of the calculus of variations: a
minimizing sequence is bounded in \(\mathcal X\) by coercivity and has a
weakly convergent subsequence (Step~1); the compact trace yields
almost-everywhere convergence of the gradients and of their traces
(Step~2); this makes the bulk and surface energies lower semicontinuous
and the limit admissible (Steps~3 and~4); and the load term is
continuous, so that the limit is a minimizer (Step~5).

Throughout the proof, \(C\) denotes a generic positive constant independent
of \(k\), whose value may change from line to line.
Let \(\{\bm\phi_k\}\subset\mathcal A\) be a minimizing sequence.
Discarding finitely many terms, we may assume
\begin{equation}\label{eq:minseq_upper}
 E[\bm\phi_k]\le E[\overline{\bm\phi}]+1.
\end{equation}

\emph{Step 1: coercivity and compactness.}
Because \(\bm\phi_k-\bm\phi_D\) has zero trace on the set \(\Gamma_D\) of
positive area, the Poincar\'e inequality yields
\begin{equation}\label{eq:poincare_p}
 \|\bm\phi_k\|_{W^{1,p}}\le C\bigl(1+\|\nabla\bm\phi_k\|_{L^p}\bigr).
\end{equation}
Hence, by \emph{(H3)} and Young's inequality, for every
\(\varepsilon>0\),
\begin{equation}\label{eq:load_absorb}
 |\mathcal W_{\mathrm{ext}}(\bm\phi_k)|\le\varepsilon\|\nabla\bm\phi_k\|_{L^p}^p
 +C_\varepsilon.
\end{equation}
Choosing \(\varepsilon=c/2\) and combining \eqref{eq:minseq_upper},
\eqref{eq:coercivity}, \(\varPhi\ge0\) and \eqref{eq:load_absorb}
gives
\begin{equation}\label{eq:FH_bound}
 \sup_k\bigl(\|\nabla\bm\phi_k\|_{L^p}^p
            +\|\nabla^2\bm\phi_k\|_{L^q}^q\bigr)<+\infty.
\end{equation}
To recover the full \(W^{2,q}\)-bound, apply the Poincar\'e--Wirtinger
inequality to \(\nabla\bm\phi_k\) minus its mean value over \(\Omega\), and
bound that mean by H\"older's inequality,
\(\bigl|\frac{1}{|\Omega|}\int_\Omega\nabla\bm\phi_k\,\rd V\bigr|
\le|\Omega|^{-1/p}\|\nabla\bm\phi_k\|_{L^p}\). On a bounded connected
Lipschitz domain this gives
\begin{equation}\label{eq:mixed_poincare}
 \|\nabla\bm\phi_k\|_{L^q}
 \le C\bigl(\|\nabla^2\bm\phi_k\|_{L^q}+\|\nabla\bm\phi_k\|_{L^p}\bigr).
\end{equation}
A second application of Poincar\'e's inequality to
\(\bm\phi_k-\bm\phi_D\) then yields
\begin{equation}\label{eq:X_bound}
 \sup_k\|\bm\phi_k\|_{\mathcal X}<+\infty.
\end{equation}
Since \(\mathcal X\) is reflexive, passing to a subsequence,
\begin{equation}\label{eq:X_weak}
 \bm\phi_k\rightharpoonup\bm\phi\qquad\text{weakly in }\mathcal X.
\end{equation}
The Dirichlet condition is preserved by continuity of the trace.

\emph{Step 2: pointwise convergence.}
Weak convergence in \(\mathcal X\) implies
\(\nabla\bm\phi_k\rightharpoonup\nabla\bm\phi\) in \(W^{1,q}(\Omega)\).
The Rellich--Kondrachov embedding \(W^{1,q}(\Omega)\hookrightarrow
L^q(\Omega)\) is compact \citep{adams2003sobolev}, and so, for every
\(q>1\), is the trace operator \(W^{1,q}(\Omega)\to L^1(\Gamma)\) on
the Lipschitz domain \(\Omega\) \citep{necas2012direct}.
Since a compact operator maps a weakly convergent sequence to a strongly
convergent one, \(\nabla\bm\phi_k\to\nabla\bm\phi\) in \(L^q(\Omega)\) and
\(\operatorname{Tr}_\Gamma(\nabla\bm\phi_k)\to
\operatorname{Tr}_\Gamma(\nabla\bm\phi)\) in \(L^1(\Gamma)\), and
multiplication by the bounded projector \(\bm I_s\) gives
\(\bm F_{s,k}:=\bigl(\operatorname{Tr}_\Gamma\nabla\bm\phi_k\bigr)\bm I_s
\to\bm F_s\) in \(L^1(S)\). After passage to a further
subsequence,
\begin{equation}\label{eq:convergences_nonneg}
 \nabla\bm\phi_k\to\nabla\bm\phi\quad\text{a.e. in }\Omega,
 \qquad
 \bm F_{s,k}\to\bm F_s\quad\mathcal H^2\text{-a.e. on }S,
 \qquad
 \nabla^2\bm\phi_k\rightharpoonup\nabla^2\bm\phi
 \quad\text{weakly in }L^q(\Omega).
\end{equation}
No growth property of \(\varPhi\) enters here.

\emph{Step 3: bulk term and admissibility.}
The extended integrand \(W\) is jointly lower semicontinuous in
\((\bm F,\bm H)\), convex in \(\bm H\) for each fixed \(\bm F\), and
nonnegative. Since \(\nabla\bm\phi_k\to\nabla\bm\phi\) in measure and
\(\nabla^2\bm\phi_k\rightharpoonup\nabla^2\bm\phi\) in \(L^q\), Ioffe's
lower semicontinuity theorem \citep{ioffe1977lower} gives
\begin{equation}\label{eq:bulk_lsc}
 \int_\Omega W(\nabla\bm\phi,\nabla^2\bm\phi)\,\rd V
 \le\liminf_{k\to\infty}
 \int_\Omega W(\nabla\bm\phi_k,\nabla^2\bm\phi_k)\,\rd V.
\end{equation}
Thus no convexity with respect to \(\bm F\) is needed: the
\(\bm F\)-dependence is handled by strong convergence, while convexity is
required only in the weakly converging variable \(\bm H\).
The right-hand side of \eqref{eq:bulk_lsc} is finite along the
minimizing sequence, by \eqref{eq:minseq_upper}, \(\varPhi\ge0\) and
\eqref{eq:load_absorb}. Consequently the left-hand side is finite.
Since \(W=+\infty\) for \(\det\bm F\le0\), it follows that
\begin{equation}\label{eq:orientation_limit}
 \det\nabla\bm\phi>0\qquad\text{a.e. in }\Omega.
\end{equation}
Hence \(\bm\phi\in\mathcal A\).

\emph{Step 4: surface term.}
By \emph{(H2)} and \eqref{eq:convergences_nonneg},
\(\varPhi(\bm X,\bm F_s)\le\liminf_k\varPhi(\bm X,\bm F_{s,k})\) for
\(\mathcal H^2\)-a.e.\ \(\bm X\in S\). Since the integrands are
nonnegative, Fatou's lemma yields
\begin{equation}\label{eq:surface_fatou}
 \int_S\varPhi(\bm X,\bm F_s)\,\rd A
 \le\liminf_{k\to\infty}\int_S\varPhi(\bm X,\bm F_{s,k})\,\rd A.
\end{equation}

\emph{Step 5: loads and conclusion.}
Since \(\mathcal W_{\mathrm{ext}}\in\mathcal X^*\), the weak convergence
\eqref{eq:X_weak} implies
\begin{equation}\label{eq:load_conv}
 \mathcal W_{\mathrm{ext}}(\bm\phi_k)\to\mathcal W_{\mathrm{ext}}(\bm\phi).
\end{equation}
The bulk and surface terms being nonnegative, the lower limit of their
sum is at least the sum of their lower limits, and
\eqref{eq:bulk_lsc}, \eqref{eq:surface_fatou} and
\eqref{eq:load_conv} give
\[
 E[\bm\phi]\le\liminf_{k\to\infty}E[\bm\phi_k]=\inf_{\mathcal A}E.
\]
Hence \(\bm\phi\) is a minimizer, and Steps~1--5 apply to every
minimizing sequence.
\end{proof}

\begin{remark}[Global injectivity]\label{rem:injectivity}
If \(q>3/2\), global injectivity can be added to the admissible class
without changing the proof. In that case
\(W^{2,q}(\Omega)\hookrightarrow W^{1,r}(\Omega)\) for some \(r>3\)
(\(r=3q/(3-q)\) if \(q<3\), so that \(q=2\) gives \(r=6\)). Hence every
\(\bm\phi\in\mathcal X\) has a representative continuous on
\(\overline\Omega\), and weak convergence in \(\mathcal X\) implies weak
convergence in \(W^{1,r}\). Let
\begin{equation}\label{eq:admissible_CN}
 \mathcal A_{\mathrm{CN}}
 :=\Bigl\{\bm\phi\in\mathcal A:
 \int_\Omega\det\nabla\bm\phi\,\rd V\le|\bm\phi(\Omega)|\Bigr\},
\end{equation}
where \(|\cdot|\) denotes the Lebesgue measure in \(\mathbb R^3\).
As shown by \citet{ciarlet1987injectivity}, for maps in \(W^{1,r}\) with
\(r>3\) and \(\det\nabla\bm\phi>0\) a.e., this condition implies
injectivity almost everywhere in \(\Omega\), and it is preserved under
weak convergence in \(W^{1,r}\). Their smoothness assumption on
\(\partial\Omega\) is not needed here, because maps in \(W^{1,r}\),
\(r>3\), carry the null set \(\partial\Omega\) to a null set.
Consequently, \cref{thm:existence_nonneg} holds with \(\mathcal A\)
replaced by \(\mathcal A_{\mathrm{CN}}\), provided that
\(\overline{\bm\phi}\in\mathcal A_{\mathrm{CN}}\), and the minimizer is
then injective almost everywhere. If \(\bm\phi_D\in
C^2(\overline\Omega;\mathbb R^3)\) is injective with
\(\det\nabla\bm\phi_D>0\) on \(\overline\Omega\), the condition holds for
\(\bm\phi_D\) with equality, so \(\overline{\bm\phi}=\bm\phi_D\) is
admissible.
\end{remark}

\subsection{Constitutive laws covered by the theorem}

The assumptions of \cref{thm:existence_nonneg} leave much freedom in
the constitutive laws. In the bulk, convexity is required only in the
second gradient, because \(\nabla\bm\phi_k\) converges strongly and only
\(\nabla^2\bm\phi_k\) weakly; no convexity, quasiconvexity or
polyconvexity in
\(\bm F\) is needed. For instance,
\(W_{\mathrm{el}}(\bm F)+\frac{\xi}{q}|\bm H|^q\) with \(\xi>0\) and
\(q>1\) satisfies \emph{(H1)} for every elastic density
\(W_{\mathrm{el}}\ge0\) that is continuous for \(\det\bm F>0\), equals
\(+\infty\) for \(\det\bm F\le0\), is lower semicontinuous, and satisfies
\(W_{\mathrm{el}}(\bm F)\ge c|\bm F|^p-C\). On the surface, every
nonnegative, lower semicontinuous law is admissible, including nonconvex
and inhomogeneous laws and constraints expressed by the value
\(+\infty\). Frame indifference is not used in the proof.

The model of \cref{sec:app} combines a compressible neo-Hookean bulk with
a quadratic second-gradient term and a surface energy that depends on the
area ratio. In the present notation, \eqref{eq:bulkEn} and
\eqref{eq:suradd} read
\begin{align}
 W(\bm F,\bm H)
 &=\frac\mu2\bigl(I_1-3-2\ln J\bigr)
   +\frac D2\left(\frac{J^2-1}{2}-\ln J\right)
   +\frac\xi2|\bm H|^2,
 \label{eq:W_existence}\\
 \varPhi(\bm X,\bm F_s)
 &=\gamma J_s+\frac{\alpha_s}{2}(J_s-1)^2,
 \qquad
 J_s=\bigl|\operatorname{cof}(\bm F_s)\bm N(\bm X)\bigr|,
 \label{eq:Phi_existence}
\end{align}
where \(I_1=|\bm F|^2\), \(J=\det\bm F\), and \(W=+\infty\) for
\(J\le0\). Here \(\mu,D>0\) are the material parameters of the bulk, \(\xi>0\) is the
second-gradient modulus, \(\gamma\ge0\) is the surface tension, and
\(\alpha_s\ge0\) is the surface stretching modulus. 
We first verify \emph{(H1)} for \(p=q=2\), followed by verification of \emph{(H2)}.

\emph{Continuity, convexity and nonnegativity:} For \(J>0\), \(W\) is
continuous; it is convex in \(\bm H\), since only its last term depends
on \(\bm H\). With the singular values \(\lambda_i\) of \(\bm F\),
\[
 I_1-3-2\ln J=\sum_i\bigl(\lambda_i^2-1-\ln\lambda_i^2\bigr),
 \qquad
 \frac{J^2-1}{2}-\ln J=\frac12\bigl(J^2-1-\ln J^2\bigr),
\]
and both are nonnegative because \(x-1-\ln x\ge0\) for \(x>0\). Hence
\(W\ge0\).

\emph{Coercivity and lower semicontinuity:} Both follow from the
rearrangement
\begin{equation}\label{eq:model_coercivity}
 \begin{aligned}
 W(\bm F,\bm H)
 &=\frac\mu2|\bm F|^2+\frac\xi2|\bm H|^2+g(J)
   -\frac{3\mu}{2}-\frac D4,\\
 g(J)&:=\frac D4J^2-\left(\mu+\frac D2\right)\ln J.
 \end{aligned}
\end{equation}
Since \(g\) is bounded below on \((0,\infty)\),
\(W\ge\frac12\min\{\mu,\xi\}\bigl(|\bm F|^2+|\bm H|^2\bigr)-C\), which is
\eqref{eq:coercivity} with \(p=q=2\). Lower semicontinuity can fail
only where \(\det\bm F=0\), because \(W\) is continuous where
\(\det\bm F>0\) and equals \(+\infty\) on the open set where
\(\det\bm F<0\). Since the first two terms in
\eqref{eq:model_coercivity} are nonnegative,
\(W\ge g(J)-\frac{3\mu}{2}-\frac D4\), and \(g(J)\to+\infty\) as
\(J\to0^+\). Hence \(W\to+\infty\) as \(\det\bm F\to0^+\), uniformly in
\(\bm F\) and \(\bm H\), and the extension by \(+\infty\) is lower
semicontinuous. Thus \emph{(H1)} holds.

\emph{Surface density:} \(\varPhi\) is nonnegative, continuous in
\(\bm F_s\), and depends measurably on \(\bm X\) through \(\bm N\), so
\emph{(H2)} holds.

Consequently, the constitutive model \eqref{eq:W_existence}--\eqref{eq:Phi_existence}
satisfies the hypotheses of \cref{thm:existence_nonneg} for
all \(\mu,D,\xi>0\) and \(\gamma,\alpha_s\ge0\). Since \(W\) is finite and
continuous for \(J>0\) and \(\varPhi\) is continuous, a finite-energy
admissible deformation exists whenever the extension \(\bm\phi_D\) belongs
to \(C^2(\overline\Omega;\mathbb R^3)\) with \(\det\nabla\bm\phi_D>0\) on
\(\overline\Omega\): one may take \(\overline{\bm\phi}=\bm\phi_D\), for
which both integrands are bounded. The total
energy then attains its minimum on \(\mathcal A\), at a deformation in
\(W^{2,2}(\Omega;\mathbb R^3)\).

The theorem concerns bounded bodies with the full placement prescribed on a
part of the boundary of positive area. The boundary-value problems studied in
\cref{sec:app,sec:num}, an infinite cylinder with \(\Gamma_D=\emptyset\) and
a finite cylinder on which only some displacement components are
prescribed, lie outside these hypotheses. For them, the result above
covers the constitutive law, not the problems as posed.

\begin{remark}[Equilibrium equations]\label{rem:EL}
\Cref{thm:existence_nonneg} does not by itself show that the minimizer
satisfies the equilibrium equations of \cref{sec:VarFormulation}, even in
weak form. For \(q=2\), as in the model above, \(\nabla\bm\phi\) need not
be continuous, and \(\det\nabla\bm\phi\) need not be bounded away from
zero. Since \(W\to+\infty\) as \(\det\bm F\to0^+\), the perturbed
deformations \(\bm\phi+\varepsilon\bm\psi\) need not have finite energy,
and the first variation cannot be computed in the usual way. If instead
\(q>3\) and \(W\) blows up like a sufficiently high negative power of
\(\det\bm F\), then \(\nabla\bm\phi\) is H\"older continuous on
\(\overline\Omega\) and \(\det\nabla\bm\phi\) is bounded below by a
positive constant \citep{healey2009injective}. Under suitable growth
conditions on the derivatives of \(W\), and if \(\varPhi(\bm X,\cdot)\)
is continuously differentiable, the weak form of the equilibrium
equations then follows. The logarithmic singularity of
\eqref{eq:W_existence} is too weak for this argument.
\end{remark}

\section{Incremental theory}\label{sec:incremental}

To detect the loss of uniqueness of a finitely deformed equilibrium state, we now linearize the governing equations of \cref{sec:VarFormulation} about that state, following the incremental, or small-on-large, approach \citep{ogden1984non,bigoni2012nonlinear,goriely2017mathematics}. The relations so obtained constitute the incremental theory of gradient hyperelasticity with surface energy.
\subsection{\texorpdfstring{Incremental kinematics}{Incremental kinematics}}
\begin{figure}[tbp]
	\centering
	\includegraphics[width=0.8\linewidth]{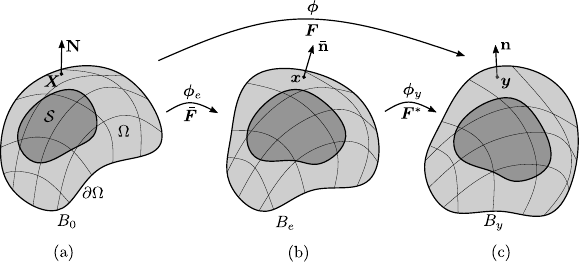}
	\caption{Three configurations of a body with an energetic surface $S$: (a) reference configuration $B_0$, (b) intermediate (finitely deformed) configuration $B_e$, (c) current configuration $B_y$. 
    }
	\label{fig:three-configurations}
\end{figure}
Following \cite{ogden1984non}, we distinguish three configurations: the reference configuration $B_0$, occupying $\Omega \subset \mathbb{R}^3$; the finitely deformed equilibrium configuration $B_e$, called intermediate; and the current (perturbed) configuration $B_y$, obtained by superimposing a small deformation on $B_e$ (\cref{fig:three-configurations}). Note that overbars denote quantities evaluated in $B_e$.

The positions of a material point in these configurations are related by
\begin{align}
  \bm{X} \in B_0, 
  \qquad \bm{x} = \bm{\phi}_e(\bm{X}) \in B_e, 
  \qquad {\bm{y}} = \bm{\phi}_y(\bm{x}) = \bm{\phi}(\bm{X}) \in B_y ,
\end{align}
where $\bm{\phi}_e : B_0\to B_e$ denotes the base deformation, $\bm{\phi}_y : B_e\to B_y$ the superimposed deformation, and $\bm{\phi}=\bm{\phi}_y\circ\bm{\phi}_e$ the resulting total deformation. The corresponding deformation gradients are denoted by
\begin{align}
  \bm{F} = \frac{\partial {\bm{y}}}{\partial \bm{X}}, 
  \qquad \bar{\bm{F}} = \frac{\partial \bm{x}}{\partial \bm{X}}, 
  \qquad \bm{F}^* = \frac{\partial {\bm{y}}}{\partial \bm{x}}. 
\end{align}
The chain rule leads to the multiplicative decomposition of the deformation gradient
\begin{align}\label{eq:FFbar}
  \bm{F} = \bm{F}^* \, \bar{\bm{F}} .
\end{align}
Let $\bar{\bm i}_s=\bm I-\bar{\bm n}\otimes\bar{\bm n}$ denote the tangent projector onto the intermediate surface, where $\bar{\bm n}$ is its outward unit normal. Introducing $\bm F_s^{*}=\bm F^{*}\bar{\bm i}_s$ and using $\bar{\bm i}_s\bar{\bm F}_s=\bar{\bm F}_s$, we obtain the surface counterpart of \eqref{eq:FFbar},
\begin{align}\label{eq:iFs}
  \bm{F}_s = \bm{F}^*_s \, \bar{\bm{F}}_s .
\end{align}

We introduce a small perturbation of the equilibrium state $B_e$ through
\begin{align}
\bm{\phi}_y(\bm{x})=\bm{x}+\epsilon \bm{u}(\bm{x}),
\end{align}
where ${\epsilon}$ is a scalar perturbation parameter and $\bm{u}$ is a displacement field on $B_e$. Differentiation of this map in the bulk and along the surface furnishes the perturbed deformation gradients 
\begin{align}
 &\bm{F}^* = \bm{I} + \epsilon \bar{\nabla} \bm{u} = \bm{I} + \epsilon \bm{\eta}, \label{eq:Fst_bulk}\\
 & \bm{F}_s^* = \bar{\bm{i}}_s + \epsilon \bar{\nabla}_s \bm{u} = \bar{\bm{i}}_s + \epsilon \bm{\eta}_s,  \label{eq:Fst_surf}
\end{align}
where $\bar{\nabla}$ and $\bar{\nabla}_s$ are the bulk and surface gradients with respect to $B_e$, and $\bm{\eta}=\bar{\nabla}\bm{u}$ and $\bm{\eta}_s=\bar{\nabla}_s\bm{u}$ are the bulk and surface incremental displacement gradients. The incremental equations follow by differentiating the equilibrium equations and boundary conditions with respect to $\epsilon$ at $\epsilon=0$, with the loads and the prescribed placement held fixed. A state at which the incremental problem with $\bm u=\bm 0$ on $\Gamma_D$ first admits a nontrivial solution $\bm u$, other than a rigid-body motion, is a candidate bifurcation point. Whether a secondary branch actually emerges there requires further analysis; for the cylinder of \cref{sec:app} it is confirmed by the post-bifurcation paths computed in \cref{sec:num}.

\subsection{\texorpdfstring{Incremental equilibrium equations}{Incremental equilibrium equations}}
Let $\bar{\bm P}$, $\bar{\bm A}$, and $\bar{\bm P}_s$ denote, respectively, the first Piola--Kirchhoff stress, the higher-order stress, and the first Piola--Kirchhoff surface stress evaluated in the intermediate configuration $B_e$, i.e., at the state $(\bar{\bm F},\bar{\bm H},\bar{\bm F}_s)$.
The intermediate equilibrium state satisfies the referential equations
\begin{alignat}{2}
& \Div\bar{\bm{Q}} + \bm{f} = \bm{0}
&& \qquad \text{in } \Omega,
\label{eq:bulk-eq_int}
\\
& \bar{\bm{A}}:(\bm{N}\otimes\bm{N}) = \bm{0}
&& \qquad \text{on } \partial\Omega,
\label{eq:edge-bc-add_int}
\\
& \bar{\bm{Q}}\bm{N}
  - \Div_s\left[\left(\bar{\bm{A}}\bm{N}\right)\bm{I}_s\right]
  = \bm{t}
&& \qquad \text{on } \Gamma_N\setminus S,
\label{eq:bulk-bc_int}
\\
& \bar{\bm{Q}}\bm{N}
  - \Div_s\left[\left(\bar{\bm{A}}\bm{N}\right)\bm{I}_s+\bar{\bm{P}}_s\right]
  = \bm{t}
&& \qquad \text{on } S\cap\Gamma_N,
\label{eq:surf-eq_int}
\\
& \sum_{\ell\in\mathcal{F}} \left[\left(\bar{\bm{A}}\bm{N}\right)\bm{V}\right]_{\ell} + \sum_{\ell\in\mathcal{F}_S} \left[\bar{\bm{P}}_s\bm{V}\right]_{\ell} = \bm{0}
&& \qquad \text{on } \mathcal C\cap\Gamma_N,
\label{eq:edge-eq_int}
\end{alignat}
where $\bar{\bm{Q}} = \bar{\bm{P}}-\Div\bar{\bm{A}}$ is the effective stress. 

The linearization proceeds from asymptotic expansions of the bulk and surface stress tensors $\bm{P}$, $\bm{A}$, and $\bm{P}_s$ in powers of the parameter~$\epsilon$,
\begin{align}
&\bm{P}    = \bar{\bm{P}}    + \epsilon \bm{P}^{[1]}    + \mathcal{O}(\epsilon^2), \label{eq:exp1} \\
&\bm{A}  = \bar{\bm{A}}  + \epsilon \bm{A}^{[1]}  + \mathcal{O}(\epsilon^2), \\
&\bm{P}_s  = \bar{\bm{P}}_s  + \epsilon \bm{P}_s^{[1]}  + \mathcal{O}(\epsilon^2), \label{eq:exp3} 
\end{align}
 in which the overbarred quantities do not depend on $\epsilon$, and the superscript $[1]$ marks the first-order perturbation, made explicit in \cref{sec:lin_const}; since $\Div$ acts on $\bm X$, the effective stress $\bm Q$ expands in the same way.

 For simplicity, the loading is assumed to be dead, that is, $\bm{f}$ and $\bm{t}$ do not depend on the deformation and hence on $\epsilon$, so that the loading terms sit entirely at zeroth order and are absent from the incremental system below. The normals, projectors and differential operators in \eqref{eq:bulk-eq}--\eqref{eq:edge-eq} are referential as well, so that $\epsilon$ enters only through the stresses, which depend on it through $\bm{F}$, $\bm{H}$, and $\bm{F}_s$.

 When the expansions are substituted into the governing equations \eqref{eq:bulk-eq}--\eqref{eq:edge-eq} associated with the perturbed configuration $B_y$, the zeroth-order terms recover the corresponding equilibrium equations  \eqref{eq:bulk-eq_int}--\eqref{eq:edge-eq_int} associated with the intermediate configuration $B_e$ and therefore cancel. Dividing the remaining terms by $\epsilon$ and passing to the limit $\epsilon\to0$ yields the incremental problem referred to $B_0$,
\begin{alignat}{2}
& \Div{\bm{Q}^{[1]}} = \bm{0}
&& \qquad \text{in } \Omega,
\label{eq:bulk-eq_inc}
\\
& {\bm{A}^{[1]}}:(\bm{N}\otimes\bm{N}) = \bm{0}
&& \qquad \text{on } \partial\Omega,
\label{eq:edge-bc-add_inc}
\\
& {\bm{Q}^{[1]}}\bm{N}
  - \Div_s\left[\left({\bm{A}^{[1]}}\bm{N}\right)\bm{I}_s\right]
  = \bm{0}
&& \qquad \text{on } \Gamma_N\setminus S,
\label{eq:bulk-bc_inc}
\\
& {\bm{Q}^{[1]}}\bm{N}
  - \Div_s\left[\left({\bm{A}^{[1]}}\bm{N}\right)\bm{I}_s+\bm{P}_s^{[1]}\right]
  = \bm{0}
&& \qquad \text{on } S\cap\Gamma_N,
\label{eq:surf-eq_inc}
\\
& \sum_{\ell\in\mathcal{F}} \left[\left({\bm{A}^{[1]}}\bm{N}\right)\bm{V}\right]_{\ell} + \sum_{\ell\in\mathcal{F}_S} \left[\bm{P}_s^{[1]}\bm{V}\right]_{\ell} = \bm{0}
&& \qquad \text{on } \mathcal C\cap\Gamma_N,
\label{eq:edge-eq_inc}
\end{alignat}
where
\bea
{\bm{Q}^{[1]}} =  {\bm{P}}^{[1]}-\Div{\bm{A}}^{[1]}.
\eea
 
\subsection{Linearization of the constitutive relations}\label{sec:lin_const}

The incremental equations involve the first-order perturbations of the three
stress measures \eqref{eq:constitutiveBulk}, which are now made explicit.
Under the superimposed deformation the arguments of the two energy densities
are perturbed by \eqref{eq:FFbar} and \eqref{eq:iFs} together with
\eqref{eq:Fst_bulk}--\eqref{eq:Fst_surf} as
\begin{align}\label{eq:args}
  \bm{F}(\epsilon)=\bar{\bm{F}}+\epsilon\,\bm{\eta}\bar{\bm{F}},
  \qquad
  \bm{H}(\epsilon)=\bar{\bm{H}}+\epsilon\,\nabla(\bm{\eta}\bar{\bm{F}}),
  \qquad
  \bm{F}_s(\epsilon)=\bar{\bm{F}}_s+\epsilon\,\bm{\eta}_s\bar{\bm{F}}_s,
\end{align}
the second of these being the referential gradient of the first. 
Substituting these expressions into the constitutive relations \eqref{eq:constitutiveBulk} and expanding in $\epsilon$, as in \eqref{eq:exp1}--\eqref{eq:exp3}, recovers the stresses of the intermediate configuration at zeroth order. The linearization retains only the terms linear in $\epsilon$, while quadratic and higher-order terms are discarded. The coefficients of the linear terms, obtained by applying the chain rule, give the first-order stress increments:
\begin{align}
  {\bm{P}}^{[1]}&=\mathbb{A}:\big(\bm{\eta}\bar{\bm{F}}\big)
    +\mathbb{B}\vdots\nabla(\bm{\eta}\bar{\bm{F}}),\label{eq:P_incr}\\[2pt]
  {\bm{A}}^{[1]}&=\mathbb{C}:\big(\bm{\eta}\bar{\bm{F}}\big)
    +\mathbb{D}\vdots\nabla(\bm{\eta}\bar{\bm{F}}),\label{eq:A_incr}\\[2pt]
  {\bm{P}}_s^{[1]}&=\mathbb{A}_s:\big(\bm{\eta}_s\bar{\bm{F}}_s\big),\label{eq:Ps_incr}
\end{align}
where the double and triple contractions are taken over the two indices of $\bm{\eta}\bar{\bm{F}}$ and the three indices of $\nabla(\bm{\eta}\bar{\bm{F}})$, respectively. The bulk tangent moduli are defined by
\begin{align}
   &\mathbb{A}  = \frac{\partial {\bm{P}}}{\partial \bm{F}}\Big|_{(\bar{\bm{F}}, \bar{\bm{H}})},
     \qquad
      \, \mathbb{B}  = \frac{\partial {\bm{P}}}{\partial \bm{H}}\Big|_{(\bar{\bm{F}}, \bar{\bm{H}})}, \label{eq:bulk_stiff1}
    \\
   & \mathbb{C} = \frac{\partial \bm{A}}{\partial \bm{F}}\Big|_{(\bar{\bm{F}}, \bar{\bm{H}})},
    \qquad  \mathbb{D} = \frac{\partial \bm{A}}{\partial \bm{H}}\Big|_{(\bar{\bm{F}}, \bar{\bm{H}})}, \label{eq:bulk_stiff2}
\end{align}
the mixed ones, $\mathbb{B}$ and $\mathbb{C}$, being transposes of each other by symmetry of the mixed second derivatives of $W$. The surface tangent modulus is defined analogously,
\begin{align} \label{eq:stiff_surf}
   &\mathbb{A}_s  = \frac{\partial {\bm{P}}_s}{\partial \bm{F}_s}\Big|_{\bar{\bm{F}}_s} = \frac{\partial^2 \varPhi }{\partial \bm{F}_s \partial \bm{F}_s}\Big|_{\bar{\bm{F}}_s},
\end{align}
where, as in the definition of $\bm{P}_s$, only the superficial part of the second derivative is retained, so that $\bm{P}_s^{[1]}$ is also superficial.

\subsection{Push-forward  to the intermediate configuration}\label{sec:push-forward}

For the application in \cref{sec:app} it is convenient to express the incremental problem \eqref{eq:bulk-eq_inc}--\eqref{eq:edge-eq_inc} in the intermediate configuration. 
To this end, we introduce the updated incremental bulk stress tensor $\bm{\chi}$, the updated incremental higher-order stress $\bm{a}^{[1]}$, and the updated incremental surface stress tensor $\bm{\chi}_s$ as
\begin{align}\label{eq:icr}
\bm{\chi} = \bar{J}^{-1} (\bm{P}^{[1]}\bar{\bm{F}}^{T} + \bm{A}^{[1]} : \bar{\bm{H}}^{T}), 
\qquad\quad  
\bm{a}^{[1]} = \bar{J}^{-1}\bm{A}^{[1]}: \left(\bar{\bm{F}}^{T} \overline{\otimes} \bar{\bm{F}}^{T}\right),
\qquad\quad  
\bm{\chi}_s = \bar{J}_s^{-1}\bm{P}^{[1]}_s\bar{\bm{F}}_s^{T},
\end{align}
in which $\bar J=\det\bar{\bm F}$ and $\bar J_s=\bar J\,|\bar{\bm F}^{-T}\bm N|$, cf.\ \eqref{eq:Js_bulk}, are the bulk and surface Jacobians of $\bm\phi_e$, $\overline{\otimes}$ denotes a nonstandard tensor product of two second-order
tensors, and the superscript $T$ on a third-order tensor denotes the cyclic shift of
its indices. In components, these and the double contraction of a third-order tensor with such a product read
\bea \label{eq:nonstd}
\left(\bm{U} \overline{\otimes} \bm{W}\right)_{ijkl} = U_{ik} W_{jl}
,
\qquad\quad
\left(\bm{G}^{T}\right)_{jki} = G_{ijk},
\qquad\quad
\big(\bm{G}:(\bm{U}\,\overline{\otimes}\,\bm{W})\big)_{ijk}=G_{imn}U_{mj}W_{nk}.
\eea
With these definitions, the Piola push-forward $\bm q^{[1]}=\bar J^{-1}\bm Q^{[1]}\bar{\bm F}^T$ of $\bm{Q}^{[1]} = {\bm{P}}^{[1]}-\Div{\bm{A}}^{[1]}$ takes the same form,
\bea
\bm{q}^{[1]} = \bm{\chi}-\overline{\ddiv}{\bm{a}}^{[1]},
\eea
where $\overline{\ddiv}$ and $\overline{\ddiv}_s$ denote the bulk and surface divergences in the intermediate configuration. In this sense, the quantities \eqref{eq:icr} are the incremental stresses updated to $B_e$.

Using the Piola identities \eqref{eq:piola_t} and Nanson's formula \eqref{eq:cof}, we push forward the incremental problem \eqref{eq:bulk-eq_inc}--\eqref{eq:edge-eq_inc} to the intermediate configuration $B_e$, obtaining
\begin{alignat}{2}
& \overline{\ddiv}{\bm{q}^{[1]}} = \bm{0}
&& \qquad \text{in } \bm{\phi}_e(\Omega),
\label{eq:bulk-eq_inc_act}
\\
& {\bm{a}^{[1]}}:(\bar{\bm{n}}\otimes\bar{\bm{n}}) = \bm{0}
&& \qquad \text{on } \bm{\phi}_e(\partial\Omega),
\label{eq:edge-bc-add_inc_act}
\\
& {\bm{q}^{[1]}}\bar{\bm{n}}
  - \overline{\ddiv}_s\left[\left({\bm{a}^{[1]}}\bar{\bm{n}}\right)\bar{\bm{i}}_s\right]
  = \bm{0}
&& \qquad \text{on } \bm{\phi}_e(\Gamma_N\setminus S),
\label{eq:bulk-bc_inc_act}
\\
& {\bm{q}^{[1]}}\bar{\bm{n}}
  - \overline{\ddiv}_s\left[\left({\bm{a}^{[1]}}\bar{\bm{n}}\right)\bar{\bm{i}}_s+{\bm{\chi}_s}\right]
  = \bm{0}
&& \qquad \text{on } \bm{\phi}_e(S\cap\Gamma_N),
\label{eq:surf-eq_inc_act}
\\
& \sum_{\ell\in\mathcal{F}} \left[\left({\bm{a}^{[1]}}\bar{\bm{n}}\right)\bar{\bm{v}}\right]_{\ell} + \sum_{\ell\in\mathcal{F}_S} \left[{\bm{\chi}_s}\bar{\bm{v}}\right]_{\ell} = \bm{0}
&& \qquad \text{on } \bm{\phi}_e(\mathcal C\cap\Gamma_N) ,
\label{eq:edge-eq_inc_act}
\end{alignat}
where $\bar{\bm v}$ is the in-plane unit normal of $\bm\phi_e(\mathcal C)$ on the image of each face. The derivation, which also uses the higher-order condition \eqref{eq:edge-bc-add_inc}, is given in \ref{app:pushforward}. Together with the incremental constitutive relations \eqref{eq:P_incr}--\eqref{eq:Ps_incr} and \eqref{eq:icr}, the system \eqref{eq:bulk-eq_inc_act}--\eqref{eq:edge-eq_inc_act} defines the incremental problem used in \cref{sec:app}.

\section{Application to \texorpdfstring{elastocapillary}{elastocapillary} beading of soft cylinders}\label{sec:app}

The framework developed in \cref{sec:VarFormulation,sec:incremental} is
now applied to the elastocapillary beading instability of an infinite cylinder made of 
a gradient-elastic material, coated by an energetic surface. Under axial stretch, the homogeneous state may bifurcate into a nonhomogeneous configuration with a periodically modulated radius, i.e., the beading configuration. \Cref{sec:hom} introduces the geometry, the homogeneous deformation, and bulk and surface constitutive laws. Equilibrium of the homogeneous state reduces to a relation between the axial and the azimuthal stretch together with an expression for the axial force. This homogeneous state serves as the intermediate configuration $B_e$ of \cref{sec:incremental}. \Cref{sec:bif} then superposes an axisymmetric incremental field on this state, specializes the incremental equations of \cref{sec:incremental} to cylindrical coordinates, and reduces the resulting system to a boundary-value problem for the radial profile, whose solvability condition yields the dispersion relation governing the bifurcation. The resulting bifurcation behavior is analyzed in
\cref{sec:results} and compared with numerical simulations in \cref{sec:num}.

\subsection{Geometry and homogeneous deformation}\label{sec:hom}
We consider an infinite soft cylinder of reference radius $A$, made of a \emph{gradient} hyperelastic material and coated by an elastic surface whose response combines classical surface tension with an additional stretching resistance. A uniform stretch is applied along the cylinder axis, while the lateral surface is free of external traction; the stresses acting there arise solely from the constitutive response of the coating. The infinite geometry and homogeneous axial loading make the geometry and the data translationally invariant in the axial coordinate \(z\).

Cylindrical coordinates are the natural choice here, with $\theta$ the angular, $Z$ the axial and $R$ the radial coordinate of the reference configuration. We use them as convected coordinates, $\theta^1=\theta$, $\theta^2=Z$ and $\theta^3=R$: they label a material point and keep their values during the deformation. The covariant base vectors of the reference and the deformed configuration are then $\bm{G}_i=\partial\bm{X}/\partial\theta^i$ and $\bm{g}_i=\partial\bm{x}/\partial\theta^i$, and the contravariant ones, $\bm{G}^i$ and $\bm{g}^i$, are their duals, $\bm{G}_i\cdot\bm{G}^j=\bm{g}_i\cdot\bm{g}^j=\delta_i^j$. Referred to the orthonormal cylindrical basis $(\bm{e}_\theta(\theta),\bm{e}_z,\bm{e}_r(\theta))$, the reference position vector becomes 
\bea
\bm{X} = Z \bm{e}_z + R \bm{e}_r.
\eea
Since $\rd\bm{e}_\theta/\rd \theta = -\bm{e}_r$ and $\rd\bm{e}_r/\rd \theta = \bm{e}_\theta$, differentiation of this expression delivers the covariant base vectors of the reference configuration together with their contravariant duals, 
\begin{align}\label{eq:G123}
\bm{G}_{1} = R \bm{e}_\theta=R^2\bm{G}^1,\qquad  \bm{G}_{2} = \bm{e}_z=\bm{G}^2, \qquad    \bm{G}_{3} = \bm{e}_r=\bm{G}^3.
\end{align}

Let the cylinder deform homogeneously, in which case the axial stretch $\lambda$ and the azimuthal stretch $a$ describe the deformation completely and the position vector in the deformed configuration is
\begin{align}
\bm{x} =z\bm{e}_z + r\bm{e}_r,
\end{align}
in which the deformed coordinates follow from the reference ones as
\begin{align}\label{eq:hom}
 z=\lambda Z,\qquad r=aR,
\end{align}
the angle $\theta$ being unchanged.
 Differentiating $\bm{x}$ in the same way gives the covariant and contravariant base vectors of the deformed configuration,
\begin{align}\label{eq:g123}
\bm{g}_{1} = r \bm{e}_\theta=r^2\bm{g}^1, \qquad\bm{g}_{2} =\lambda \bm{e}_z=\lambda^2\bm{g}^2, \qquad  \bm{g}_{3} =a \bm{e}_r=a^2\bm{g}^3.
\end{align} 
In the present geometry the reference and deformed unit normals of the lateral surface $R=A$ coincide,
\begin{align}
\bm{N} = \bar{\bm{n}} = \bm{e}_r.
\end{align}
The reference and deformed tangential projectors therefore coincide as well,
\begin{align}
\bm{I}_s=\bar{\bm{i}}_s=\bm{I}-\bar{\bm{n}}\otimes\bar{\bm{n}}=\bm{e}_\theta\otimes\bm{e}_\theta+\bm{e}_z\otimes\bm{e}_z.
\end{align}

 Relations \eqref{eq:G123} and \eqref{eq:g123} allow all subsequent expressions to be written in the orthonormal basis $(\bm{e}_\theta,\bm{e}_z,\bm{e}_r)$. This is convenient because the resulting components have a direct physical interpretation, and the distinction between covariant and contravariant indices ceases to matter as they coincide in an orthonormal basis. Henceforth, components refer to this basis, with indices $1$, $2$, and $3$ corresponding to $\theta$, $z$ and $r$; a comma denotes partial differentiation with respect to the coordinates $z$ and $r$ of $B_e$. Under this convention the bulk and surface deformation gradients of the homogeneous deformation read
 \begin{align}\label{eq:homgrad}
 &\bar{\bm{F}} =\bm{g}_i \otimes \bm{G}^i = a \bm{e}_\theta \otimes \bm{e}_\theta + \lambda \bm{e}_z \otimes \bm{e}_z  + a \bm{e}_r \otimes \bm{e}_r,\\ 
 &\bar{\bm{F}}_s = \bar{\bm{F}}\bm{I}_s =a \bm{e}_\theta \otimes \bm{e}_\theta + \lambda \bm{e}_z \otimes \bm{e}_z.
 \end{align}

The bulk solid is taken to be made of a compressible neo-Hookean hyperelastic material \citep[see, e.g.,][]{bakiler2023surface,yu2025incremental} supplemented by a simple gradient term, with strain energy function
\begin{align} \label{eq:bulkEn}
W=\frac{\mu}{2}(I_1-3-2\ln J)+ \frac{D}{2}\Big(\frac{J^2-1}{2}-\ln J\Big) + \frac{\xi}{2} \bm{H}\vdots \bm{H},
\end{align}
with $\mu$ and $D$ the Lam\'{e} parameters, $I_1=\tr(\bm{C})$ the first invariant of the right Cauchy--Green tensor, $\bm{C}=\bm{F}^T\bm{F}$, and $J=\det(\bm{F})$ the Jacobian determinant; $\xi$ is the second-gradient modulus. Since $\bm{H}$ is a second gradient with respect to the reference coordinates, $\mu$ and $D$ carry the dimension of stress while $\xi$ carries that of stress times length squared, so that $\ell_\xi=\sqrt{\xi/\mu}$ is an intrinsic bulk length scale introduced by the gradient term. Compressibility is characterized by the Poisson ratio $\nu$, given here by $\nu={D}/{(2(D+\mu))}$.
Evaluating the constitutive relation~\eqref{eq:constitutiveBulk} at the homogeneous deformation \eqref{eq:hom} gives the first Piola--Kirchhoff stress
\begin{align}\label{eq:Pbar}
\begin{split}
\bar{\bm{P}}&=\Big[\mu\Big(a-\frac{1}{a}\Big)+\frac{D}{2}\Big(a^3\lambda^2-\frac{1}{a}\Big)\Big]\bm{e}_\theta\otimes\bm{e}_\theta+\Big[\mu\Big(\lambda-\frac{1}{\lambda}\Big)+\frac{D}{2}\Big(a^4\lambda-\frac{1}{\lambda}\Big)\Big]\bm{e}_z\otimes\bm{e}_z\\
&\quad+\Big[\mu\Big(a-\frac{1}{a}\Big)+\frac{D}{2}\Big(a^3\lambda^2-\frac{1}{a}\Big)\Big]\bm{e}_r\otimes\bm{e}_r.
\end{split}
\end{align}

Because the deformation is homogeneous, $\bar{\bm{H}}=\bm{0}$; consequently, the higher-order stress $\bar{\bm{A}}$ vanishes, and the second-gradient contribution does not affect the homogeneous state.

The coating is described by a surface tension model enhanced by stretching resistance, characterized by the surface energy density
\begin{align}
\varPhi=\gamma J_s+\frac{\alpha_s}{2}(J_s-1)^2, \label{eq:suradd}
\end{align}
in which $J_s$ denotes the surface Jacobian determinant, $\gamma$ the surface tension parameter and $\alpha_s$ the surface stretching modulus.

By the identity
\bea \label{eq:djdF}
\frac{\partial J_s}{\partial \bm{F}_s} = J_s \bm{F}_s^{-T},
\eea
the surface stress is obtained as
\bea
\bar{\bm{P}}_s = \left( \gamma  + \alpha_s \left( \bar{J}_s-1\right) \right) \bar{J}_s\bar{\bm{F}}_s^{-T} = 
\left( \gamma  + \alpha_s \left( a \lambda-1\right) \right) \left[ \lambda \bm{e}_\theta\otimes\bm{e}_\theta+a\bm{e}_z\otimes\bm{e}_z \right].\label{eq:Psbar}
\eea

Because the cylinder is infinite, the surface $S$ carrying the surface energy is the entire lateral boundary, so that $\mathcal{C}=\emptyset$; the edge condition \eqref{eq:edge-eq_int} is consequently vacuous here, as is its incremental counterpart \eqref{eq:edge-eq_inc_act} in the analysis that follows. Here $\Gamma_D=\emptyset$ and $\Gamma_N=\Gamma=S$. Since $\bar{\bm P}$ is uniform and $\bar{\bm A}=\bm 0$, the bulk equation \eqref{eq:bulk-eq_int} with $\bm f=\bm 0$ and the higher-order condition \eqref{eq:edge-bc-add_int} hold identically. What remains is the boundary condition \eqref{eq:surf-eq_int}, which, in the absence of external traction, reduces to
\bea
\bar{\bm{P}}\bm{N}- \Div_s\left[\bar{\bm{P}}_s\right]
  = \bm{0}.
\eea
Although the components of the surface stress are constant, its surface divergence is nonzero because the circumferential basis vector $\bm e_\theta$ varies along the circumference. In particular, $\Div_s\bar{\bm P}_s=-(\bar P_s)_{11}\bm e_r/A$, so that only the radial component of the surface equilibrium equation is nontrivial,
\bea\label{eq:bccc}
\bar{P}_{33}= -\frac{1}{A}(\bar{P}_s)_{11},
\eea
and with \eqref{eq:Pbar} and \eqref{eq:Psbar} it becomes a relation between the stretches $a$ and $\lambda$,
\bea \label{eq:alambda}
\Big[\mu\Big(a-\frac{1}{a}\Big)+\frac{D}{2}\Big(a^3\lambda^2-\frac{1}{a}\Big)\Big] + 
\frac{1}{A} \lambda
\left( \gamma  + \alpha_s \left( a \lambda-1\right) \right) = 0.
\eea
After multiplication by $a$, \eqref{eq:alambda} becomes a quartic in $a$ whose coefficients change sign exactly once for $\alpha_s\ge0$. By Descartes' rule of signs it has exactly one positive root, so that \eqref{eq:alambda} defines $a=a(\lambda)$ uniquely for every $\lambda>0$.

The resultant axial force, finally, collects the bulk contribution over the reference cross-section and the surface contribution along its circumference,
\begin{align}
F_z &= \pi A^2\bar P_{22}+2\pi A(\bar P_s)_{22} \nonumber\\
&= \pi A^2\Big[\mu\Big(\lambda-\frac{1}{\lambda}\Big)+\frac{D}{2}\Big(a^4\lambda-\frac{1}{\lambda}\Big)\Big]+2\pi Aa\big(\gamma+\alpha_s(a\lambda-1)\big). \label{eq:ax_force}
\end{align}
\subsection{Linear bifurcation analysis}\label{sec:bif}
With the intermediate configuration fully determined, the incremental problem introduced in  \cref{sec:incremental} can now be written out explicitly. We consider an axisymmetric, torsion-free perturbation of the form
\begin{align}
\bm{u}=v(z,r)\bm{e}_z+u(z,r)\bm{e}_r,
\end{align}
with $v(z,r)$ and $u(z,r)$ denoting, respectively, the longitudinal and radial components of the displacement increment. Differentiation supplies the incremental displacement gradient
\begin{align}
\bm{\eta}=\bar{\nabla}\bm{u}=\frac{\partial \bm{u}}{\partial\theta^i}\otimes \bm{g}^i=\frac{u}{r}\bm{e}_\theta\otimes\bm{e}_\theta+v_{,z}\bm{e}_z\otimes\bm{e}_z+v_{,r}\bm{e}_z\otimes \bm{e}_r+u_{,z}\bm{e}_r\otimes\bm{e}_z+u_{,r}\bm{e}_r\otimes\bm{e}_r,
\end{align}
in which the derivatives $\partial/{\partial\theta^i}$ follow from the chain rule together with the relation \eqref{eq:hom}.

Evaluation of the incremental stresses \eqref{eq:P_incr} and \eqref{eq:A_incr} additionally requires the second incremental displacement gradient $\bar{\nabla}\bm{\eta}$, which is provided in \ref{app:cylinder} together with the remaining intermediate results of this subsection.

Inserting the energy \eqref{eq:bulkEn} into \eqref{eq:bulk_stiff1} and \eqref{eq:bulk_stiff2} gives the bulk moduli. For the present energy the mixed moduli $\mathbb{B}$ and $\mathbb{C}$ vanish identically, while $\mathbb{D}$ is equal to $\xi$ times the identity tensor of third order. Since also $\bar{\bm{H}}=\bm{0}$, the incremental stresses \eqref{eq:P_incr} and \eqref{eq:A_incr} reduce to
\begin{align}
{\bm{P}}^{[1]} &= \mathbb{A} : \big( {\bm{\eta}} \, \bar{{\bm{F}}} \big),
\\
{\bm{A}}^{[1]} &= \mathbb{D} \vdots \nabla(\bm{\eta}\bar{\bm{F}}) = \mathbb{D} \vdots \left(\bar{\nabla}\bm{\eta} : \bar{\bm{F}} \overline{\otimes} \bar{\bm{F}} \right) = \xi \bar{\nabla}\bm{\eta} : \bar{\bm{F}} \overline{\otimes} \bar{\bm{F}},
\end{align}
and, by \eqref{eq:icr}, their updated counterparts are
\begin{align}
\bm{\chi} &= \bar{J}^{-1} \bm{P}^{[1]}\bar{\bm{F}}^{T} = \frac{1}{\bar{J}} \left(\mathbb{A} : \big( {\bm{\eta}} \, \bar{{\bm{F}}} \big)\right) \bar{\bm{F}}^{T},
\\
\bm{a}^{[1]} &= \frac{1}{\bar{J}}\bm{A}^{[1]}: \left(\bar{\bm{F}}^{T} \overline{\otimes} \bar{\bm{F}}^{T}\right) = \frac{1}{\bar{J}} \left( \xi \bar{\nabla}\bm{\eta} : \bar{\bm{F}} \overline{\otimes} \bar{\bm{F}} \right): \left(\bar{\bm{F}}^{T} \overline{\otimes} \bar{\bm{F}}^{T}\right) = \frac{\xi}{\bar{J}} \bar{\nabla}\bm{\eta} : \bar{\bm{F}}\bar{\bm{F}}^T \overline{\otimes} \bar{\bm{F}}\bar{\bm{F}}^T.
\end{align}

Their nonzero components in the orthonormal basis, and those of the effective incremental stress $\bm{q}^{[1]} = \bm{\chi}-\overline{\ddiv}{\bm{a}}^{[1]}$, are listed in \ref{app:cylinder}.

For the axisymmetric field considered here, the azimuthal component of the incremental equilibrium equation \eqref{eq:bulk-eq_inc_act} is satisfied identically, and its axial and radial components read
\begin{align}
&\frac{\partial q^{[1]}_{22}}{\partial z}+\frac{\partial q^{[1]}_{23}}{\partial r}+\frac{q^{[1]}_{23}}{r}=0, \label{eq:iequi1_rev}\\
&\frac{\partial q^{[1]}_{32}}{\partial z}+\frac{\partial q^{[1]}_{33}}{\partial r}+\frac{q^{[1]}_{33}-q^{[1]}_{11}}{r}=0. \label{eq:iequi2_rev}
\end{align}

Insertion of the effective incremental stress \eqref{eq:chicomp} into \eqref{eq:iequi1_rev} and \eqref{eq:iequi2_rev} converts equilibrium into a coupled differential system for two unknown displacement fields $v(z,r)$, $u(z,r)$:
\begin{align}
&
\frac{\frac{D}{2}a^4\lambda^2+\mu+\frac{D}{2}}{a^2\lambda r}\,u_{,z}
+\frac{\frac{D}{2}a^4\lambda^2+\mu+\frac{D}{2}}{a^2\lambda}\,u_{,zr}
+\left(\frac{\mu}{\lambda r}-\frac{a^2\xi}{\lambda r^3}\right)v_{,r}
+\frac{\frac{D}{2}a^4\lambda^2+\mu\lambda^2+\mu+\frac{D}{2}}{a^2\lambda}\,v_{,zz}
+\left(\frac{\mu}{\lambda}+\frac{a^2\xi}{\lambda r^2}\right)v_{,rr}
\notag\\
&\qquad
-\frac{2\lambda\xi}{r}\,v_{,zzr}
-\frac{\lambda^3\xi}{a^2}\,v_{,zzzz}
-\frac{2a^2\xi}{\lambda r}\,v_{,rrr}
-2\lambda\xi\,v_{,zzrr}
-\frac{a^2\xi}{\lambda}\,v_{,rrrr}
=0, \label{eq:equiuv1}
\\[2mm]
&
\left[
-\frac{\frac{D}{2}a^4\lambda^2+\mu a^2+\mu+\frac{D}{2}}{a^2\lambda r^2}
+\frac{3a^2\xi}{\lambda r^4}
\right]u
+\left[
\frac{\frac{D}{2}a^4\lambda^2+\mu a^2+\mu+\frac{D}{2}}{a^2\lambda r}
-\frac{3a^2\xi}{\lambda r^3}
\right]u_{,r}
+\left(\frac{\lambda\mu}{a^2}+\frac{2\lambda\xi}{r^2}\right)u_{,zz}
\notag\\
&\qquad
+\left[
\frac{\frac{D}{2}a^4\lambda^2+\mu a^2+\mu+\frac{D}{2}}{a^2\lambda}
+\frac{3a^2\xi}{\lambda r^2}
\right]u_{,rr}
-\frac{2\lambda\xi}{r}\,u_{,zzr}
-\frac{\lambda^3\xi}{a^2}\,u_{,zzzz}
\notag\\
&\qquad
-\frac{2a^2\xi}{\lambda r}\,u_{,rrr}
-2\lambda\xi\,u_{,zzrr}
-\frac{a^2\xi}{\lambda}\,u_{,rrrr}
+\frac{\frac{D}{2}a^4\lambda^2+\mu+\frac{D}{2}}{a^2\lambda}\,v_{,zr}
=0.\label{eq:equiuv2}
\end{align}

For the cylinder, only the boundary conditions \eqref{eq:edge-bc-add_inc_act} and \eqref{eq:surf-eq_inc_act} apply, since $S=\Gamma$ and $\mathcal{C}=\emptyset$. With $\bar{\bm n}=\bm e_r$, the first involves only the components $a^{[1]}_{233}$ and $a^{[1]}_{333}$ of \eqref{eq:a_coeffs}, and for $\xi>0$ it reduces to
\begin{equation}
u_{,rr} = 0, \qquad v_{,rr} = 0 \qquad \text{at } r=aA. \label{eq:bc_ort}
\end{equation}

Condition \eqref{eq:surf-eq_inc_act}, on the other hand, involves the incremental deformation of the lateral surface, and through it the incremental surface stress $\bm{\chi}_s$. The argument of the surface divergence in \eqref{eq:surf-eq_inc_act} is denoted here by $\bm{\tau}=\left({\bm{a}^{[1]}}\bar{\bm{n}}\right)\bar{\bm{i}}_s+{\bm{\chi}_s}$; the components of $\bm{\chi}_s$ and $\bm{\tau}$ are derived in \ref{app:cylinder}.

In this notation, the azimuthal component of \eqref{eq:surf-eq_inc_act} is again satisfied identically, and its axial and radial components read
\begin{alignat}{2}
\frac{\partial \tau_{22}}{\partial z} &= q^{[1]}_{23} &\qquad& \text{at } r=aA, \label{eq:bc_final1_rev}\\
\frac{\partial \tau_{32}}{\partial z}-\frac{\tau_{11}}{aA} &= q^{[1]}_{33} && \text{at } r=aA. \label{eq:bc_final2_rev}
\end{alignat}
Introducing the incremental bulk and surface stress expressions \eqref{eq:chicomp} and \eqref{eq:tau_str} into \eqref{eq:bc_final1_rev} and \eqref{eq:bc_final2_rev} supplies two further conditions on the unknown displacements $u(z,r)$ and $v(z,r)$ at $r=aA$:
\begin{align}
& \left[ \frac{\gamma-\alpha_s+2a\alpha_s\lambda}{r} -\frac{\frac{D}{2}-\frac{D}{2}a^4\lambda^2+\mu}{a^2\lambda} \right]u_{,z} -\left(\frac{\mu}{\lambda}+\frac{a^2\xi}{\lambda r^2}\right)v_{,r} +a\alpha_s\lambda\,v_{,zz} +\frac{a^2\xi}{\lambda r}\,v_{,rr} \notag\\ &\qquad +2\lambda\xi\,v_{,zzr} +\frac{a^2\xi}{\lambda}\,v_{,rrr} =0, \label{eq:bs_subs1} \\[2mm] 
& \left( \frac{3a^2\xi}{\lambda r^3} -\frac{Da^2\lambda}{r} -\frac{a\alpha_s\lambda}{r^2} \right)u -\left[ \frac{\frac{D}{2}a^4\lambda^2+\mu a^2+\mu+\frac{D}{2}}{a^2\lambda} +\frac{3a^2\xi}{\lambda r^2} \right]u_{,r} +\left(\gamma-\alpha_s+a\alpha_s\lambda\right)u_{,zz} +\frac{a^2\xi}{\lambda r}\,u_{,rr} \notag\\ &\qquad +2\lambda\xi\,u_{,zzr} +\frac{a^2\xi}{\lambda}\,u_{,rrr} -\left[ Da^2\lambda +\frac{\gamma-\alpha_s+2a\alpha_s\lambda}{r} \right]v_{,z} =0. \label{eq:bs_subs2}
\end{align}

Equations \eqref{eq:equiuv1} and \eqref{eq:equiuv2}, together with the boundary conditions \eqref{eq:bc_ort}, \eqref{eq:bs_subs1}, and \eqref{eq:bs_subs2} at $r=aA$ and the regularity conditions on the axis $r=0$, define the boundary-value problem of the linear bifurcation analysis.

Since the problem is invariant under axial translations, it admits normal modes
\begin{align}
u(z,r)=f(r)e^{\ri k z},\quad v(z,r)=g(r)e^{\ri k z},
\end{align}
in which $k>0$ denotes the axial wavenumber in $B_e$. Substitution of this ansatz into \eqref{eq:equiuv1} and \eqref{eq:equiuv2} yields two coupled ordinary differential equations in $r$. Eliminating $g(r)$ reduces the system to the eighth-order equation \eqref{eq:diff1} for $f(r)$, and the four boundary conditions \eqref{eq:bc_ort}, \eqref{eq:bs_subs1}, and \eqref{eq:bs_subs2} reduce to conditions on $f$ alone:
\begin{align}
\mathcal{L}_1\mathcal{L}_2\mathcal{L}_3\mathcal{L}_4 f(r) &= 0,\label{eq:diff1}\\
f''&=0 \qquad \text{at } r=aA,\label{eq:bcff3}\\
f^{(5)}
+l_{1}f^{(4)}
+l_{2}f'''
+l_{3}f''
+l_{4}f'
+l_{5}f&=0\qquad \text{at } r=aA,\label{eq:bcff4}
\\
f^{(7)}
+l_{6} f^{(6)}
+l_{7}f^{(5)}
+l_{8}f^{(4)}
+l_{9}f'''
+l_{10}f''
+l_{11}f'
+l_{12}f&=0 \qquad \text{at } r=aA,\label{eq:bcff1}\\
f^{(7)}
+l_{13} f^{(6)}
+l_{14}f^{(5)}
+l_{15}f^{(4)}
+l_{16}f'''
+l_{17}f''
+l_{18}f'
+l_{19}f&=0\qquad \text{at } r=aA,\label{eq:bcff2}
\end{align}
where
\bea
\begin{aligned}
\mathcal{L}_i &= \Big(\frac{\rd^2}{\rd r^2}+\frac{1}{r}\frac{\rd}{\rd r}-\Big(\frac{1}{r^2}+k^2 \kappa_i^2\Big)\Big),
\\[1mm]
\kappa_1
&=
\frac{\sqrt{\xi k^2\lambda^2+\mu}}
     {ak\sqrt{\xi}},
\\[1mm]
\kappa_2
&=
\frac{\lambda}{a},
\\[1mm]
\kappa_3
&=
\frac{
\sqrt{
D+2\mu-\sqrt{\mathcal R}
+2a^2\mu
+Da^4\lambda^2
+4a^2k^2\lambda^2\xi
}}
{2a^2k\sqrt{\xi}},
\\[1mm]
\kappa_4
&=
\frac{
\sqrt{
D+2\mu+\sqrt{\mathcal R}
+2a^2\mu
+Da^4\lambda^2
+4a^2k^2\lambda^2\xi
}}
{2a^2k\sqrt{\xi}},
\end{aligned}
\eea
with
\begin{equation}
\mathcal R=\big[D(a^4\lambda^2+1)+2\mu(a^2+1)\big]^2+8\xi a^2k^2(\lambda^2-a^2)\big[D(a^4\lambda^2+1)+2\mu\big].
\end{equation}
Conditions \eqref{eq:bcff3} and \eqref{eq:bcff4} follow from \eqref{eq:bc_ort}, and \eqref{eq:bcff1} and \eqref{eq:bcff2} from \eqref{eq:bs_subs1} and \eqref{eq:bs_subs2}, respectively. Their coefficients $l_i$ depend on $k$, $a$, $\lambda$, $A$, and the material parameters and are too lengthy to be reported here.

Provided the four values $\kappa_i^2$ are distinct, the general solution of
\eqref{eq:diff1} that is regular at the axis $r=0$ is
\begin{align}
f(r)=c_1 \mathrm{I}_1(k \kappa_1 r)+c_2 \mathrm{I}_1(k \kappa_2 r) + c_3 \mathrm{I}_1(k \kappa_3 r)+c_4 \mathrm{I}_1(k \kappa_4 r),
\end{align}
where $c_1$, $c_2$, $c_3$, and $c_4$ are arbitrary constants and $\mathrm{I}_1(x)$ denotes the modified Bessel function of the first kind. Imposing the boundary conditions \eqref{eq:bcff3}--\eqref{eq:bcff2} yields a homogeneous linear system for these constants. A nontrivial solution exists only if the coefficient matrix is singular. The resulting determinant condition gives the dispersion equation
\begin{align}\label{eq:Omega}
\Delta(k,a,\lambda)=0.
\end{align}

Like the coefficients $l_i$, $\Delta$ is too lengthy to be written explicitly here. With $a=a(\lambda)$ from \eqref{eq:alambda}, the dispersion equation \eqref{eq:Omega} becomes a relation between $k$ and $\lambda$, the dispersion relation of the bifurcation. Its solutions form the neutral curve in the $(k,\lambda)$-plane, and its infimum over $k>0$ gives the onset of bifurcation.

In the long-wave limit $k\to0$, the neutral mode reduces to a uniform increment of the two stretches along $a=a(\lambda)$, on which the second gradient vanishes. The dispersion equation \eqref{eq:Omega} then reduces to
\begin{equation}\label{eq:onset_k0}
\frac{\rd F_z}{\rd \lambda} = 0 \qquad \text{along } a = a(\lambda),
\end{equation}
a stationary point of the axial force \eqref{eq:ax_force}, independent of $\xi$. For the parameters of \cref{sec:results}, the two ends of the neutral curve lie at $\lambda=2.8678$ and $\lambda=3.5953$.
\section{Results}\label{sec:results}
The numerical illustration that follows adopts $A=1$ and $\mu=1$, which amounts to measuring all lengths in units of $A$ and all stresses in units of $\mu$. The bulk Lamé parameter is fixed at $D=49$, corresponding to $\nu=0.49$  in the small-strain limit, while the surface parameters are $\gamma=5.5$ and $\alpha_s = 48$. This choice is deliberate, since in the absence of the second-gradient enhancement, the onset of bifurcation occurs at a finite wavenumber. For other material parameters, the neutral stretch may have its infimum in the long-wave limit $k\to0$. This occurs, e.g., in the case of constant surface tension ($\alpha_s=0$) and, according to \citet{bakiler2023surface} and \citet{yu2025incremental}, also for surface elasticity with surface Lam\'e moduli. In such cases, the second-gradient enhancement does not alter the onset, because the long-wave limit \eqref{eq:onset_k0} is independent of $\xi$. The finite critical wavenumber obtained here for $\xi=0$ therefore stems from the particular surface law \eqref{eq:suradd}, which penalizes only changes of area, together with the large stretching modulus adopted.

The second-gradient modulus $\xi$ is then varied in order to expose its effect on the onset of bifurcation. The values examined are $0$, $0.01$, $0.1$, $0.2$, and $0.4$, corresponding to the ratios $\ell_\xi/A=0$, $0.100$, $0.316$, $0.447$, and $0.632$, where  $\ell_\xi=\sqrt{\xi/\mu}$ is the intrinsic length introduced in \cref{sec:hom}.
At $\xi=0$ the problem is of lower order, and the dispersion equation \eqref{eq:Omega} is not applicable. The values reported for this case are therefore evaluated at $\xi=10^{-6}$, which approximates the limit $\xi\to0$ to an accuracy exceeding the number of digits reported below.

The resulting neutral curves of \cref{sec:bif} are shown in \cref{fig:cyl_equil_tikz}(a), one for each value of $\xi$. For comparison, \cref{fig:cyl_equil_tikz}(b) shows the axial force $F_z$ of the homogeneous state as a function of $\lambda$, obtained by substituting $a(\lambda)$ from \eqref{eq:alambda} into \eqref{eq:ax_force}.

\definecolor{xiA}{RGB}{0,114,189}
\definecolor{xiB}{RGB}{237,177,32}
\definecolor{xiC}{RGB}{119,172,48}
\definecolor{xiD}{RGB}{162,20,47}
\definecolor{xiE}{RGB}{217,83,25}
\begin{figure}[tbp]
	\centering
	\subfloat[]{%
		\tikzsetnextfilename{fig_neutral_curves}%
		\begin{tikzpicture}
			\begin{axis}[
				width=0.41\textwidth, height=0.37\textwidth,
				xlabel={$k$}, ylabel={$\lambda$},
				ylabel style={rotate=-90},
				xmin=0, xmax=1.12, ymin=1.5, ymax=4.0,
				xtick distance=0.2, ytick distance=0.5,
				grid=both,
				major grid style={line width=0.2pt, draw=black!15},
				minor grid style={line width=0.2pt, draw=black!7},
				minor tick num=1, tick align=outside,
				tick label style={font=\scriptsize}, label style={font=\small},
				legend style={at={(1.03,0.5)}, anchor=west, draw=black!40,
					fill=white, font=\tiny, cells={anchor=west}, row sep=-1pt},
				legend cell align=left,
				every axis plot post/.append style={line join=round, line cap=round},
			]
				\addplot[xiA, line width=0.9pt] table {paper_3/tikz_data/neutral_xi0_b1.dat};
				\addlegendentry{$\xi = 0$}
				\addplot[xiB, line width=0.9pt] table {paper_3/tikz_data/neutral_xi0p01_b1.dat};
				\addlegendentry{$\xi = 0.01$}
				\addplot[xiC, line width=0.9pt] table {paper_3/tikz_data/neutral_xi0p1_b1.dat};
				\addlegendentry{$\xi = 0.1$}
				\addplot[xiD, line width=0.9pt] table {paper_3/tikz_data/neutral_xi0p2_b1.dat};
				\addlegendentry{$\xi = 0.2$}
				\addplot[xiE, line width=0.9pt] table {paper_3/tikz_data/neutral_xi0p4_b1.dat};
				\addlegendentry{$\xi = 0.4$}
				\addplot[only marks, mark=*, mark size=1.5pt, xiA, forget plot] table {paper_3/tikz_data/min_xi0.dat};
				\addplot[only marks, mark=*, mark size=1.5pt, xiB, forget plot] table {paper_3/tikz_data/min_xi0p01.dat};
				\addplot[only marks, mark=*, mark size=1.5pt, xiC, forget plot] table {paper_3/tikz_data/min_xi0p1.dat};
				\addplot[only marks, mark=*, mark size=1.5pt, xiD, forget plot] table {paper_3/tikz_data/min_xi0p2.dat};
				\addplot[only marks, mark=o, mark size=2.4pt, xiE, forget plot] table {paper_3/tikz_data/min_xi0p4.dat};
			\end{axis}
		\end{tikzpicture}}\quad
	\subfloat[]{%
		\label{fig:cyl_equil_tikz_b}
		\tikzsetnextfilename{fig_hom_response}%
		\begin{tikzpicture}
			\begin{axis}[
				width=0.41\textwidth, height=0.37\textwidth,
				xlabel={$\lambda$}, ylabel={$F_z$},
				ylabel style={rotate=-90},
				xmin=0, xmax=5, ymin=0, ymax=42,
				xtick distance=1, ytick distance=10,
				grid=both,
				major grid style={line width=0.2pt, draw=black!15},
				minor grid style={line width=0.2pt, draw=black!7},
				minor tick num=1, tick align=outside,
				tick label style={font=\scriptsize}, label style={font=\small},
				restrict y to domain=-20:60, unbounded coords=jump,
				legend style={at={(1.03,0.5)}, anchor=west, draw=black!40,
					fill=white, font=\tiny, cells={anchor=west}, row sep=-1pt},
				legend cell align=left,
				every axis plot post/.append style={line join=round, line cap=round},
			]
				\addplot[black, line width=0.7pt, forget plot]
					table {paper_3/tikz_data/homforce.dat};
				\addplot[only marks, mark=*, mark size=2pt, xiA] table {paper_3/tikz_data/onset_xi0.dat};
				\addlegendentry{$\xi = 0$}
				\addplot[only marks, mark=*, mark size=2pt, xiB] table {paper_3/tikz_data/onset_xi0p01.dat};
				\addlegendentry{$\xi = 0.01$}
				\addplot[only marks, mark=*, mark size=2pt, xiC] table {paper_3/tikz_data/onset_xi0p1.dat};
				\addlegendentry{$\xi = 0.1$}
				\addplot[only marks, mark=*, mark size=2pt, xiD] table {paper_3/tikz_data/onset_xi0p2.dat};
				\addlegendentry{$\xi = 0.2$}
				\addplot[only marks, mark=o, mark size=3pt, xiE] table {paper_3/tikz_data/onset_xi0p4.dat};
				\addlegendentry{$\xi = 0.4$}
			\end{axis}
		\end{tikzpicture}}
	\caption{(a) Neutral curves $\lambda$ versus $k$, one for each $\xi$,
		with the onset marked. (b) The same onsets on the force--stretch
		curve $F_z(\lambda)$ of the homogeneous state. For $\xi=0.2$ and $0.4$
		the onset is approached at the long-wave end $k\to0$; the two onsets
		coincide to seven digits, so the marker for $\xi=0.4$ is open.
		}
	\label{fig:cyl_equil_tikz}
\end{figure}
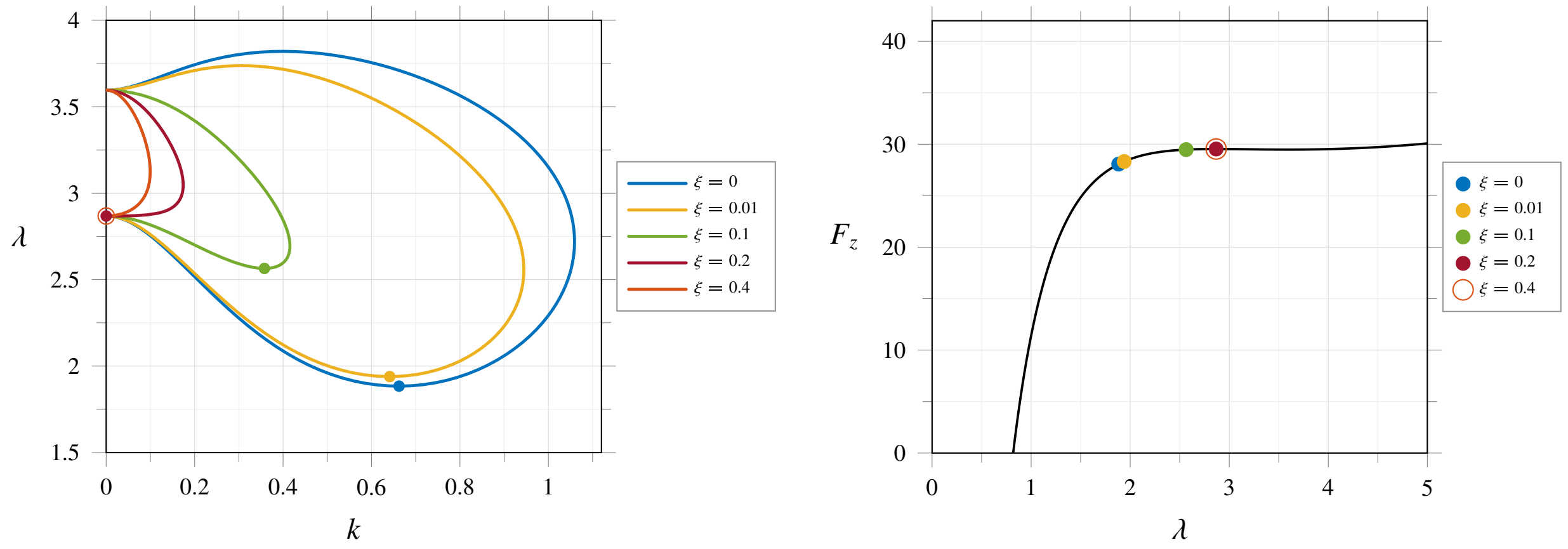

We assume that the loading is controlled by the axial stretch $\lambda$ and starts from the homogeneous state of zero axial force, $F_z=0$, which here corresponds to $\lambda=0.82$ (\cref{fig:cyl_equil_tikz}(b)). Since all computed neutral curves lie  above this value of stretch, the initial state is stable against axisymmetric bifurcation. As $\lambda$ is increased monotonically, the onset is reached at the infimum of the respective neutral curve, marked in both panels of \cref{fig:cyl_equil_tikz}. The second and fifth columns of \cref{tab:bif_onset} report the critical axial stretches and corresponding wavenumbers.

\begin{table}[ht]
\centering
\pgfplotstableread[col sep=space]{paper_3/tikz_data/pairs_onset.dat}\onsettable
\pgfplotstabletypeset[
	every head row/.style={
		before row={\toprule
			& \multicolumn{3}{c}{$\lambda_{\mathrm{b}}$}
			& \multicolumn{3}{c}{$k_{\mathrm{b}}$}
			& \multicolumn{2}{c}{waves along the specimen} \\
			\cmidrule(lr){2-4}\cmidrule(lr){5-7}\cmidrule(lr){8-9}},
		after row=\midrule},
	every last row/.style={after row=\bottomrule},
	col sep=space,
	columns={xi, lamAnalyt, lamNum, relDiff, kAnalyt, kNum, kRelDiff,
		nAnalyt, nNum},
	columns/xi/.style={column name=$\xi$, fixed, precision=2, column type=c},
	columns/lamAnalyt/.style={column name=analyt., fixed, fixed zerofill,
		precision=4, column type/.add={@{\hspace{9pt}}}{}},
	columns/lamNum/.style={column name=num., fixed, fixed zerofill,
		precision=4},
	columns/relDiff/.style={column name=diff.\ [\%], fixed, fixed zerofill,
		precision=2},
	columns/kAnalyt/.style={column name=analyt., fixed, precision=4,
		column type/.add={@{\hspace{9pt}}}{}},
	columns/kNum/.style={column name=num., fixed, fixed zerofill,
		precision=4},
	columns/kRelDiff/.style={column name=diff.\ [\%], string type,
		column type=r,
		assign cell content/.code={%
			\pgfkeyssetvalue{/pgfplots/table/@cell content}{\relcell{##1}}}},
	columns/nAnalyt/.style={column name=analyt., fixed, precision=2,
		column type/.add={@{\hspace{9pt}}}{}},
	columns/nNum/.style={column name=num., fixed, fixed zerofill,
		precision=2},
]{\onsettable}
\caption{Analytically and numerically determined bifurcation stretches
$\lambda_{\mathrm{b}}$ and critical wavenumbers $k_{\mathrm{b}}$ for different
values of the second-gradient modulus $\xi$, each followed by its relative
difference. The last two columns give the number of waves the wavenumber makes
along the deformed specimen. A tilde marks the
rows in which the analytical wavenumber is zero, leaving nothing to compare
against. In these two rows the computed wavenumber is the lower end
$\pi/(\lambda_{\mathrm{b}}L)$ of the range searched by the fit, half a wave over
the specimen, and is not resolved beyond it.}
\label{tab:bif_onset}
\end{table}

Two trends are apparent. First, the onset is delayed as $\xi$ increases: a larger axial stretch $\lambda$ is required to reach bifurcation. Second, the critical wavenumber decreases at the same time, so the emerging pattern has a longer wavelength. Both trends are consistent with the regularizing effect of the second-gradient term, which penalizes rapid spatial variations of the deformation and therefore suppresses short-wavelength modes.

For $\xi=0.2$ and $0.4$, the onset is approached in the long-wave limit, $k\to0$. Along this neutral branch, the leading-order incremental strain is locally homogeneous. Its spatial variation tends to zero, so the second-gradient contribution to the onset condition vanishes for fixed finite $\xi$ as $k\to0$. The limiting threshold is therefore governed by \eqref{eq:onset_k0}, and, for the present parameters, coincides with the first local maximum of the homogeneous force--stretch curve at $\lambda=2.8678$. Since this curve is independent of $\xi$, increasing $\xi$ further cannot raise the critical stretch once the long-wave limit governs the onset.

Under prescribed axial stretch, however, a spatially uniform increment of the axial stretch is inadmissible. The threshold is therefore not attained by the $k=0$ mode, but is approached by nonuniform beading modes with arbitrarily long wavelengths.

\section{Numerical validation}\label{sec:num}
\subsection{Simulation setup}

The formulation was implemented in the open-source finite element software OOFEM \citep{patzak2012oofem}. The numerical simulations have two objectives. First, to assess how closely the bifurcation onset and the corresponding critical mode of a finite cylinder agree with the predictions derived in \cref{sec:app,sec:results} for an infinitely long cylinder. Second, to trace the nonlinear, spatially nonhomogeneous equilibrium branches beyond onset, including both their stable and unstable segments.

The numerical model consists of a finite cylinder with reference radius $A=1$ and length $L=20$. This length limits the influence of the ends at a manageable computational cost. The finite length and the end conditions can nevertheless shift the onset and alter the selected mode relative to the prediction for the infinite cylinder. Owing to the axisymmetry of the geometry and of the considered modes, the problem is solved on a meridional section of the cylinder.

We use a conforming displacement formulation, in which the second-gradient energy requires continuity of the first derivatives across element interfaces. Standard $C^0$ finite elements therefore cannot be used, and the meridional section is discretized using isogeometric analysis (IGA). The domain is represented by a single tensor-product B-spline patch of degree 2 with $6\times101$ control points, that is, 4 radial and 99 axial knot spans. The interior knots are simple, which gives $C^1$ continuity, and the end knots have multiplicity 3, so that the basis interpolates the boundary control points. The lateral surface is discretized by a quadratic IGA surface patch sharing the boundary degrees of freedom with the bulk patch. Thus, the bulk and surface contributions to the residual and the tangent stiffness are assembled into a single nonlinear system.

On the axis, the radial displacement is constrained, while the axial displacement remains free. Both ends are free to move radially. The axial displacement is fixed at one end, whereas at the other end all control points share a common axial displacement, whose value is determined by the loading procedure of \cref{sec:path}. These conditions eliminate the rigid-body motion. The control net and the boundary conditions are shown in \cref{fig:geometry}.
\begin{figure}[htbp]
	\centering
	{\tikzsetfigurename{geom_scheme_panel_}\input{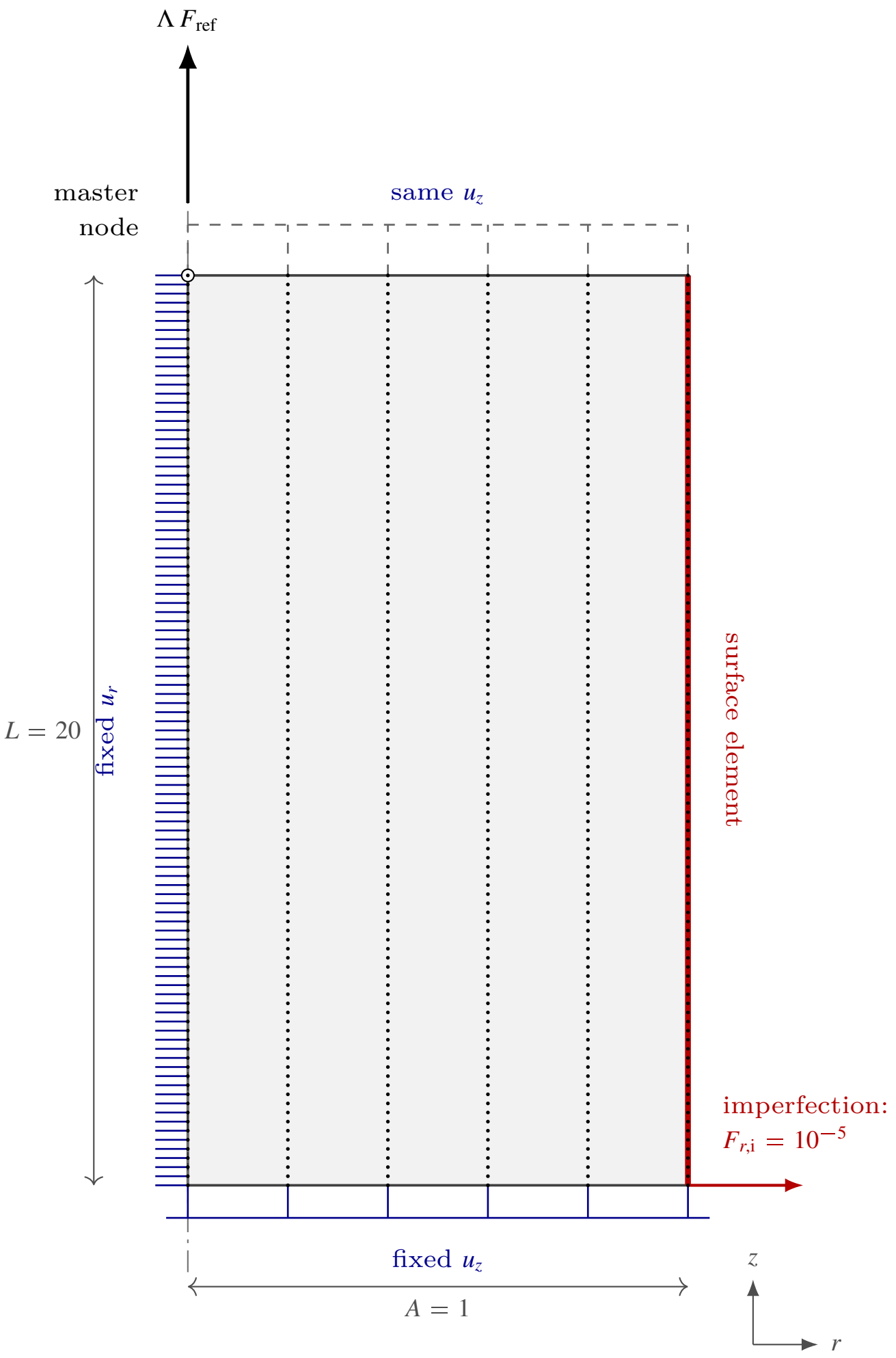}}
	\caption{Axisymmetric numerical model: the undeformed meridional
		section, the corresponding control net and the imposed boundary conditions. The
		radial scale is exaggerated relative to the axial one.}
	\label{fig:geometry}
\end{figure}

The constitutive parameters and the values of $\xi$ are those of \cref{sec:results}; a separate simulation is run for each $\xi$.

\subsection{Path following, stability and the extracted quantities}\label{sec:path}

The equilibrium paths are followed by Crisfield's arc-length method \citep{crisfield1981fast}, which allows the continuation to pass through turning points and snap-backs. The common axial displacement of the loaded end is driven by a reference axial force scaled by a load multiplier, which is treated as an additional unknown alongside the displacement degrees of freedom. The arc-length constraint is imposed on the combined increment of the displacement vector and the load multiplier, so neither the axial force nor the stretch is prescribed along the equilibrium path. 

Even though the equilibrium points on the homogeneous branch beyond the critical stretch are unstable, nothing forces the solver to depart from this branch. 
In order to trigger the bifurcation, an imperfection is therefore introduced into the model: an outward radial nodal force of $10^{-5}$ per radian is applied at the control point located at the outer surface at the fixed end of the cylinder, see \cref{fig:geometry}. The force is more than five orders of magnitude smaller than the axial force at the onset of bifurcation, so it perturbs the homogeneous state without appreciably affecting the predicted onset. The imperfection is removed once the bifurcation point has been crossed, so that the subsequent nonhomogeneous path corresponds to equilibrium of the unperturbed problem.

The tangent stiffness matrix assembled in the arc-length method is the same as under force control, with the common axial displacement of the loaded end among the unknowns. The computed equilibria do not depend on whether the force or the stretch is prescribed, but their stability does. Consistent with \cref{sec:results}, the stretch is regarded as prescribed, so stability is assessed at fixed stretch, that is, the row and column associated with the axial end displacement are removed from the matrix, while radial end motion remains admissible. The loading is conservative, so positive definiteness of this reduced matrix is sufficient for a strict local minimum of the total potential energy. %
The matrix is symmetric and is factorized at every converged state using a pivoted $\bm{L}\bm{D}\bm{L}^{\mathrm{T}}$ decomposition, with $\bm{L}$ unit lower triangular and $\bm{D}$ block diagonal with symmetric $1\times1$ and $2\times2$ blocks. By Sylvester's law of inertia, the reduced matrix and $\bm{D}$ have the same numbers of positive, negative, and zero eigenvalues, which are determined from the individual blocks of $\bm{D}$. A state is classified as stable if all blocks of $\bm{D}$ are positive definite and as unstable if at least one block has a negative eigenvalue. The classification is restricted to axisymmetric perturbations; non-axisymmetric modes are not examined.

For the comparison with the analytical prediction, each computed equilibrium path is characterized by two quantities: the stretch $\lambda_{\mathrm b}$ at which the solution departs from the homogeneous branch and the wavenumber of the emerging mode. The onset is identified as the first computed state at which the axial force departs from the homogeneous force \eqref{eq:ax_force} by more than the scatter of the first steps. The wavenumber is determined by fitting the deformed outer profile at the onset by a harmonic function. 

\subsection{Simulation results}
The results are summarized in \bifref, with one row for each value of $\xi$. The left panel in each row shows the deformed configurations at four states of the computed equilibrium path, one shortly before the onset and three after it, with the stretch indicated beneath each configuration. The right panel plots the axial force against the stretch. Each converged state is represented by a marker, colored blue for stable states and red for unstable states in the sense of \cref{sec:path}; the markers for the four configurations shown on the left are circled. The black line shows the homogeneous state of \cref{fig:cyl_equil_tikz}(b), restricted to the vicinity of the bifurcation. All deformed configurations are shown at the same scale, whereas the axis ranges in the right panels differ between rows to make each nonhomogeneous path clearly visible.

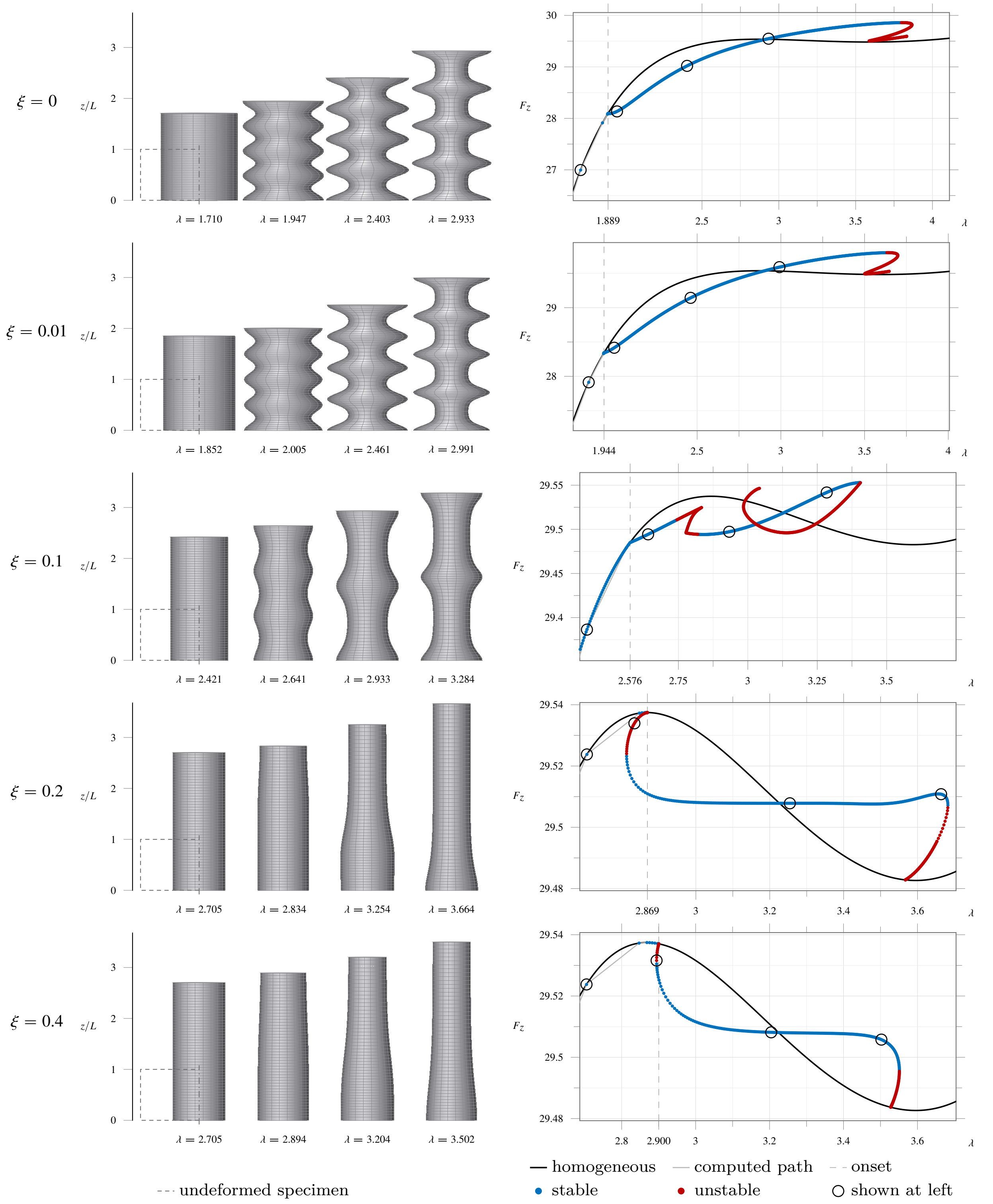
\begin{figure}[p]
	\centering
	\sbox{\scaleaxisbox}{{\tikzsetfigurename{same_axis_}	\begin{tikzpicture}[baseline={(current axis.south)}]
		\begin{axis}[
			width=1pt, height=0.19\textwidth,
			scale only axis,
			xmin=-0.0359, xmax=0.0359, ymin=-0.0068, ymax=3.6853,
			axis lines=left, tick align=outside,
			clip=false,
			axis line style={-},
			tick label style={font=\tiny},
			xlabel near ticks, ylabel near ticks,
			label style={font=\tiny},
			ylabel style={font=\tiny, rotate=-90},
			xtick={-0.025,0,0.025}, ytick={0,1,2,3},
			axis x line=none, ylabel={$z/L$},
		]
		\end{axis}
	\end{tikzpicture}%
}}%
	\sbox{\keyshapebox}{{\tikzsetfigurename{same_key_shape_}	\begin{tikzpicture}[baseline={(current bounding box.south)}]
		\begin{axis}[
			hide axis, width=1pt, height=1pt, scale only axis,
			xmin=0, xmax=1, ymin=0, ymax=1,
			legend columns=1, legend cell align=left,
			legend style={at={(0,0)}, anchor=north west, draw=none, fill=none,
				font=\footnotesize, inner sep=0pt,
				/tikz/every even column/.append style={column sep=2mm}},
			legend image code/.code={\draw[mark repeat=2, mark phase=2, #1]
				plot coordinates {(0pt,0pt) (4pt,0pt) (8pt,0pt)};},
		]
			\addplot[draw=black!55, dashed, line width=0.4pt, dash pattern=on 2pt off 1.6pt] coordinates {(9,9)};
			\addlegendentry{undeformed specimen}
		\end{axis}
	\end{tikzpicture}%
}}%
	\sbox{\keycurvebox}{{\tikzsetfigurename{same_key_curve_}	\begin{tikzpicture}[baseline={(current bounding box.south)}]
		\begin{axis}[
			hide axis, width=1pt, height=1pt, scale only axis,
			xmin=0, xmax=1, ymin=0, ymax=1,
			legend columns=3, legend cell align=left,
			legend style={at={(0,0)}, anchor=north west, draw=none, fill=none,
				font=\footnotesize, inner sep=0pt,
				/tikz/every even column/.append style={column sep=2mm}},
			legend image code/.code={\draw[mark repeat=2, mark phase=2, #1]
				plot coordinates {(0pt,0pt) (4pt,0pt) (8pt,0pt)};},
		]
			\addplot[black, line width=0.7pt] coordinates {(9,9)};
			\addlegendentry{homogeneous}
			\addplot[gray!55, line width=0.5pt] coordinates {(9,9)};
			\addlegendentry{computed path}
			\addplot[dashed, draw=black!35, line width=0.3pt] coordinates {(9,9)};
			\addlegendentry{onset}
			\addplot[only marks, mark=*, mark size=0.55pt, color=matlabblue, mark size=1.4pt] coordinates {(9,9)};
			\addlegendentry{stable}
			\addplot[only marks, mark=*, mark size=0.55pt, color=red!75!black, mark size=1.4pt] coordinates {(9,9)};
			\addlegendentry{unstable}
			\addplot[only marks, mark=o, mark size=2.6pt, black, line width=0.5pt] coordinates {(9,9)};
			\addlegendentry{shown at left}
		\end{axis}
	\end{tikzpicture}%
}}%
	\begin{tabular}{@{}c@{\hspace{1mm}}c@{\hspace{4mm}}c@{\hspace{2mm}}c@{}}
		\rowlab{0} &
		\usebox{\scaleaxisbox} &
		{\tikzsetfigurename{same_shade_xi0_fullfwdprod_panels_}	\begin{tabular}[b]{@{}c@{}}%
		\begin{tikzpicture}[baseline={(current axis.south)}, trim axis left]
			\begin{axis}[
				width=0.085\textwidth, height=0.19\textwidth,
				scale only axis, view={0}{90},
				xmin=-0.0359, xmax=0.0359, ymin=-0.0068, ymax=3.6853,
				axis lines=left,
				axis line style={draw=none},
				tick style={draw=none},
				xtick=\empty, ytick=\empty,
				clip=false,
				colormap name=shine,
				point meta min=0, point meta max=1,
				unbounded coords=jump,
			]
				\addplot3[surf, shader=interp, point meta=explicit,
					mesh/cols=50, mesh/ordering=rowwise]
					table[x index=0, y index=1, z index=2, meta index=3] {paper_3/tikz_data/same_shade_xi0_fullfwdprod_p01.dat};
				\addplot3[surf, shader=interp, point meta=explicit,
					mesh/cols=49, mesh/ordering=rowwise]
					table[x index=0, y index=1, z index=2, meta index=3] {paper_3/tikz_data/same_shade_xi0_fullfwdprod_p01_cap.dat};
				\addplot[black!45, line width=0.15pt]
					table[x index=0, y index=1] {paper_3/tikz_data/same_shade_xi0_fullfwdprod_p01_wire.dat};
				\draw[draw=black!55, dashed, line width=0.4pt, dash pattern=on 2pt off 1.6pt]
					(axis cs:0,0) -- (axis cs:-0.05000,0) -- (axis cs:-0.05000,1) -- (axis cs:0,1);
				\draw[draw=black!55, line width=0.4pt, dash pattern=on 3pt off 1.4pt on 0.6pt off 1.4pt]
					(axis cs:0,-0.08000) -- (axis cs:0,1.08000);
				\node[anchor=north, font=\tiny, inner sep=0pt, yshift=-2.4mm]
					at (rel axis cs:0.5,0) {$\lambda = 1.710$};
			\end{axis}
		\end{tikzpicture}%
	\end{tabular}%
	\begin{tabular}[b]{@{}c@{}}%
		\begin{tikzpicture}[baseline={(current axis.south)}]
			\begin{axis}[
				width=0.085\textwidth, height=0.19\textwidth,
				scale only axis, view={0}{90},
				xmin=-0.0359, xmax=0.0359, ymin=-0.0068, ymax=3.6853,
				axis lines=left,
				axis line style={draw=none},
				tick style={draw=none},
				xtick=\empty, ytick=\empty,
				clip=false,
				colormap name=shine,
				point meta min=0, point meta max=1,
				unbounded coords=jump,
			]
				\addplot3[surf, shader=interp, point meta=explicit,
					mesh/cols=50, mesh/ordering=rowwise]
					table[x index=0, y index=1, z index=2, meta index=3] {paper_3/tikz_data/same_shade_xi0_fullfwdprod_p02.dat};
				\addplot3[surf, shader=interp, point meta=explicit,
					mesh/cols=49, mesh/ordering=rowwise]
					table[x index=0, y index=1, z index=2, meta index=3] {paper_3/tikz_data/same_shade_xi0_fullfwdprod_p02_cap.dat};
				\addplot[black!45, line width=0.15pt]
					table[x index=0, y index=1] {paper_3/tikz_data/same_shade_xi0_fullfwdprod_p02_wire.dat};
				\node[anchor=north, font=\tiny, inner sep=0pt, yshift=-2.4mm]
					at (rel axis cs:0.5,0) {$\lambda = 1.947$};
			\end{axis}
		\end{tikzpicture}%
	\end{tabular}%
	\begin{tabular}[b]{@{}c@{}}%
		\begin{tikzpicture}[baseline={(current axis.south)}]
			\begin{axis}[
				width=0.085\textwidth, height=0.19\textwidth,
				scale only axis, view={0}{90},
				xmin=-0.0359, xmax=0.0359, ymin=-0.0068, ymax=3.6853,
				axis lines=left,
				axis line style={draw=none},
				tick style={draw=none},
				xtick=\empty, ytick=\empty,
				clip=false,
				colormap name=shine,
				point meta min=0, point meta max=1,
				unbounded coords=jump,
			]
				\addplot3[surf, shader=interp, point meta=explicit,
					mesh/cols=50, mesh/ordering=rowwise]
					table[x index=0, y index=1, z index=2, meta index=3] {paper_3/tikz_data/same_shade_xi0_fullfwdprod_p03.dat};
				\addplot3[surf, shader=interp, point meta=explicit,
					mesh/cols=49, mesh/ordering=rowwise]
					table[x index=0, y index=1, z index=2, meta index=3] {paper_3/tikz_data/same_shade_xi0_fullfwdprod_p03_cap.dat};
				\addplot[black!45, line width=0.15pt]
					table[x index=0, y index=1] {paper_3/tikz_data/same_shade_xi0_fullfwdprod_p03_wire.dat};
				\node[anchor=north, font=\tiny, inner sep=0pt, yshift=-2.4mm]
					at (rel axis cs:0.5,0) {$\lambda = 2.403$};
			\end{axis}
		\end{tikzpicture}%
	\end{tabular}%
	\begin{tabular}[b]{@{}c@{}}%
		\begin{tikzpicture}[baseline={(current axis.south)}]
			\begin{axis}[
				width=0.085\textwidth, height=0.19\textwidth,
				scale only axis, view={0}{90},
				xmin=-0.0359, xmax=0.0359, ymin=-0.0068, ymax=3.6853,
				axis lines=left,
				axis line style={draw=none},
				tick style={draw=none},
				xtick=\empty, ytick=\empty,
				clip=false,
				colormap name=shine,
				point meta min=0, point meta max=1,
				unbounded coords=jump,
			]
				\addplot3[surf, shader=interp, point meta=explicit,
					mesh/cols=50, mesh/ordering=rowwise]
					table[x index=0, y index=1, z index=2, meta index=3] {paper_3/tikz_data/same_shade_xi0_fullfwdprod_p04.dat};
				\addplot3[surf, shader=interp, point meta=explicit,
					mesh/cols=49, mesh/ordering=rowwise]
					table[x index=0, y index=1, z index=2, meta index=3] {paper_3/tikz_data/same_shade_xi0_fullfwdprod_p04_cap.dat};
				\addplot[black!45, line width=0.15pt]
					table[x index=0, y index=1] {paper_3/tikz_data/same_shade_xi0_fullfwdprod_p04_wire.dat};
				\node[anchor=north, font=\tiny, inner sep=0pt, yshift=-2.4mm]
					at (rel axis cs:0.5,0) {$\lambda = 2.933$};
			\end{axis}
		\end{tikzpicture}%
	\end{tabular}%
} &
		{\tikzsetfigurename{pairs_curve_xi0_fullfwdprod_}	\begin{tikzpicture}[baseline={(current axis.south)}]
		\begin{axis}[
			width=0.38\textwidth, height=0.19\textwidth,
			scale only axis,
			xlabel={$\lambda$}, ylabel={$F_z$},
			xlabel style={at={(ticklabel cs:1)}, anchor=base west,
				xshift=0.8mm, yshift=3.9mm, font=\tiny},
			ylabel style={rotate=-90},
			xmin=1.6617, xmax=4.1106, ymin=26.4009, ymax=30.0554,
			tick label style={font=\tiny},
			label style={font=\tiny},
			xtick={2.5,3,3.5,4},
			grid=both, minor tick num=1,
			major grid style={line width=0.2pt, draw=black!15},
			minor grid style={line width=0.2pt, draw=black!7},
			tick align=outside,
			scaled y ticks=false,
			y tick label style={/pgf/number format/fixed},
			every axis plot post/.append style={line join=round, line cap=round},
			restrict y to domain=11.7830:44.6733,
			unbounded coords=jump,
			axis line style={draw=black!55},
			extra x ticks={1.8886},
			extra x tick labels={$1.889$},
			extra x tick style={grid=major, major grid style={dashed, draw=black!35, line width=0.3pt},
				tick label style={font=\tiny}},
		]
			\addplot[black, line width=0.7pt, forget plot]
				table[x index=0, y index=1] {paper_3/tikz_data/pairs_curve_xi0_fullfwdprod_hom.dat};
			\addplot[gray!55, line width=0.5pt, forget plot]
				table[x index=0, y index=1] {paper_3/tikz_data/pairs_curve_xi0_fullfwdprod.dat};
			\addplot[only marks, mark=*, mark size=0.55pt, color=matlabblue]
				table[x index=0, y index=1] {paper_3/tikz_data/pairs_curve_xi0_fullfwdprod_stable.dat};
			\addplot[only marks, mark=*, mark size=0.55pt, color=red!75!black]
				table[x index=0, y index=1] {paper_3/tikz_data/pairs_curve_xi0_fullfwdprod_unstable.dat};
			\addplot[only marks, mark=o, mark size=2.6pt, black, line width=0.5pt]
				table[x index=0, y index=1] {paper_3/tikz_data/pairs_curve_xi0_fullfwdprod_marks.dat};
		\end{axis}
	\end{tikzpicture}%
} \\[1.5mm]
		\rowlab{0.01} &
		\usebox{\scaleaxisbox} &
		{\tikzsetfigurename{same_shade_xi001_fullfwdprod_panels_}	\begin{tabular}[b]{@{}c@{}}%
		\begin{tikzpicture}[baseline={(current axis.south)}, trim axis left]
			\begin{axis}[
				width=0.085\textwidth, height=0.19\textwidth,
				scale only axis, view={0}{90},
				xmin=-0.0359, xmax=0.0359, ymin=-0.0068, ymax=3.6853,
				axis lines=left,
				axis line style={draw=none},
				tick style={draw=none},
				xtick=\empty, ytick=\empty,
				clip=false,
				colormap name=shine,
				point meta min=0, point meta max=1,
				unbounded coords=jump,
			]
				\addplot3[surf, shader=interp, point meta=explicit,
					mesh/cols=50, mesh/ordering=rowwise]
					table[x index=0, y index=1, z index=2, meta index=3] {paper_3/tikz_data/same_shade_xi001_fullfwdprod_p01.dat};
				\addplot3[surf, shader=interp, point meta=explicit,
					mesh/cols=49, mesh/ordering=rowwise]
					table[x index=0, y index=1, z index=2, meta index=3] {paper_3/tikz_data/same_shade_xi001_fullfwdprod_p01_cap.dat};
				\addplot[black!45, line width=0.15pt]
					table[x index=0, y index=1] {paper_3/tikz_data/same_shade_xi001_fullfwdprod_p01_wire.dat};
				\draw[draw=black!55, dashed, line width=0.4pt, dash pattern=on 2pt off 1.6pt]
					(axis cs:0,0) -- (axis cs:-0.05000,0) -- (axis cs:-0.05000,1) -- (axis cs:0,1);
				\draw[draw=black!55, line width=0.4pt, dash pattern=on 3pt off 1.4pt on 0.6pt off 1.4pt]
					(axis cs:0,-0.08000) -- (axis cs:0,1.08000);
				\node[anchor=north, font=\tiny, inner sep=0pt, yshift=-2.4mm]
					at (rel axis cs:0.5,0) {$\lambda = 1.852$};
			\end{axis}
		\end{tikzpicture}%
	\end{tabular}%
	\begin{tabular}[b]{@{}c@{}}%
		\begin{tikzpicture}[baseline={(current axis.south)}]
			\begin{axis}[
				width=0.085\textwidth, height=0.19\textwidth,
				scale only axis, view={0}{90},
				xmin=-0.0359, xmax=0.0359, ymin=-0.0068, ymax=3.6853,
				axis lines=left,
				axis line style={draw=none},
				tick style={draw=none},
				xtick=\empty, ytick=\empty,
				clip=false,
				colormap name=shine,
				point meta min=0, point meta max=1,
				unbounded coords=jump,
			]
				\addplot3[surf, shader=interp, point meta=explicit,
					mesh/cols=50, mesh/ordering=rowwise]
					table[x index=0, y index=1, z index=2, meta index=3] {paper_3/tikz_data/same_shade_xi001_fullfwdprod_p02.dat};
				\addplot3[surf, shader=interp, point meta=explicit,
					mesh/cols=49, mesh/ordering=rowwise]
					table[x index=0, y index=1, z index=2, meta index=3] {paper_3/tikz_data/same_shade_xi001_fullfwdprod_p02_cap.dat};
				\addplot[black!45, line width=0.15pt]
					table[x index=0, y index=1] {paper_3/tikz_data/same_shade_xi001_fullfwdprod_p02_wire.dat};
				\node[anchor=north, font=\tiny, inner sep=0pt, yshift=-2.4mm]
					at (rel axis cs:0.5,0) {$\lambda = 2.005$};
			\end{axis}
		\end{tikzpicture}%
	\end{tabular}%
	\begin{tabular}[b]{@{}c@{}}%
		\begin{tikzpicture}[baseline={(current axis.south)}]
			\begin{axis}[
				width=0.085\textwidth, height=0.19\textwidth,
				scale only axis, view={0}{90},
				xmin=-0.0359, xmax=0.0359, ymin=-0.0068, ymax=3.6853,
				axis lines=left,
				axis line style={draw=none},
				tick style={draw=none},
				xtick=\empty, ytick=\empty,
				clip=false,
				colormap name=shine,
				point meta min=0, point meta max=1,
				unbounded coords=jump,
			]
				\addplot3[surf, shader=interp, point meta=explicit,
					mesh/cols=50, mesh/ordering=rowwise]
					table[x index=0, y index=1, z index=2, meta index=3] {paper_3/tikz_data/same_shade_xi001_fullfwdprod_p03.dat};
				\addplot3[surf, shader=interp, point meta=explicit,
					mesh/cols=49, mesh/ordering=rowwise]
					table[x index=0, y index=1, z index=2, meta index=3] {paper_3/tikz_data/same_shade_xi001_fullfwdprod_p03_cap.dat};
				\addplot[black!45, line width=0.15pt]
					table[x index=0, y index=1] {paper_3/tikz_data/same_shade_xi001_fullfwdprod_p03_wire.dat};
				\node[anchor=north, font=\tiny, inner sep=0pt, yshift=-2.4mm]
					at (rel axis cs:0.5,0) {$\lambda = 2.461$};
			\end{axis}
		\end{tikzpicture}%
	\end{tabular}%
	\begin{tabular}[b]{@{}c@{}}%
		\begin{tikzpicture}[baseline={(current axis.south)}]
			\begin{axis}[
				width=0.085\textwidth, height=0.19\textwidth,
				scale only axis, view={0}{90},
				xmin=-0.0359, xmax=0.0359, ymin=-0.0068, ymax=3.6853,
				axis lines=left,
				axis line style={draw=none},
				tick style={draw=none},
				xtick=\empty, ytick=\empty,
				clip=false,
				colormap name=shine,
				point meta min=0, point meta max=1,
				unbounded coords=jump,
			]
				\addplot3[surf, shader=interp, point meta=explicit,
					mesh/cols=50, mesh/ordering=rowwise]
					table[x index=0, y index=1, z index=2, meta index=3] {paper_3/tikz_data/same_shade_xi001_fullfwdprod_p04.dat};
				\addplot3[surf, shader=interp, point meta=explicit,
					mesh/cols=49, mesh/ordering=rowwise]
					table[x index=0, y index=1, z index=2, meta index=3] {paper_3/tikz_data/same_shade_xi001_fullfwdprod_p04_cap.dat};
				\addplot[black!45, line width=0.15pt]
					table[x index=0, y index=1] {paper_3/tikz_data/same_shade_xi001_fullfwdprod_p04_wire.dat};
				\node[anchor=north, font=\tiny, inner sep=0pt, yshift=-2.4mm]
					at (rel axis cs:0.5,0) {$\lambda = 2.991$};
			\end{axis}
		\end{tikzpicture}%
	\end{tabular}%
} &
		{\tikzsetfigurename{pairs_curve_xi001_fullfwdprod_}	\begin{tikzpicture}[baseline={(current axis.south)}]
		\begin{axis}[
			width=0.38\textwidth, height=0.19\textwidth,
			scale only axis,
			xlabel={$\lambda$}, ylabel={$F_z$},
			xlabel style={at={(ticklabel cs:1)}, anchor=base west,
				xshift=0.8mm, yshift=3.9mm, font=\tiny},
			ylabel style={rotate=-90},
			xmin=1.7596, xmax=4.0071, ymin=27.2079, ymax=29.9484,
			tick label style={font=\tiny},
			label style={font=\tiny},
			xtick={2.5,3,3.5,4},
			grid=both, minor tick num=1,
			major grid style={line width=0.2pt, draw=black!15},
			minor grid style={line width=0.2pt, draw=black!7},
			tick align=outside,
			scaled y ticks=false,
			y tick label style={/pgf/number format/fixed},
			every axis plot post/.append style={line join=round, line cap=round},
			restrict y to domain=16.2459:40.9104,
			unbounded coords=jump,
			axis line style={draw=black!55},
			extra x ticks={1.9437},
			extra x tick labels={$1.944$},
			extra x tick style={grid=major, major grid style={dashed, draw=black!35, line width=0.3pt},
				tick label style={font=\tiny}},
		]
			\addplot[black, line width=0.7pt, forget plot]
				table[x index=0, y index=1] {paper_3/tikz_data/pairs_curve_xi001_fullfwdprod_hom.dat};
			\addplot[gray!55, line width=0.5pt, forget plot]
				table[x index=0, y index=1] {paper_3/tikz_data/pairs_curve_xi001_fullfwdprod.dat};
			\addplot[only marks, mark=*, mark size=0.55pt, color=matlabblue]
				table[x index=0, y index=1] {paper_3/tikz_data/pairs_curve_xi001_fullfwdprod_stable.dat};
			\addplot[only marks, mark=*, mark size=0.55pt, color=red!75!black]
				table[x index=0, y index=1] {paper_3/tikz_data/pairs_curve_xi001_fullfwdprod_unstable.dat};
			\addplot[only marks, mark=o, mark size=2.6pt, black, line width=0.5pt]
				table[x index=0, y index=1] {paper_3/tikz_data/pairs_curve_xi001_fullfwdprod_marks.dat};
		\end{axis}
	\end{tikzpicture}%
} \\[1.5mm]
		\rowlab{0.1} &
		\usebox{\scaleaxisbox} &
		{\tikzsetfigurename{same_shade_xi01_fullfwdprod_panels_}	\begin{tabular}[b]{@{}c@{}}%
		\begin{tikzpicture}[baseline={(current axis.south)}, trim axis left]
			\begin{axis}[
				width=0.085\textwidth, height=0.19\textwidth,
				scale only axis, view={0}{90},
				xmin=-0.0359, xmax=0.0359, ymin=-0.0068, ymax=3.6853,
				axis lines=left,
				axis line style={draw=none},
				tick style={draw=none},
				xtick=\empty, ytick=\empty,
				clip=false,
				colormap name=shine,
				point meta min=0, point meta max=1,
				unbounded coords=jump,
			]
				\addplot3[surf, shader=interp, point meta=explicit,
					mesh/cols=50, mesh/ordering=rowwise]
					table[x index=0, y index=1, z index=2, meta index=3] {paper_3/tikz_data/same_shade_xi01_fullfwdprod_p01.dat};
				\addplot3[surf, shader=interp, point meta=explicit,
					mesh/cols=49, mesh/ordering=rowwise]
					table[x index=0, y index=1, z index=2, meta index=3] {paper_3/tikz_data/same_shade_xi01_fullfwdprod_p01_cap.dat};
				\addplot[black!45, line width=0.15pt]
					table[x index=0, y index=1] {paper_3/tikz_data/same_shade_xi01_fullfwdprod_p01_wire.dat};
				\draw[draw=black!55, dashed, line width=0.4pt, dash pattern=on 2pt off 1.6pt]
					(axis cs:0,0) -- (axis cs:-0.05000,0) -- (axis cs:-0.05000,1) -- (axis cs:0,1);
				\draw[draw=black!55, line width=0.4pt, dash pattern=on 3pt off 1.4pt on 0.6pt off 1.4pt]
					(axis cs:0,-0.08000) -- (axis cs:0,1.08000);
				\node[anchor=north, font=\tiny, inner sep=0pt, yshift=-2.4mm]
					at (rel axis cs:0.5,0) {$\lambda = 2.421$};
			\end{axis}
		\end{tikzpicture}%
	\end{tabular}%
	\begin{tabular}[b]{@{}c@{}}%
		\begin{tikzpicture}[baseline={(current axis.south)}]
			\begin{axis}[
				width=0.085\textwidth, height=0.19\textwidth,
				scale only axis, view={0}{90},
				xmin=-0.0359, xmax=0.0359, ymin=-0.0068, ymax=3.6853,
				axis lines=left,
				axis line style={draw=none},
				tick style={draw=none},
				xtick=\empty, ytick=\empty,
				clip=false,
				colormap name=shine,
				point meta min=0, point meta max=1,
				unbounded coords=jump,
			]
				\addplot3[surf, shader=interp, point meta=explicit,
					mesh/cols=50, mesh/ordering=rowwise]
					table[x index=0, y index=1, z index=2, meta index=3] {paper_3/tikz_data/same_shade_xi01_fullfwdprod_p02.dat};
				\addplot3[surf, shader=interp, point meta=explicit,
					mesh/cols=49, mesh/ordering=rowwise]
					table[x index=0, y index=1, z index=2, meta index=3] {paper_3/tikz_data/same_shade_xi01_fullfwdprod_p02_cap.dat};
				\addplot[black!45, line width=0.15pt]
					table[x index=0, y index=1] {paper_3/tikz_data/same_shade_xi01_fullfwdprod_p02_wire.dat};
				\node[anchor=north, font=\tiny, inner sep=0pt, yshift=-2.4mm]
					at (rel axis cs:0.5,0) {$\lambda = 2.641$};
			\end{axis}
		\end{tikzpicture}%
	\end{tabular}%
	\begin{tabular}[b]{@{}c@{}}%
		\begin{tikzpicture}[baseline={(current axis.south)}]
			\begin{axis}[
				width=0.085\textwidth, height=0.19\textwidth,
				scale only axis, view={0}{90},
				xmin=-0.0359, xmax=0.0359, ymin=-0.0068, ymax=3.6853,
				axis lines=left,
				axis line style={draw=none},
				tick style={draw=none},
				xtick=\empty, ytick=\empty,
				clip=false,
				colormap name=shine,
				point meta min=0, point meta max=1,
				unbounded coords=jump,
			]
				\addplot3[surf, shader=interp, point meta=explicit,
					mesh/cols=50, mesh/ordering=rowwise]
					table[x index=0, y index=1, z index=2, meta index=3] {paper_3/tikz_data/same_shade_xi01_fullfwdprod_p03.dat};
				\addplot3[surf, shader=interp, point meta=explicit,
					mesh/cols=49, mesh/ordering=rowwise]
					table[x index=0, y index=1, z index=2, meta index=3] {paper_3/tikz_data/same_shade_xi01_fullfwdprod_p03_cap.dat};
				\addplot[black!45, line width=0.15pt]
					table[x index=0, y index=1] {paper_3/tikz_data/same_shade_xi01_fullfwdprod_p03_wire.dat};
				\node[anchor=north, font=\tiny, inner sep=0pt, yshift=-2.4mm]
					at (rel axis cs:0.5,0) {$\lambda = 2.933$};
			\end{axis}
		\end{tikzpicture}%
	\end{tabular}%
	\begin{tabular}[b]{@{}c@{}}%
		\begin{tikzpicture}[baseline={(current axis.south)}]
			\begin{axis}[
				width=0.085\textwidth, height=0.19\textwidth,
				scale only axis, view={0}{90},
				xmin=-0.0359, xmax=0.0359, ymin=-0.0068, ymax=3.6853,
				axis lines=left,
				axis line style={draw=none},
				tick style={draw=none},
				xtick=\empty, ytick=\empty,
				clip=false,
				colormap name=shine,
				point meta min=0, point meta max=1,
				unbounded coords=jump,
			]
				\addplot3[surf, shader=interp, point meta=explicit,
					mesh/cols=50, mesh/ordering=rowwise]
					table[x index=0, y index=1, z index=2, meta index=3] {paper_3/tikz_data/same_shade_xi01_fullfwdprod_p04.dat};
				\addplot3[surf, shader=interp, point meta=explicit,
					mesh/cols=49, mesh/ordering=rowwise]
					table[x index=0, y index=1, z index=2, meta index=3] {paper_3/tikz_data/same_shade_xi01_fullfwdprod_p04_cap.dat};
				\addplot[black!45, line width=0.15pt]
					table[x index=0, y index=1] {paper_3/tikz_data/same_shade_xi01_fullfwdprod_p04_wire.dat};
				\node[anchor=north, font=\tiny, inner sep=0pt, yshift=-2.4mm]
					at (rel axis cs:0.5,0) {$\lambda = 3.284$};
			\end{axis}
		\end{tikzpicture}%
	\end{tabular}%
} &
		{\tikzsetfigurename{pairs_curve_xi01_fullfwdprod_}	\begin{tikzpicture}[baseline={(current axis.south)}]
		\begin{axis}[
			width=0.38\textwidth, height=0.19\textwidth,
			scale only axis,
			xlabel={$\lambda$}, ylabel={$F_z$},
			xlabel style={at={(ticklabel cs:1)}, anchor=base west,
				xshift=0.8mm, yshift=3.9mm, font=\tiny},
			ylabel style={rotate=-90},
			xmin=2.3950, xmax=3.7499, ymin=29.3512, ymax=29.5643,
			tick label style={font=\tiny},
			label style={font=\tiny},
			xtick={2.75,3,3.25,3.5},
			grid=both, minor tick num=1,
			major grid style={line width=0.2pt, draw=black!15},
			minor grid style={line width=0.2pt, draw=black!7},
			tick align=outside,
			scaled y ticks=false,
			y tick label style={/pgf/number format/fixed},
			every axis plot post/.append style={line join=round, line cap=round},
			restrict y to domain=28.4987:30.4168,
			unbounded coords=jump,
			axis line style={draw=black!55},
			extra x ticks={2.5763},
			extra x tick labels={$2.576$},
			extra x tick style={grid=major, major grid style={dashed, draw=black!35, line width=0.3pt},
				tick label style={font=\tiny}},
		]
			\addplot[black, line width=0.7pt, forget plot]
				table[x index=0, y index=1] {paper_3/tikz_data/pairs_curve_xi01_fullfwdprod_hom.dat};
			\addplot[gray!55, line width=0.5pt, forget plot]
				table[x index=0, y index=1] {paper_3/tikz_data/pairs_curve_xi01_fullfwdprod.dat};
			\addplot[only marks, mark=*, mark size=0.55pt, color=matlabblue]
				table[x index=0, y index=1] {paper_3/tikz_data/pairs_curve_xi01_fullfwdprod_stable.dat};
			\addplot[only marks, mark=*, mark size=0.55pt, color=red!75!black]
				table[x index=0, y index=1] {paper_3/tikz_data/pairs_curve_xi01_fullfwdprod_unstable.dat};
			\addplot[only marks, mark=o, mark size=2.6pt, black, line width=0.5pt]
				table[x index=0, y index=1] {paper_3/tikz_data/pairs_curve_xi01_fullfwdprod_marks.dat};
		\end{axis}
	\end{tikzpicture}%
} \\[1.5mm]
		\rowlab{0.2} &
		\usebox{\scaleaxisbox} &
		{\tikzsetfigurename{same_shade_xi02_fullfwdprod_panels_}	\begin{tabular}[b]{@{}c@{}}%
		\begin{tikzpicture}[baseline={(current axis.south)}, trim axis left]
			\begin{axis}[
				width=0.085\textwidth, height=0.19\textwidth,
				scale only axis, view={0}{90},
				xmin=-0.0359, xmax=0.0359, ymin=-0.0068, ymax=3.6853,
				axis lines=left,
				axis line style={draw=none},
				tick style={draw=none},
				xtick=\empty, ytick=\empty,
				clip=false,
				colormap name=shine,
				point meta min=0, point meta max=1,
				unbounded coords=jump,
			]
				\addplot3[surf, shader=interp, point meta=explicit,
					mesh/cols=50, mesh/ordering=rowwise]
					table[x index=0, y index=1, z index=2, meta index=3] {paper_3/tikz_data/same_shade_xi02_fullfwdprod_p01.dat};
				\addplot3[surf, shader=interp, point meta=explicit,
					mesh/cols=49, mesh/ordering=rowwise]
					table[x index=0, y index=1, z index=2, meta index=3] {paper_3/tikz_data/same_shade_xi02_fullfwdprod_p01_cap.dat};
				\addplot[black!45, line width=0.15pt]
					table[x index=0, y index=1] {paper_3/tikz_data/same_shade_xi02_fullfwdprod_p01_wire.dat};
				\draw[draw=black!55, dashed, line width=0.4pt, dash pattern=on 2pt off 1.6pt]
					(axis cs:0,0) -- (axis cs:-0.05000,0) -- (axis cs:-0.05000,1) -- (axis cs:0,1);
				\draw[draw=black!55, line width=0.4pt, dash pattern=on 3pt off 1.4pt on 0.6pt off 1.4pt]
					(axis cs:0,-0.08000) -- (axis cs:0,1.08000);
				\node[anchor=north, font=\tiny, inner sep=0pt, yshift=-2.4mm]
					at (rel axis cs:0.5,0) {$\lambda = 2.705$};
			\end{axis}
		\end{tikzpicture}%
	\end{tabular}%
	\begin{tabular}[b]{@{}c@{}}%
		\begin{tikzpicture}[baseline={(current axis.south)}]
			\begin{axis}[
				width=0.085\textwidth, height=0.19\textwidth,
				scale only axis, view={0}{90},
				xmin=-0.0359, xmax=0.0359, ymin=-0.0068, ymax=3.6853,
				axis lines=left,
				axis line style={draw=none},
				tick style={draw=none},
				xtick=\empty, ytick=\empty,
				clip=false,
				colormap name=shine,
				point meta min=0, point meta max=1,
				unbounded coords=jump,
			]
				\addplot3[surf, shader=interp, point meta=explicit,
					mesh/cols=50, mesh/ordering=rowwise]
					table[x index=0, y index=1, z index=2, meta index=3] {paper_3/tikz_data/same_shade_xi02_fullfwdprod_p02.dat};
				\addplot3[surf, shader=interp, point meta=explicit,
					mesh/cols=49, mesh/ordering=rowwise]
					table[x index=0, y index=1, z index=2, meta index=3] {paper_3/tikz_data/same_shade_xi02_fullfwdprod_p02_cap.dat};
				\addplot[black!45, line width=0.15pt]
					table[x index=0, y index=1] {paper_3/tikz_data/same_shade_xi02_fullfwdprod_p02_wire.dat};
				\node[anchor=north, font=\tiny, inner sep=0pt, yshift=-2.4mm]
					at (rel axis cs:0.5,0) {$\lambda = 2.834$};
			\end{axis}
		\end{tikzpicture}%
	\end{tabular}%
	\begin{tabular}[b]{@{}c@{}}%
		\begin{tikzpicture}[baseline={(current axis.south)}]
			\begin{axis}[
				width=0.085\textwidth, height=0.19\textwidth,
				scale only axis, view={0}{90},
				xmin=-0.0359, xmax=0.0359, ymin=-0.0068, ymax=3.6853,
				axis lines=left,
				axis line style={draw=none},
				tick style={draw=none},
				xtick=\empty, ytick=\empty,
				clip=false,
				colormap name=shine,
				point meta min=0, point meta max=1,
				unbounded coords=jump,
			]
				\addplot3[surf, shader=interp, point meta=explicit,
					mesh/cols=50, mesh/ordering=rowwise]
					table[x index=0, y index=1, z index=2, meta index=3] {paper_3/tikz_data/same_shade_xi02_fullfwdprod_p03.dat};
				\addplot3[surf, shader=interp, point meta=explicit,
					mesh/cols=49, mesh/ordering=rowwise]
					table[x index=0, y index=1, z index=2, meta index=3] {paper_3/tikz_data/same_shade_xi02_fullfwdprod_p03_cap.dat};
				\addplot[black!45, line width=0.15pt]
					table[x index=0, y index=1] {paper_3/tikz_data/same_shade_xi02_fullfwdprod_p03_wire.dat};
				\node[anchor=north, font=\tiny, inner sep=0pt, yshift=-2.4mm]
					at (rel axis cs:0.5,0) {$\lambda = 3.254$};
			\end{axis}
		\end{tikzpicture}%
	\end{tabular}%
	\begin{tabular}[b]{@{}c@{}}%
		\begin{tikzpicture}[baseline={(current axis.south)}]
			\begin{axis}[
				width=0.085\textwidth, height=0.19\textwidth,
				scale only axis, view={0}{90},
				xmin=-0.0359, xmax=0.0359, ymin=-0.0068, ymax=3.6853,
				axis lines=left,
				axis line style={draw=none},
				tick style={draw=none},
				xtick=\empty, ytick=\empty,
				clip=false,
				colormap name=shine,
				point meta min=0, point meta max=1,
				unbounded coords=jump,
			]
				\addplot3[surf, shader=interp, point meta=explicit,
					mesh/cols=50, mesh/ordering=rowwise]
					table[x index=0, y index=1, z index=2, meta index=3] {paper_3/tikz_data/same_shade_xi02_fullfwdprod_p04.dat};
				\addplot3[surf, shader=interp, point meta=explicit,
					mesh/cols=49, mesh/ordering=rowwise]
					table[x index=0, y index=1, z index=2, meta index=3] {paper_3/tikz_data/same_shade_xi02_fullfwdprod_p04_cap.dat};
				\addplot[black!45, line width=0.15pt]
					table[x index=0, y index=1] {paper_3/tikz_data/same_shade_xi02_fullfwdprod_p04_wire.dat};
				\node[anchor=north, font=\tiny, inner sep=0pt, yshift=-2.4mm]
					at (rel axis cs:0.5,0) {$\lambda = 3.664$};
			\end{axis}
		\end{tikzpicture}%
	\end{tabular}%
} &
		{\tikzsetfigurename{pairs_curve_xi02_fullfwdprod_}	\begin{tikzpicture}[baseline={(current axis.south)}]
		\begin{axis}[
			width=0.38\textwidth, height=0.19\textwidth,
			scale only axis,
			xlabel={$\lambda$}, ylabel={$F_z$},
			xlabel style={at={(ticklabel cs:1)}, anchor=base west,
				xshift=0.8mm, yshift=3.9mm, font=\tiny},
			ylabel style={rotate=-90},
			xmin=2.6858, xmax=3.7044, ymin=29.4793, ymax=29.5407,
			tick label style={font=\tiny},
			label style={font=\tiny},
			xtick={3,3.2,3.4,3.6},
			grid=both, minor tick num=1,
			major grid style={line width=0.2pt, draw=black!15},
			minor grid style={line width=0.2pt, draw=black!7},
			tick align=outside,
			scaled y ticks=false,
			y tick label style={/pgf/number format/fixed},
			every axis plot post/.append style={line join=round, line cap=round},
			restrict y to domain=29.2338:29.7863,
			unbounded coords=jump,
			axis line style={draw=black!55},
			extra x ticks={2.8688},
			extra x tick labels={$2.869$},
			extra x tick style={grid=major, major grid style={dashed, draw=black!35, line width=0.3pt},
				tick label style={font=\tiny}},
		]
			\addplot[black, line width=0.7pt, forget plot]
				table[x index=0, y index=1] {paper_3/tikz_data/pairs_curve_xi02_fullfwdprod_hom.dat};
			\addplot[gray!55, line width=0.5pt, forget plot]
				table[x index=0, y index=1] {paper_3/tikz_data/pairs_curve_xi02_fullfwdprod.dat};
			\addplot[only marks, mark=*, mark size=0.55pt, color=matlabblue]
				table[x index=0, y index=1] {paper_3/tikz_data/pairs_curve_xi02_fullfwdprod_stable.dat};
			\addplot[only marks, mark=*, mark size=0.55pt, color=red!75!black]
				table[x index=0, y index=1] {paper_3/tikz_data/pairs_curve_xi02_fullfwdprod_unstable.dat};
			\addplot[only marks, mark=o, mark size=2.6pt, black, line width=0.5pt]
				table[x index=0, y index=1] {paper_3/tikz_data/pairs_curve_xi02_fullfwdprod_marks.dat};
		\end{axis}
	\end{tikzpicture}%
} \\[1.5mm]
		\rowlab{0.4} &
		\usebox{\scaleaxisbox} &
		{\tikzsetfigurename{same_shade_xi04_fullfwdprod_panels_}	\begin{tabular}[b]{@{}c@{}}%
		\begin{tikzpicture}[baseline={(current axis.south)}, trim axis left]
			\begin{axis}[
				width=0.085\textwidth, height=0.19\textwidth,
				scale only axis, view={0}{90},
				xmin=-0.0359, xmax=0.0359, ymin=-0.0068, ymax=3.6853,
				axis lines=left,
				axis line style={draw=none},
				tick style={draw=none},
				xtick=\empty, ytick=\empty,
				clip=false,
				colormap name=shine,
				point meta min=0, point meta max=1,
				unbounded coords=jump,
			]
				\addplot3[surf, shader=interp, point meta=explicit,
					mesh/cols=50, mesh/ordering=rowwise]
					table[x index=0, y index=1, z index=2, meta index=3] {paper_3/tikz_data/same_shade_xi04_fullfwdprod_p01.dat};
				\addplot3[surf, shader=interp, point meta=explicit,
					mesh/cols=49, mesh/ordering=rowwise]
					table[x index=0, y index=1, z index=2, meta index=3] {paper_3/tikz_data/same_shade_xi04_fullfwdprod_p01_cap.dat};
				\addplot[black!45, line width=0.15pt]
					table[x index=0, y index=1] {paper_3/tikz_data/same_shade_xi04_fullfwdprod_p01_wire.dat};
				\draw[draw=black!55, dashed, line width=0.4pt, dash pattern=on 2pt off 1.6pt]
					(axis cs:0,0) -- (axis cs:-0.05000,0) -- (axis cs:-0.05000,1) -- (axis cs:0,1);
				\draw[draw=black!55, line width=0.4pt, dash pattern=on 3pt off 1.4pt on 0.6pt off 1.4pt]
					(axis cs:0,-0.08000) -- (axis cs:0,1.08000);
				\node[anchor=north, font=\tiny, inner sep=0pt, yshift=-2.4mm]
					at (rel axis cs:0.5,0) {$\lambda = 2.705$};
			\end{axis}
		\end{tikzpicture}%
	\end{tabular}%
	\begin{tabular}[b]{@{}c@{}}%
		\begin{tikzpicture}[baseline={(current axis.south)}]
			\begin{axis}[
				width=0.085\textwidth, height=0.19\textwidth,
				scale only axis, view={0}{90},
				xmin=-0.0359, xmax=0.0359, ymin=-0.0068, ymax=3.6853,
				axis lines=left,
				axis line style={draw=none},
				tick style={draw=none},
				xtick=\empty, ytick=\empty,
				clip=false,
				colormap name=shine,
				point meta min=0, point meta max=1,
				unbounded coords=jump,
			]
				\addplot3[surf, shader=interp, point meta=explicit,
					mesh/cols=50, mesh/ordering=rowwise]
					table[x index=0, y index=1, z index=2, meta index=3] {paper_3/tikz_data/same_shade_xi04_fullfwdprod_p02.dat};
				\addplot3[surf, shader=interp, point meta=explicit,
					mesh/cols=49, mesh/ordering=rowwise]
					table[x index=0, y index=1, z index=2, meta index=3] {paper_3/tikz_data/same_shade_xi04_fullfwdprod_p02_cap.dat};
				\addplot[black!45, line width=0.15pt]
					table[x index=0, y index=1] {paper_3/tikz_data/same_shade_xi04_fullfwdprod_p02_wire.dat};
				\node[anchor=north, font=\tiny, inner sep=0pt, yshift=-2.4mm]
					at (rel axis cs:0.5,0) {$\lambda = 2.894$};
			\end{axis}
		\end{tikzpicture}%
	\end{tabular}%
	\begin{tabular}[b]{@{}c@{}}%
		\begin{tikzpicture}[baseline={(current axis.south)}]
			\begin{axis}[
				width=0.085\textwidth, height=0.19\textwidth,
				scale only axis, view={0}{90},
				xmin=-0.0359, xmax=0.0359, ymin=-0.0068, ymax=3.6853,
				axis lines=left,
				axis line style={draw=none},
				tick style={draw=none},
				xtick=\empty, ytick=\empty,
				clip=false,
				colormap name=shine,
				point meta min=0, point meta max=1,
				unbounded coords=jump,
			]
				\addplot3[surf, shader=interp, point meta=explicit,
					mesh/cols=50, mesh/ordering=rowwise]
					table[x index=0, y index=1, z index=2, meta index=3] {paper_3/tikz_data/same_shade_xi04_fullfwdprod_p03.dat};
				\addplot3[surf, shader=interp, point meta=explicit,
					mesh/cols=49, mesh/ordering=rowwise]
					table[x index=0, y index=1, z index=2, meta index=3] {paper_3/tikz_data/same_shade_xi04_fullfwdprod_p03_cap.dat};
				\addplot[black!45, line width=0.15pt]
					table[x index=0, y index=1] {paper_3/tikz_data/same_shade_xi04_fullfwdprod_p03_wire.dat};
				\node[anchor=north, font=\tiny, inner sep=0pt, yshift=-2.4mm]
					at (rel axis cs:0.5,0) {$\lambda = 3.204$};
			\end{axis}
		\end{tikzpicture}%
	\end{tabular}%
	\begin{tabular}[b]{@{}c@{}}%
		\begin{tikzpicture}[baseline={(current axis.south)}]
			\begin{axis}[
				width=0.085\textwidth, height=0.19\textwidth,
				scale only axis, view={0}{90},
				xmin=-0.0359, xmax=0.0359, ymin=-0.0068, ymax=3.6853,
				axis lines=left,
				axis line style={draw=none},
				tick style={draw=none},
				xtick=\empty, ytick=\empty,
				clip=false,
				colormap name=shine,
				point meta min=0, point meta max=1,
				unbounded coords=jump,
			]
				\addplot3[surf, shader=interp, point meta=explicit,
					mesh/cols=50, mesh/ordering=rowwise]
					table[x index=0, y index=1, z index=2, meta index=3] {paper_3/tikz_data/same_shade_xi04_fullfwdprod_p04.dat};
				\addplot3[surf, shader=interp, point meta=explicit,
					mesh/cols=49, mesh/ordering=rowwise]
					table[x index=0, y index=1, z index=2, meta index=3] {paper_3/tikz_data/same_shade_xi04_fullfwdprod_p04_cap.dat};
				\addplot[black!45, line width=0.15pt]
					table[x index=0, y index=1] {paper_3/tikz_data/same_shade_xi04_fullfwdprod_p04_wire.dat};
				\node[anchor=north, font=\tiny, inner sep=0pt, yshift=-2.4mm]
					at (rel axis cs:0.5,0) {$\lambda = 3.502$};
			\end{axis}
		\end{tikzpicture}%
	\end{tabular}%
} &
		{\tikzsetfigurename{pairs_curve_xi04_fullfwdprod_}	\begin{tikzpicture}[baseline={(current axis.south)}]
		\begin{axis}[
			width=0.38\textwidth, height=0.19\textwidth,
			scale only axis,
			xlabel={$\lambda$}, ylabel={$F_z$},
			xlabel style={at={(ticklabel cs:1)}, anchor=base west,
				xshift=0.8mm, yshift=3.9mm, font=\tiny},
			ylabel style={rotate=-90},
			xmin=2.6868, xmax=3.7044, ymin=29.4793, ymax=29.5407,
			tick label style={font=\tiny},
			label style={font=\tiny},
			xtick={2.8,3,3.2,3.4,3.6},
			grid=both, minor tick num=1,
			major grid style={line width=0.2pt, draw=black!15},
			minor grid style={line width=0.2pt, draw=black!7},
			tick align=outside,
			scaled y ticks=false,
			y tick label style={/pgf/number format/fixed},
			every axis plot post/.append style={line join=round, line cap=round},
			restrict y to domain=29.2338:29.7863,
			unbounded coords=jump,
			axis line style={draw=black!55},
			extra x ticks={2.9004},
			extra x tick labels={$2.900$},
			extra x tick style={grid=major, major grid style={dashed, draw=black!35, line width=0.3pt},
				tick label style={font=\tiny}},
		]
			\addplot[black, line width=0.7pt, forget plot]
				table[x index=0, y index=1] {paper_3/tikz_data/pairs_curve_xi04_fullfwdprod_hom.dat};
			\addplot[gray!55, line width=0.5pt, forget plot]
				table[x index=0, y index=1] {paper_3/tikz_data/pairs_curve_xi04_fullfwdprod.dat};
			\addplot[only marks, mark=*, mark size=0.55pt, color=matlabblue]
				table[x index=0, y index=1] {paper_3/tikz_data/pairs_curve_xi04_fullfwdprod_stable.dat};
			\addplot[only marks, mark=*, mark size=0.55pt, color=red!75!black]
				table[x index=0, y index=1] {paper_3/tikz_data/pairs_curve_xi04_fullfwdprod_unstable.dat};
			\addplot[only marks, mark=o, mark size=2.6pt, black, line width=0.5pt]
				table[x index=0, y index=1] {paper_3/tikz_data/pairs_curve_xi04_fullfwdprod_marks.dat};
		\end{axis}
	\end{tikzpicture}%
} \\[1.5mm]
		& & \multicolumn{1}{@{}l@{}}{\usebox{\keyshapebox}} &
		\usebox{\keycurvebox}
	\end{tabular}
	\setlength{\abovecaptionskip}{4pt}%
	\caption{Equilibrium configurations and force--stretch responses, with one row for each value of $\xi$. Left: four deformed configurations along the computed equilibrium path, one before the onset and three after it. The half section superimposed on the first configuration shows the undeformed specimen. The bodies are obtained by revolving the meridional solution about the symmetry axis. Circumferential lines mark every second axial knot line; meridians are included for visualization only. All twenty bodies are shown at the same scale, indicated by the axis on the left, with radial dimensions magnified by a factor of approximately 23 relative to axial dimensions. Right: axial force versus stretch near the bifurcation, with the homogeneous equilibrium branch shown in black.
}

	\label{fig:bif_page}
\end{figure}

\subsubsection{Onset of bifurcation}
The computed onsets are collected in \cref{tab:bif_onset} beside the
analytical ones of \cref{sec:results}. The differences in onset stretches are below $0.5\,\%$ for $\xi\le0.2$ and reach $1.1\,\%$ at
$\xi=0.4$.

For $\xi=0$, $0.01$ and $0.1$, the analytical critical modes imply
$3.97$, $3.96$ and $2.92$ waves over the specimen. The computed modes
contain the nearest integer number of waves, namely, $4$, $4$ and $3$ waves, and their wavenumbers exceed the analytical ones by at most $2.3\,\%$. These small differences are consistent with the selection of modes compatible with the finite specimen and its end conditions.

For $\xi=0.2$ and $0.4$, the analytical threshold is reached only in the
long-wave limit $k\to0$, which a finite specimen cannot reach. The
computed mode is then a single half-wave over the specimen, with
$k=\pi/(\lambda L)$. Imposing this wavenumber in the neutral relation of
the infinite cylinder and solving for $\lambda$ gives onset stretches of $2.8691$ for
$\xi=0.2$ and $2.9002$ for $\xi=0.4$, compared with the computed values $2.8688$ and $2.9004$. This close agreement indicates that the larger onset shift at $\xi=0.4$ is primarily a finite-length effect. At this value of $\xi$, the neutral stretch increases more strongly away from $k=0$, so the finite half-wave requires a larger increase in stretch relative to the long-wave threshold.

\subsubsection{\texorpdfstring{Post-bifurcation}{Post-bifurcation} behavior}
The post-bifurcation problem can admit multiple nonhomogeneous equilibria at the same prescribed stretch. The continuation follows the path selected by the imposed imperfection. Thus, the paths in \bifref illustrate possible post-bifurcation responses; they are not necessarily paths of the lowest energy.

For $\xi=0$ and $0.01$, the number of waves selected at onset is preserved along the computed path. Initially, the amplitude grows throughout the specimen. At larger stretches the bulges stop growing and further elongation occurs primarily through stretching of the unmodulated segments between them. The bulges therefore move farther apart rather than multiply or grow further.

For $\xi = 0.2$ and $0.4$, no bulge forms. Instead, the specimen develops a thick and a thin segment of nearly uniform radius, separated by a transition zone. This zone occupies
about a quarter of the length at $\xi=0.2$ and nearly half of it at
$\xi=0.4$. Further stretching moves this zone along the specimen, so that the
thin segment lengthens at the expense of the thick one, while both radii
remain nearly constant. This behavior is consistent with the long-wave onset predicted in \cref{sec:bif} and with the interpretation of beading as separation into two coexisting deformation states \citep{xuan2017plateau}.

The case $\xi=0.1$ lies between these two regimes. The three-wave
pattern that emerges at the onset does not persist, but reorganizes into a two-wave pattern along the unstable portion of the path. Beyond this portion, it evolves as for $\xi=0$ and $0.01$. Its response folds back on itself more than once, a clear sign of the non-uniqueness noted above.

The axial force is consistent with these shapes. At every $\xi$ the branch
leaves the homogeneous curve downwards, so the modulated specimen carries a
lower force than the homogeneous one at the same stretch, and crosses back
above it further along the path. The deviation from the homogeneous response decreases markedly as $\xi$ grows. At the two largest values the force stays nearly
constant once the transition zone has formed. This plateau corresponds to the
advance of the transition zone described above.

The cases also differ in the stability of the branches. At $\xi=0$ and
$0.01$ the bifurcation is supercritical: the emerging branch is stable from the onset onwards and can be followed continuously under increasing stretch. It loses
stability only at much larger stretch, where the force passes a maximum and the
path folds back. For $\xi=0.2$ and $0.4$ the bifurcation is
subcritical. That is, the emerging branch is unstable and initially extends toward smaller stretches before turning and regaining stability. A specimen
stretched through $\lambda_{\mathrm b}$ therefore cannot follow the branch and
must jump to a distant stable state, which the static computation does not
determine. The second gradient thus changes the character of the bifurcation as
well as the onset and the wavelength. In the cases computed here, the
bifurcation becomes subcritical once the critical mode is the long-wave one. Note that for the first-gradient problem, a subcritical beading transition has been found by
weakly nonlinear analysis \citep{taffetani2015beading}.

\section{Conclusion}\label{sec:con}

In this work, we have developed a comprehensive theoretical and computational framework for finite-strain second-gradient hyperelasticity coupled with elastic surface energies. Starting from a generic variational formulation, we derived the coordinate-free governing equilibrium equations along with the corresponding higher-order bulk, surface, and edge boundary conditions.
We then proved, under the stated assumptions, the existence of a minimizer for the coupled bulk--surface functional. The second-gradient term provides compactness of the boundary traces of the deformation gradients. This yields almost-everywhere convergence of the surface deformation gradients along a subsequence of each bounded-energy sequence. The compactness thus no longer has to
come from the surface term itself. For nonnegative, lower semicontinuous
surface densities, this pointwise convergence alone yields lower
semicontinuity of the surface functional. Consequently, neither positive
one-homogeneity nor convexity or polyconvexity of the surface density is
needed, no upper growth bound is imposed, and no measure-theoretic reading
of the surface integral is required, in contrast to the first-gradient
setting. The admissible class of surface models is thereby widened, and
it contains the constitutive choice made in the application for all
nonnegative values of its surface parameters.
To systematically study material instabilities and pattern formation, we developed an incremental (small-on-large) perturbation theory and performed a push-forward of the linearized field equations to the finitely deformed intermediate configuration.

The theoretical framework was applied to investigate the elastocapillary beading instability of soft cylindrical rods under axial stretch. Through a linear bifurcation analysis, we demonstrated the interplay between bulk gradient elasticity and surface stretching resistance. The analytical results reveal that the second-gradient modulus $\xi$ acts as an effective regularizing mechanism and influences the instability in three ways:
\begin{itemize}
    \item It delays the onset of bifurcation, requiring a larger axial stretch $\lambda$ to initiate instability.
    \item It penalizes sharp spatial gradients, thereby suppressing short-wavelength modes and shifting the critical wavenumber $k$ toward smaller values (longer wavelengths).
    \item For sufficiently large values of $\xi$, the critical mode transitions to the long-wave limit $k\to0$, with the onset at the maximum of the axial force. Beyond this threshold, the critical stretch saturates because the second-gradient energy contribution is no longer activated in that limit.
\end{itemize}

To validate the theoretical predictions and capture post-bifurcation behavior, the framework was implemented in the open-source software OOFEM using $C^1$-continuous isogeometric analysis. The equilibrium paths were followed by arc-length continuation, and the stability with respect to axisymmetric perturbations was assessed at each computed state. That allowed the nonhomogeneous post-bifurcation paths to be traced and the post-bifurcation shapes to be followed, ranging from beaded configurations at small $\xi$ to a thick and a thin segment separated by a transition zone at large $\xi$. The numerically determined critical stretches and wavelengths agreed
closely with the analytical predictions, and the largest differences were
traced to the finite length of the specimen. The computed branches also show that the second-gradient modulus changes the character of the bifurcation; namely, it is supercritical at small $\xi$ and turns subcritical once the critical mode is the long-wave one.

Overall, the proposed methodology offers a robust foundation for studying surface- and gradient-driven instabilities in soft materials and flexible structures. Future extensions could incorporate explicit surface curvature energy (bending resistance) into the incremental formulation or generalize the framework to fully three-dimensional, non-axisymmetric bifurcation modes.

\section*{Acknowledgements}
This work was supported by the CSF-FWF  project 26-21585K. MK was further supported by the Strategy AV21 program ``AI: Artificial Intelligence for Science and Society'' and by the EU/M\v{S}MT \v{C}R  under the ROBOPROX project (reg.~no.~CZ.02.01.01/00/22\_008/0004590). M\v{S} was further supported by the Grant Agency of the Czech Technical University in Prague, grant No.~SGS26/038/OHK1/1T/11. MH was further supported by internal funding of the Czech Technical University in Prague.

\appendix
\section{\texorpdfstring{Push-forward of the incremental problem}{Push-forward of the incremental problem}}\label{app:pushforward}
\begin{fixblock}
This appendix derives the incremental problem \eqref{eq:bulk-eq_inc_act}--\eqref{eq:edge-eq_inc_act} in the intermediate configuration from its referential form \eqref{eq:bulk-eq_inc}--\eqref{eq:edge-eq_inc}. Throughout, the identities of \cref{sec:2} are applied to the map $\bm\phi_e$. In particular, Nanson's formula \eqref{eq:cof} and \eqref{eq:Js_bulk} give
\begin{equation}\label{eq:nanson_e}
\bm N=|\bar{\bm F}^{-T}\bm N|\,\bar{\bm F}^T\bar{\bm n}=\frac{\bar J_s}{\bar J}\,\bar{\bm F}^T\bar{\bm n}.
\end{equation}

To derive \eqref{eq:bulk-eq_inc_act} from \eqref{eq:bulk-eq_inc}, we apply the Piola identity \eqref{eq:piola_t} twice, to $\bm Q^{[1]}$ and to $\bm A^{[1]}$, and then move $\bar{\bm F}^T$ under the inner divergence by the product rule; the $\bm x$-derivative of $\bar{\bm F}$ produces the term with $\bar{\bm H}^T$:
\begin{align}
\Div{({\bm{P}}^{[1]}-\Div{\bm{A}}^{[1]})} &=\bar{J} ~\overline{\ddiv}{ \left(\frac{1}{\bar{J}}{\bm{P}}^{[1]}\bar{\bm{F}}^{T} -\frac{1}{\bar{J}} \Div{(\bm{A}}^{[1]}) \bar{\bm{F}}^{T}\right)} =
\bar{J} ~\overline{\ddiv}{ \left(\frac{1}{\bar{J}}{\bm{P}}^{[1]}\bar{\bm{F}}^{T} - \overline{\ddiv}\left( \frac{1}{\bar{J}} {\bm{A}}^{[1]} \bar{\bm{F}}^{T}\right)\bar{\bm{F}}^{T} \right)} \nonumber \\
&=
\bar{J} ~\overline{\ddiv}{ \left(\frac{1}{\bar{J}}{\bm{P}}^{[1]}\bar{\bm{F}}^{T} - \overline{\ddiv}\left( \frac{1}{\bar{J}} {\bm{A}}^{[1]}: \left(\bar{\bm{F}}^{T} \overline{\otimes}\bar{\bm{F}}^{T}\right)\right) +  \frac{1}{\bar{J}}{\bm{A}}^{[1]} : \bar{\bm{H}}^T    \right)}=\bar{J} ~ \overline{\ddiv} \left(\bm{\chi}-\overline{\ddiv}{\bm{a}}^{[1]} \right).\label{eq:push_bulk}
\end{align}
The argument of the outer divergence in the first line is the push-forward $\bar J^{-1}\bm Q^{[1]}\bar{\bm F}^T$, so \eqref{eq:push_bulk} also proves the form of $\bm q^{[1]}$ stated in \cref{sec:push-forward}. Since $\bar J>0$, \eqref{eq:bulk-eq_inc} is equivalent to \eqref{eq:bulk-eq_inc_act}.

Boundary condition \eqref{eq:edge-bc-add_inc_act} follows from \eqref{eq:edge-bc-add_inc}, since \eqref{eq:nanson_e} and the definition \eqref{eq:icr} of $\bm a^{[1]}$ give $\bm A^{[1]}:(\bm N\otimes\bm N)=(\bar J_s^2/\bar J)\,\bm a^{[1]}:(\bar{\bm n}\otimes\bar{\bm n})$.

We now derive \eqref{eq:surf-eq_inc_act}; \eqref{eq:bulk-bc_inc_act} follows in the same way without the surface stress. The first term of \eqref{eq:surf-eq_inc} becomes
\begin{align}
\bm{Q}^{[1]} \bm{N} &= \left( {\bm{P}}^{[1]} -\Div{\bm{A}}^{[1]} \right) \bm{N} =\frac{1}{\bar{J}} {\bm{P}}^{[1]} \bar{\bm{F}}^T \bar{\bm{n}} |\bar{\bm{F}}^{-T} {\bm{N}}| \bar{J} - \overline{\ddiv}\left( \frac{1}{\bar{J}}{\bm{A}}^{[1]}\bar{\bm{F}}^T \right)\bar{\bm{F}}^T \bar{\bm{n}} |\bar{\bm{F}}^{-T} {\bm{N}}|\bar{J} \nonumber \\
&=\bar{J}_s \left[ \frac{1}{\bar{J}} {\bm{P}}^{[1]} \bar{\bm{F}}^T - \overline{\ddiv}\left( \frac{1}{\bar{J}}{\bm{A}}^{[1]}\bar{\bm{F}}^T \right)\bar{\bm{F}}^T  \right]\bar{\bm{n}} =
\bar{J}_s \left[\bm{\chi}-\overline{\ddiv}{\bm{a}}^{[1]} \right]\bar{\bm{n}}
=\bar{J}_s \bm{q}^{[1]} \bar{\bm{n}}, \label{eq:show_bulk}
\end{align}
where $\bm N$ is replaced by \eqref{eq:nanson_e}, $\Div\bm A^{[1]}$ is transformed by the Piola identity \eqref{eq:piola_t}, and the bracket is rewritten as in the second line of \eqref{eq:push_bulk}. The second term of \eqref{eq:surf-eq_inc} yields
\begin{align}
\Div_s\left[\left({\bm{A}^{[1]}}\bm{N}\right)\bm{I}_s\right] &=
\bar{J}_s \overline{\ddiv}_s\left[\frac{1}{\bar{J}_s}\left({\bm{A}^{[1]}}\bm{N}\right)\bm{I}_s \bar{\bm{F}}_s^T \right]=
\bar{J}_s \overline{\ddiv}_s\left[\frac{1}{\bar{J}_s}\left({\bm{A}^{[1]}}\bm{N}\right)\bar{\bm{F}}^T \bar{\bm{i}}_s \right]
\nonumber \\
&=
\bar{J}_s \overline{\ddiv}_s\left[\frac{1}{\bar{J}_s \bar{J}}\left({\bm{A}^{[1]}}\bar{\bm{F}}^T \bar{\bm{n}} |\bar{\bm{F}}^{-T} {\bm{N}}| \bar{J}\right)\bar{\bm{F}}^T \bar{\bm{i}}_s \right]
=
\bar{J}_s \overline{\ddiv}_s\left[\frac{1}{\bar{J}}\left({\bm{A}^{[1]}}\bar{\bm{F}}^T \bar{\bm{n}} \right)\bar{\bm{F}}^T \bar{\bm{i}}_s \right] = \bar{J}_s \overline{\ddiv}_s\left[\left({\bm{a}^{[1]}}\bar{\bm{n}}\right)\bar{\bm{i}}_s\right],
\end{align}
where the surface Piola identity \eqref{eq:piola_t} gives the first equality, the identity below the second, \eqref{eq:nanson_e} the third and fourth, and the definition \eqref{eq:icr} of $\bm a^{[1]}$ the last. The identity reads
\begin{align}\label{eq:AN_identity}
\left({\bm{A}^{[1]}}\bm{N}\right)\bm{I}_s \bar{\bm{F}}_s^T =
\left({\bm{A}^{[1]}}\bm{N}\right)\bm{I}_s \bar{\bm{F}}_s^T \bar{\bm{i}}_s  =
\left({\bm{A}^{[1]}}\bm{N}\right)\bm{I}_s \bar{\bm{F}}^T \bar{\bm{i}}_s=
\left({\bm{A}^{[1]}}\bm{N}\right)\left(\bm{I}-\bm{N}\otimes\bm{N}\right)\bar{\bm{F}}^T \bar{\bm{i}}_s =
\left({\bm{A}^{[1]}}\bm{N}\right)\bar{\bm{F}}^T \bar{\bm{i}}_s,
\end{align}
which rests on \eqref{eq:FIs} and \eqref{eq41} and, in its last step, on the higher-order condition \eqref{eq:edge-bc-add_inc}. The third term becomes $\Div_s\bm P_s^{[1]}=\bar J_s\,\overline{\ddiv}_s\bm\chi_s$ by the surface Piola identity \eqref{eq:piola_t} and the definition \eqref{eq:icr} of $\bm\chi_s$, $\bm P_s^{[1]}$ being superficial. Division by $\bar J_s>0$ gives \eqref{eq:surf-eq_inc_act}.

The edge condition \eqref{eq:edge-eq_inc} is transferred by the line-element counterpart of Nanson's formula \eqref{eq:cof} taken on the surface,
\begin{align} \label{eq:cof_line}
\bar{\bm{v}} \,\rd \bar{s} = \bar{J}_s \bar{\bm{F}}_s^{-T} \bm{V} \,\rd s,
\end{align}
which, for each face adjoining $\mathcal C$, ties the oriented line element $\bm V\,\rd s$ to its counterpart $\bar{\bm v}\,\rd\bar s$ on the image of that face. Here $\rd s$ and $\rd\bar s$ denote the arc-length elements of $\mathcal C$ and of $\bm\phi_e(\mathcal C)$, $\bm V$ is the in-plane unit normal of the face boundary introduced with \eqref{eq:surf_div_thm}, and $\bar{\bm v}$ is its counterpart, lying in the image face and pointing out of it. Multiplying \eqref{eq:cof_line} by $\bar{\bm F}_s^T$ and using $\bar{\bm F}_s^T\bar{\bm F}_s^{-T}=\bm I_s$ and $\bm I_s\bm V=\bm V$ gives $\bm V\,\rd s=\bar J_s^{-1}\bar{\bm F}_s^T\bar{\bm v}\,\rd\bar s$; inserted into the second sum of \eqref{eq:edge-eq_inc}, this yields
\begin{align}
\bm{P}_s^{[1]}\bm{V} \,\rd s = \frac{1}{\bar{J}_s}\bm{P}_s^{[1]}\bar{\bm{F}}_s^{T} \bar{\bm{v}}\,\rd \bar{s} = \bm{\chi}_s \bar{\bm{v}}\,\rd \bar{s},
\end{align}
where the definition \eqref{eq:icr} of $\bm{\chi}_s$ is used. The term with $\bm A^{[1]}$ is transferred in the same way. Once \eqref{eq:cof_line} has been applied, the identity \eqref{eq:AN_identity} together with $\bar{\bm i}_s\bar{\bm v}=\bar{\bm v}$ replaces $\bar{\bm F}_s^T$ by $\bar{\bm F}^T$, and \eqref{eq:nanson_e} cancels $\bar J_s$ and leaves the factor $\bar J^{-1}$ of $\bm a^{[1]}$ in \eqref{eq:icr}:
\begin{align}
\left({\bm{A}^{[1]}}\bm{N}\right)\bm{V} \,\rd s
&= \frac{1}{\bar{J}_s}\left({\bm{A}^{[1]}}\bm{N}\right)\bar{\bm{F}}_s^{T} \bar{\bm{v}}\,\rd \bar{s}
= \frac{1}{\bar{J}_s}\left({\bm{A}^{[1]}}\bm{N}\right)\bar{\bm{F}}^{T} \bar{\bm{v}}\,\rd \bar{s}
\nonumber \\
&= \frac{1}{\bar{J}}\left({\bm{A}^{[1]}}\bar{\bm{F}}^{T}\bar{\bm{n}}\right)\bar{\bm{F}}^{T} \bar{\bm{v}}\,\rd \bar{s}
= \left({\bm{a}^{[1]}}\bar{\bm{n}}\right)\bar{\bm{v}}\,\rd \bar{s}.
\end{align}
Both transformations act one face at a time. The ratio $\rd\bar s/\rd s$ is the stretch of the tangent of $\mathcal C$ and is therefore common to the adjoining faces, so the two sums carry over unchanged and \eqref{eq:edge-eq_inc_act} follows.
\end{fixblock}

\section{\texorpdfstring{Incremental fields of the cylinder}{Incremental fields of the cylinder}}\label{app:cylinder}
This appendix lists the intermediate results of \cref{sec:bif}: the second incremental displacement gradient, the components of the updated incremental stresses and of the effective incremental stress $\bm{q}^{[1]}$, and the surface quantities entering the boundary condition \eqref{eq:surf-eq_inc_act}. All components refer to the orthonormal basis $(\bm e_\theta,\bm e_z,\bm e_r)$ of \cref{sec:hom}, the indices 1, 2 and 3 standing for $\theta$, $z$ and $r$.

The second incremental displacement gradient, which enters \eqref{eq:P_incr} and \eqref{eq:A_incr}, reads
\bea
\begin{aligned}
\bar{\nabla}\bm{\eta}={}&\frac{\partial \bm{\eta}}{\partial\theta^i}\otimes \bm{g}^i=
\frac{u_{,z}}{r}\,\bm e_\theta\otimes\bm e_z\otimes\bm e_\theta
+\frac{r u_{,r}-u}{r^2}\,\bm e_\theta\otimes\bm e_r\otimes\bm e_\theta
+\frac{v_{,r}}{r}\,\bm e_z\otimes\bm e_\theta\otimes\bm e_\theta
+\frac{r u_{,r}-u}{r^2}\,\bm e_r\otimes\bm e_\theta\otimes\bm e_\theta
\\
&+\frac{u_{,z}}{r}\,\bm e_\theta\otimes\bm e_\theta\otimes\bm e_z
+v_{,zz}\,\bm e_z\otimes\bm e_z\otimes\bm e_z
+v_{,zr}\,\bm e_z\otimes\bm e_r\otimes\bm e_z
+u_{,zz}\,\bm e_r\otimes\bm e_z\otimes\bm e_z
+u_{,zr}\,\bm e_r\otimes\bm e_r\otimes\bm e_z
\\
&+\frac{r u_{,r}-u}{r^2}\,\bm e_\theta\otimes\bm e_\theta\otimes\bm e_r
+v_{,zr}\,\bm e_z\otimes\bm e_z\otimes\bm e_r
+v_{,rr}\,\bm e_z\otimes\bm e_r\otimes\bm e_r
+u_{,zr}\,\bm e_r\otimes\bm e_z\otimes\bm e_r
+u_{,rr}\,\bm e_r\otimes\bm e_r\otimes\bm e_r .
\end{aligned}
\eea

The nonzero components $\chi_{ij}$ of the updated incremental stress tensor $\bm{\chi}$ take the form
\begin{align}
\begin{split}
&\chi_{11} = \frac{2\mu(a^2+1)+D(a^4\lambda^2+1)}{2a^2\lambda }\frac{u}{r}+ D a^2\lambda \Big(u_{,r}+v_{,z}\Big) ,\\ 
&\chi_{22} = \frac{2\mu(\lambda^2+1)+D(a^4\lambda^2+1)}{2a^2\lambda}v_{,z}+Da^2\lambda\Big(u_{,r}+\frac{u}{r}\Big),
\\ 
&\chi_{33} = \frac{2\mu(a^2+1)+D(a^4\lambda^2+1)}{2a^2\lambda}u_{,r}+Da^2\lambda\Big(v_{,z}+\frac{u}{r}\Big),
\\ 
&\chi_{23} = \frac{2\mu+D(1-a^4\lambda^2)}{2a^2\lambda}u_{,z}+\frac{\mu}{\lambda}v_{,r},
\\ 
&\chi_{32} = \frac{2\mu+D(1-a^4\lambda^2)}{2a^2\lambda}v_{,r}+\frac{\mu\lambda }{a^2}u_{,z}. 
\end{split}
\end{align}
The nonzero components of the updated incremental higher-order stress tensor $\bm{a}^{[1]}$ are
\bea \label{eq:a_coeffs}
\begin{aligned}
a^{[1]}_{121}=a^{[1]}_{112}
&=\xi\frac{\lambda u_{,z}}{r},
\\
a^{[1]}_{131}=a^{[1]}_{311}=a^{[1]}_{113}
&=\xi\frac{a^2\left(ru_{,r}-u\right)}{\lambda r^2},
\\
a^{[1]}_{211}
&=\xi\frac{a^2v_{,r}}{\lambda r},
\\
a^{[1]}_{222}
&=\xi\frac{\lambda^3}{a^2}v_{,zz},
\\
a^{[1]}_{232}=a^{[1]}_{223}
&=\xi\lambda v_{,zr},
\\
a^{[1]}_{322}
&=\xi\frac{\lambda^3}{a^2}u_{,zz},
\\
a^{[1]}_{332}=a^{[1]}_{323}
&=\xi\lambda u_{,zr},
\\
a^{[1]}_{233}
&=\xi\frac{a^2}{\lambda}v_{,rr},
\\
a^{[1]}_{333}
&=\xi\frac{a^2}{\lambda}u_{,rr}.
\end{aligned}
\eea
Assembling the two sets according to $\bm{q}^{[1]} = \bm{\chi}-\overline{\ddiv}{\bm{a}}^{[1]}$ gives the components of the effective incremental stress as
\begin{equation}\label{eq:chicomp}
\begin{aligned}
q^{[1]}_{11}
={}&
\frac{1}{a^2\lambda}
\left[
D a^4\lambda^2\left(u_{,r}+v_{,z}\right)
+\frac{u}{r}
\left(
\frac{D a^4\lambda^2}{2}
+\mu a^2+\mu+\frac{D}{2}
\right)
\right]
-\frac{\lambda\xi}{r}u_{,zz}
-\frac{a^2\xi}{\lambda r^3}
\left(r^2u_{,rr}+ru_{,r}-u\right),
\\[2mm]
q^{[1]}_{22}
={}&
\frac{1}{a^2\lambda}
\left[
\left(
\frac{D a^4\lambda^2}{2}
+\mu\lambda^2+\mu+\frac{D}{2}
\right)v_{,z}
+D a^4\lambda^2
\left(u_{,r}+\frac{u}{r}\right)
\right]
-\frac{\lambda\xi}{a^2}
\left(a^2v_{,zrr}+\lambda^2v_{,zzz}\right)
-\frac{\lambda\xi}{r}v_{,zr},
\\[2mm]
q^{[1]}_{23}
={}&
\frac{\mu}{\lambda}v_{,r}
+\frac{1}{a^2\lambda}
\left(
\frac{D}{2}+\mu-\frac{D a^4\lambda^2}{2}
\right)u_{,z}
-\frac{\xi}{\lambda}
\left(a^2v_{,rrr}+\lambda^2v_{,zzr}\right)
+\frac{a^2\xi}{\lambda r^2}
\left(v_{,r}-rv_{,rr}\right),
\\[2mm]
q^{[1]}_{32}
={}&
\frac{\lambda\mu}{a^2}u_{,z}
+\frac{1}{a^2\lambda}
\left(
\frac{D}{2}+\mu-\frac{D a^4\lambda^2}{2}
\right)v_{,r}
-\frac{\lambda\xi}{a^2}
\left(a^2u_{,zrr}+\lambda^2u_{,zzz}\right)
+\frac{\lambda\xi}{r^2}
\left(u_{,z}-ru_{,zr}\right),
\\[2mm]
q^{[1]}_{33}
={}&
\frac{1}{a^2\lambda}
\left[
\left(
\frac{D a^4\lambda^2}{2}
+\mu a^2+\mu+\frac{D}{2}
\right)u_{,r}
+D a^4\lambda^2
\left(v_{,z}+\frac{u}{r}\right)
\right]
-\frac{\xi}{\lambda}
\left(a^2u_{,rrr}+\lambda^2u_{,zzr}\right)
-\frac{a^2\xi}{\lambda r^3}
\left(2u+r^2u_{,rr}-2ru_{,r}\right).
\end{aligned}
\end{equation}

For the boundary condition \eqref{eq:surf-eq_inc_act}, restricting the bulk displacement gradient to the lateral surface gives
\begin{align}
\bm{\eta}_s=\bm{\eta}|_{r=aA}\bar{\bm{i}}_s=\frac{u}{aA}\bm{e}_\theta\otimes\bm{e}_\theta+v_{,z}\bm{e}_z\otimes\bm{e}_z+u_{,z}\bm{e}_r\otimes\bm{e}_z,
\end{align}
the displacement functions $u$ and $v$ being evaluated at $r=aA$.

Differentiation of \eqref{eq:djdF} supplies the second derivative of the
surface Jacobian; contracted with an arbitrary second-order tensor $\bm{T}$, it
gives
\bea \label{eq:d2Js}
\frac{\partial^2 J_s}{\partial \bm{F}_s \partial \bm{F}_s}: \bm{T} = J_s \left[\bm{F}_s^{-T} \left(\bm{F}_s^{-T}:\bm{T}  \right)
-\bm{F}_s^{-T}\bm{T}^{T}\bm{F}_s^{-T} +(\bm{n}\otimes\bm{n})\bm{T}\bm{F}_s^{-1}\bm{F}_s^{-T}
\right].
\eea
The surface modulus $\mathbb{A}_s$ introduced in
\eqref{eq:stiff_surf} then follows from the surface energy \eqref{eq:suradd} as
\bea \label{eq:As_expl}
\mathbb{A}_s = \left( \gamma + \alpha_s \left( J_s - 1 \right) \right)
\frac{\partial^2 J_s}{\partial \bm{F}_s \partial \bm{F}_s}
+ \alpha_s J_s^2  \bm{F}_s^{-T} \otimes \bm{F}_s^{-T} .
\eea

Evaluated at $\bar{\bm F}_s$ and inserted, together with $\bm\eta_s$, into \eqref{eq:Ps_incr} and \eqref{eq:icr}, this gives the nonzero components $(\chi_s)_{ij}$ of the updated incremental surface stress tensor $\bm{\chi}_s$,
\bea\label{eq:chis_comp}
\begin{aligned}
(\chi_s)_{11} &= \left(\gamma-\alpha_s+2a\alpha_s\lambda\right)v_{,z} +\frac{a\alpha_s\lambda}{r}\,u, \\ 
(\chi_s)_{22} &= a\alpha_s\lambda\,v_{,z} +\frac{\gamma-\alpha_s+2a\alpha_s\lambda}{r}\,u, \\ 
(\chi_s)_{32} &= \left(\gamma-\alpha_s+a\alpha_s\lambda\right)u_{,z}, \end{aligned} 
\eea
evaluated at $r=aA$. The tensor $\bm{\tau}=\left(\bm{a}^{[1]}\bar{\bm{n}}\right)\bar{\bm{i}}_s+\bm{\chi}_s$ of \cref{sec:bif} has, by \eqref{eq:a_coeffs} and \eqref{eq:chis_comp}, the nonzero components
\bea \label{eq:tau_str}
\begin{aligned} 
\tau_{11} &=\left(\gamma-\alpha_s+2a\alpha_s\lambda\right)v_{,z} +\frac{a\alpha_s\lambda}{r}\,u+ \frac{a^2\xi}{\lambda r^2} \left(ru_{,r}-u\right), \\ 
\tau_{22} &= a\alpha_s\lambda\,v_{,z} +\frac{\gamma-\alpha_s+2a\alpha_s\lambda}{r}\,u+\lambda\xi\,v_{,zr}, \\ 
\tau_{32} &= \left(\gamma-\alpha_s+a\alpha_s\lambda\right)u_{,z}+\lambda\xi\,u_{,zr}. 
\end{aligned} 
\eea

\section*{Declaration of generative AI and AI-assisted technologies in the manuscript preparation process}
During the preparation of this work, the authors used Claude (Anthropic)
and Codex (OpenAI) in order to improve the language and structure of the
text, to edit the LaTeX source, and to check derivations and
cross-references. After using these tools, the authors reviewed and
edited the content as needed and take full responsibility for the
content of the published article.

\bibliographystyle{model5-names}
\bibliography{mybibfile}

\end{document}